\documentclass[12pt]{article}
\usepackage{amssymb}
\begin{document}
\title{\LARGE\bf 
Algebras satisfying Triality and $S_{4}$-symmetry \\}
\date{}
\author{\large Noriaki Kamiya$^1$ and Susumu Okubo$^2$\\ \\
$^1$Department of Mathematics, 
University of Aizu, \\
Aizuwakamatsu, Japan\\
$^2$Department of Physics and Astronomy, University of Rochester,\\
Rochester, N.Y,  U.S.A}
\maketitle
\thispagestyle{empty}
\vskip 2mm
{\bf Abstract} 
\par
\vskip 2mm
We give a review of recent works for nonassociative algebras,
especially Lie algebras satisfying the triality relation.
They are also intimately related to
$S_{4}$ (symmetric group of $4$-objects)
symmetry of the Lie algebras.
\par
\vskip 2mm
{\bf Keywords}
\par
\vskip 2mm
Structurable algebras, Freudenthal-Kantor triple systems,
Lie algebras and superalgebras, Triality, $S_{4}$-symmetry.
\vskip 5mm
{\bf Contents}
\par
\vskip 1mm
\begin{description}
\item[\S1.]
Symmetric triality algebras
\item[\S2.]
Examples of normal triality algebras
\item[\S3.]
Lie algebras satisfying triality
\item[\S4.]
Lie algebras satisfying tetrahedral symmetry
\item[\S5.]
Prestructurable algebras
\item[\S6.]
Kantor triple systems and $A$-ternary algebras
\item[\S7.]
Lie algebras and superalgebras associated with
$(\varepsilon,\delta)$
Freudenthal-Kantor triple systems
\item[\S8.]
$BC_{1}$graded Lie algebra of type $B_{1}$
\item[\S9.]
Final comments
\item[\ ]
References
\end{description}
\par
\noindent
E-mail adresses:
\par
kamiya@u-aizu.ac.jp\ (N.Kamiya)
\par
okubo@pas.rochester.edu\ (S.Okubo)
\par
\newpage
{\bf 1. Symmetric Triality Algebras}
\par
\vskip 3mm
This note is intented to be a brief review of recent works related to
 triality relations as well as to $A_{4}$ or $S_{4}$-symmetry
(alternative or symmetric group of $4$-objects)
satisfied by some algebras,
especially, Lie algebras.
\par
More precisely,
we introduce the notion of 
triality algebras and
describe constructions 
of Lie algebras
or superalgebras
from them.
\par
Let $A$ be an algebra over a field $F$ with bi-linear
 product denoted by juxtaposition $xy$.
Suppose that some $t_{j}\in\  {\rm  End}\  A$
for $j=0,1,2$
satisfy the symmetric triality relation
$$
t_{j}(xy)=(t_{j+1}x)y+x(t_{j+2}y)\eqno(1.1)
$$
for any $x,y\in A,$
where indices $j$ are defined modulo $3$, i.e.
$$
t_{j\pm 3}=t_{j}.
$$
We then call the triple
 $t=(t_{1},t_{2},t_{3})\in ({\rm End}\ A)^{3}$ 
be a symmetric Lie-related triple 
([O.05]) and set
$$
s\circ Lrt(A)=\{t=(t_{1},t_{2},t_{3})|t_{j}(xy)=(t_{j+1}x)y+x(t_{j+2}y)\}.\eqno(1.2)
$$
For any two
 $t, t^{'}\in s\circ Lrt(A),\ T_{j,k}\in \ {\rm End}\ A$ defined by
$$
T_{j,k}=[t_{j},t^{'}_{k}]:=t_{j}t^{'}_{k}-t^{'}_{k}t_{j},\ (j,k=0,1,2) \eqno(1.3a)
$$
then satisfy
$$
T_{j,k}(xy)=(T_{j+1,k+1}x)y+x(T_{j+2,k+2}y). \eqno(1.3b)
$$
Especially $s\circ Lrt(A)$ is a Lie algebra with respect to 
the component-wise commutation relation.
 Moreover, it is endowed with a natural order $3$ 
automorphysm $\theta$ given by 
$$
\theta(t_{0},t_{1},t_{2})=(t_{2},t_{0},t_{1}).
$$
It is a generalization of the derivation Lie algebra
$$
Der(A)=\{d|d(xy)=(dx)y+x(dy),\ \forall x,y\in A,\ d\in {\rm End} A\}.
$$
For any constants $\lambda_{j}\in F$
 satisfying the condition
 $\lambda_{j\pm3}=\lambda_{j}$,
 then 
$t^{'}=(t^{'}_{0},t^{'}_{1},t^{'}_{2})$ 
defined by
$$
t^{'}_{j}=\sum^{2}_{k=0}\lambda_{j-k}t_{k}
$$
belongs also to $t^{'}\in s\circ Lrt(A),$ 
when we note
$$
t^{'}=(t^{'}_{0},t^{'}_{1},t^{'}_{2})=
\sum^{2}_{j=0}\lambda_{j}\theta^{j}(t_{0},t_{1},t_{2}).
$$
Especially, 
for the choice of 
$\lambda_{0}=\lambda_{1}=\lambda_{2}=1,$
$$
d=t_{0}+t_{1}+t_{2} \eqno(1.4)
$$
is a derivation of $A$, i.e. $d\in {\rm Der}\ (A).$
\par
Suppose now that $A$ is also 
involutive with the involution map
 $x\rightarrow {\overline  x}$,
 satisfying
$$
\overline{\overline x}=x,\ \bar{x}\bar{y}=\overline{yx}.\eqno(1.5)
$$
For any
 $Q\in \ {\rm End}\  A,$ 
we introduce
 ${\overline Q}\in \  {\rm End}\ A$
 by 
$$
\overline{Qx}={\overline Q}{\bar x}.\eqno(1.6)
$$
We then note that for any two 
$Q_{1},Q_{2}\in \ {\rm End}\ A,$ 
we have 
$$
\overline{Q_{1}Q_{2}}=
{\overline Q_{1}}{\overline Q_{2}}.\eqno(1.7)
$$
Taking the involution of
 Eq.(1.1)
 and then letting
$x\leftrightarrow {\overline y}.$
it gives 
$$
{\overline t}_{j}(xy)=({\overline t}_{j+2}x)y+x({\overline t}_{j+1}y)
\eqno(1.8)$$
so that
$$
({\overline t}_{0},{\overline t}_{2},{\overline t}_{1})\in
s\circ Lrt(A).\eqno(1.9)
$$
If we introduce $\sigma \in \ {\rm End}(s\circ Lrt(A))$
 by
$\sigma(t_{0},t_{1},t_{2})=({\overline t_{0}},{\overline t_{2}},{\overline t_{1}}),$
then $\sigma$ and $\theta$
 generate automorphism  group
 $S_{3}$ of
 $s\circ Lrt(A)$,
 since we have
 $\theta \sigma\theta =\sigma,$ and $\theta^{3}=\sigma^{2}=id.$
\par
We next introduce the second bi-linear
 product in the vector space of
 $A$ by
$$
x\star y:=\overline{ xy}={\bar y}\ {\bar x}.\eqno(1.10)
$$
Then the resulting algebra
$A^ {*}$
 which we call the conjugate algebra of
 $A$ is also involutive, i.e.,
$$
\overline{x\star y}={\bar y}\star {\bar x}(=xy)\eqno(1.11)
$$
where Eq.(1.8)
 is rewritten as the Lie-related triple relation
([A-F.93])).
$$
{\overline t}_{j}(x\star y)=
(t_{j+1}x)\star y+x\star (t_{j+2}y).\eqno(1.12)
$$
A reason for introducing 
$A^{*}$
 is due to the following consideration.
\par
If $A$ is a unital algebra over the field 
$F$ of characteristic not $2$, 
then it is easy to show that we have 
$t_{0}=t_{1}=t_{2}$ for $s\circ Lrt(A)$
 and hence
 $s\circ Lrt(A)\simeq {\rm Der}(A),$
since for the unit element
$e$
of $A$,
it holds
$
t_{j}(xe)=
t_{j+1}(x)e+
xt_{j+2}(e)$
and
$t_{j}(e)=0,\ 
j=0,1,2.$
\par
However,
 $A^{*}$ 
could be unital,
satisfying
 $e\star x=x\star e=x.$
 Then the relation is immediately translated into
 $A$ to yield
$$
ex=xe={\bar x}.\eqno(1.13)
$$
We call
 $e\in A$
 satisfying Eq.(1.13) be the para-unit of
 $A$.
\par
Introducing multiplication operators in
 $A$
 and 
$A^{*}$
 by
$$
L(x)y=xy,\quad R(x)y=yx,\eqno(1.14a)
$$
$$
l(x)y=x\star y,\quad r(x)y=y\star x,\eqno(1.14b)
$$
they satisfy
$$
L(x)R(y)=r({\bar x})r(y)\eqno(1.15a)
$$
$$
R(x)L(y)=l({\bar x})l(y).\eqno(1.15b)
$$
\par
We note that an algebra $A$ could have more than
one involution. Moreover, it is often easier to deal with
$A$ rather than $A^{*}$
and
we will discuss mostly relations involving $A$
in this section,
although they can be readily translated into those of 
$A^{*}.$
\par
\vskip 3mm
{\bf Lemma 1.1}
\par
\vskip 3mm
For any $(t_{0},t_{1},t_{2})\in s\circ Lrt(A)$,
we have
$$
[t_{j},L(x)R(y)]=
L(x)R(t_{j+1}y)+L(t_{j+1}x)R(y)\eqno(1.16a)
$$
$$
[t_{j},R(x)L(y)]=
R(x)L(t_{j+2}y)+R(t_{j+2}x)L(y).\eqno(1.16b)
$$
\par
\vskip 3mm
{\bf Proof}
\par
\vskip 3mm
We can rewrite Eq.(1.1) as
$$
t_{j}L(x)=L(x)t_{j+2}+L(t_{j+1}x),\eqno(1.17a)
$$
$$
t_{j}R(y)=
R(y)t_{j+1}+R(t_{j+2}y).\eqno(1.17b)
$$
Multiplying $R(y)$ to Eq.(1.17a) from the right and
$L(x)$ to Eq.(1.17b)
from the left,
we obtain
$$
\begin{array}{l}
t_{j}L(x)R(y)=L(x)t_{j+2}R(y)+L (t_{j+1}x)R(y),\\
L(x)t_{j}R(y)=L(x)R(y)t_{j+1}+L(x)R(t_{j+2}y).
\end{array}
$$
Letting 
$j \to j+2$
in the 2nd relation and adding it to the 
first one
this yield Eq.(1.16a).
Similarly from Eqs.(1.17),
we find
$$
\begin{array}{l}
R(y)t_{j}L(x)=
R(y)L(x)t_{j+2}+R(y)L(t_{j+1}x),\\
t_{j}R(y)L(x)=
R(y)t_{j+1}L(x)+R(t_{j+2}y)L(x).
\end{array}
$$
Letting $j\rightarrow j+1$
in the first relation and adding it to the
second one, we obtain Eq.(1.16b).$\square$
\par
\vskip 3mm
{\bf Def.1.2}
\par
\vskip 3mm
Let $A$ be an algebra which possess 
$d_{j}(x,y)\in \mbox{End}\ A$
for $j=0,1,2$
and for
$x,y\in {\it A},$
satisfying
\par
\noindent
(1)
$$
d_{j}(y,x)=-d_{j}(x,y)\eqno(1.18a)$$
(2)
$$
d_{1}(x,y)=R(y)L(x)-R(x)L(y)\eqno(1.18b)$$
$$
d_{2}(x,y)=L(y)R(x)-L(x)R(y)\eqno(1.18c)$$
(3)
$$
(d_{0}(x,y),d_{1}(x,y),d_{2}(x,y))\in s\circ Lrt({\it A}),$$
i.e, we have
$$
d_{j}(x,y)(uv)=(d_{j+1}(x,y)u)v+u(d_{j+2}(x,y)v).\eqno(1.19)
$$
We call $A$ then be a regular triality algebra.
Note that a explicit form for $d_{0}(x,y)$
is {\it not}
specified at all.
\par
If $A$ satisfies further
\par
\noindent
(4)
$$[d_{j}(u,v),d_{k}(x,y)]=
d_{k}(d_{j-k}(u,v)x,y)+
d_{k}(x,d_{j-k}(u,v)y),\eqno(1.20)$$
and
\par
\noindent
(5)
$$
d_{0}(x,y)z+
d_{0}(y,z)x+
d_{0}(z,x)y=0\eqno(1.21)$$
in addition for any
$j,k=0,1,2$
and any $u,v,x,y\in A,$
then $A$ is called a 
pre-normal triality algebra.
\par
The reason for introducing these definitions is due to 
the following considerations.
To this end,
we introduce
\par
\vskip 3mm
{\bf Condition (B)}
\par
\vskip 3mm
We have $AA=A.$
\par
\vskip 3mm
{\bf Condition (C)}
\par
\vskip 3mm
If some $b\in A$
satisfies either
$bA=0$
or 
$Ab=0,$
then $b=0.$
\par
We can now prove:
\par
\vskip 3mm
{\bf Proposition 1.3}
\par
\vskip 3mm
Let $A$ be a regular triality algebra satisfying
the condition (C).
Then,
$A$ is a pre-normal triality algebra.
More generally,
we obtain the followings:
\par
If either condition (B) or
(C)
holds valid,
we have
\par
\begin{description}
\item[(1)]
$$
[t_{j},d_{k}(x,y)]=
d_{k}(t_{j-k}x,y)+
d_{k}(x,t_{j-k}y)\eqno(1.22)$$
for any $t=(t_{0}, t_{1}, t_{2}) \in \ s\ \circ Lrt\ (A).$
Especially for a choice of
$t_{j}=d_{j}(u,v),$
this implies the
validity of
Eq.(1.20).
\item[(2)]
$d_{0}(x,y)$
is uniquely
determined by Eqs.(1.18)
and (1.19).
\item[(3)]
If $A$
is involutive in addition with the 
involution map
$x\rightarrow {\bar x},$
we have
$$
\overline{d_{j}(x,y)}=d_{3-j}({\bar x},{\bar y}).\eqno(1.23)
$$
\par
\item[(4)]
 Finally,
if we assume the condition
(C),
then
$d_{0}(x,y)$
satisfies Eq.(1.21).
\par
\end{description}
\vskip 3mm
{\bf Proof}
\par
\vskip 3mm
For a proof of this Proposition, we first set
$$
D_{j,k}:=
[t_{j},d_{k}(x,y)]-
d_{k}(t_{j-k}x,y)-
d_{k}(x,t_{j-k}y).\eqno(1.24)$$
Then,
Lemma 1.1 immediately gives
$D_{j,1}=D_{j,2}=0$
identically
for any
$j=0,1,2.$
Moreover Eq.(1.3b)
for $t^{'}_{k}=d_{k}(x,y)$
together with
Eq.(1.19)
leads to
$$
D_{j,k}(uv)=
(D_{j+1,k+1}u)v+u(D_{j+2,k+2}v).\eqno(1.25)$$
Setting $k=0,$
we find
$D_{j,0}(uv)=0$
which gives $D_{j,0} =0$ if the condition
(B) holds.
If we choose $k=1$
or $2$,
then Eq.(1.25) implies
$$
u(D_{j+2,0}v)=0=
(D_{j+1,0}u)v$$
for any
$j=0,1,2,$
and for any $u,v\in A$.
Therefore,
under the condition (C),
this gives
$D_{j,0}=0$
again,
proving the validity of
Eq.(1.22).
\par
Next,
the uniqueness of 
$d_{0}(x,y)$
can be similarly proven as follows.
Suppose that
Eq.(1.18)
and (1.19)
allow the second solution for
$d_{0}(x,y)$
 which we write as 
$d^{'}_{0}(x,y).$
Then,
$$
(D_{0},D_{1},D_{2}):=
(d_{0}(x,y)-d^{'}_{0}(x,y),0,0)\in s\circ Lrt(A)$$
so that
$$
D_{j}(uv)=
(D_{j+1}u)v+u(D_{j+2}v).$$
Choosing $j=0,1$
or $2,$ and
repeating the same
reasoning,
this gives
$D_{0}=0,$
i,e.,
$d^{'}_{0}(x,y)=d_{0}(x,y).$
\par
If $A$
is involutive,
we have
$$
\overline{ L(x)}=
R({\bar x}),\ 
\overline{ R(x)}=
L({\bar x}),\eqno(1.26)
$$
so that
$$
\overline{ d_{1}(x,y)}=
d_{2}({\bar x},{\bar y}),
\ \mbox{and}\ 
\overline{ d_{2}(x,y)}=
d_{1}({\bar x},{\bar y})$$
which satisfy Eq.(1.23) for
$j=1$
and $2$.
In order to show its validity for
$j=0,$
we set
$$
\tilde{D}_{j}:=
\overline{d_{j}(x,y)}-
d_{3-j}({\overline x},{\overline y})\quad
(j=0,1,2)
$$
so that 
$\tilde{D}_{1}=\tilde{D}_{2}=0.$
Moreover,
Eqs.(1.1)
and (1.8)
imply now that we have
$$
\tilde{D}_{j}(xy)=
(\tilde{D}_{j+2}x)y+
x(\tilde{D}_{j+1}y).$$
Repeating again the
same augument,
we obtain
$\tilde{D}_{0}=0.$
\par
Finally,
let us set 
$$
\Lambda(x)=
\left(
\begin{array}{cc}
0,&R(x)\\
L(x),&0
\end{array}
\right),
D_{j}(x,y)=
\left(
\begin{array}{cc}
d_{j}(x,y),&0\\
0,&d_{j+1}(x,y)
\end{array}
\right).
\eqno(1.27)$$
Then,
Eq.(1.19)
with Eqs.(1.18)
are equivalent to the validity of
$$
[D_{j}(x,y),\Lambda(z)]=
\Lambda(d_{j+2}(x,y)z)\eqno(1.28a)
$$
if we note Eqs.(1.17) for
$t_{j}=d_{j}(u,v).$
Further,
we see
$$
[\Lambda(x),\Lambda(y)]=
-D_{1}(x,y)\eqno(1.28b)$$
so that
$$
[\Lambda(z),[\Lambda(x),\Lambda(y)]]=
\Lambda(d_{0}(x,y)z).\eqno(1.28c)$$
If we set
$$
w=d_{0}(x,y)z+
d_{0}(y,z)x+
d_{0}(z,x)y,$$
then the Jacobi identity among
$\Lambda(x)^{'}$s
leads to
$\Lambda(w)=0,$
or $R(w)=L(w)=0$
so that we have
$w=0$
under the condition (C).
This completes the proof of Proposition 1.3.$\square$
\par
We also note the Eq.(1.20)
gives
$$
[D_{j}(u,v),D_{k}(x,y)]=
D_{k}(d_{j-k}(u,v)x,y)+
D_{k}(x,d_{j-k}(u,v)y).\eqno(1.29)$$
Therefore,
$\Lambda(z)$
and $D_{j}(x,y)$
form a Lie algebra,
although we will not go into details.
\par
\vskip 3mm
Let
$$
g(A)=d_{0}(A,A)+d_{1}(A,A)+
d_{2}(A,A).\eqno(1.30)$$
Then,
Eqs.(1.20) and
(1.22)
imply that $g(A)$
is a Lie algebra
which is a ideal of the larger Lie algebra
$s\circ Lrt(A).$
Moreover,
if we set
$j=k$
in Eq.(1.20),
we obtain
$$
[d_{j}(u,v),d_{j}(x,y)]=
d_{j}(d_{0}(u,v)x,y)+
d_{j}(x,d_{0}(u,v)y)
\eqno(1.31)
$$
so that
$d_{j}(A,A)$
for each $j=0,1,2$
is also a Lie algebra which
is a ideal of $g(A)$.
Therefore if
$g(A)$
is simple,
and
if $d_{j}(A,A)\not= 0,$
then we must have
$$
g(A)=d_{0}(A,A)=
d_{1}(A,A)=
d_{2}(A,A).$$
\par
\vskip 3mm
{\bf Corollary 1.4}
\par
\vskip 3mm
Let 
$A$ be a pre-normal triality algebra.
Then,
the triple product defined by
$$
[xyz]:=d_{0}(x,y)z$$
is a Lie triple product,
i.e.,
it satisfies
$$
\begin{array}{cl}
(i)&[x,y,z]=-[y,x,z]\\
(ii)&[x,y,z]+[y,z,x]+[z,x,y]=0\\
(iii)&[u,v,[x,y,z]]=
[[u,v,x],y,z]+
[x,[u,v,y],z]+
[x,y,[u,v,z]].
\end{array}
$$
\par
\vskip 3mm
{\bf Proof}
\par
\vskip 3mm
First,
(i) follows trivially since
$d_{0}(y,x)=
-d_{0}(x,y),$
while
(ii)
is a 
consequence of Eq.(1.21).
Finally,
(iii)
is equivalent 
to the validity of Eq.(1.31) for $j=0.$
$\square$
\par
We next set
$$
D(x,y):=
d_{0}(x,y)+
d_{1}(x,y)+d_{2}(x,y).\eqno(1.32)$$
Then,
$D(x,y)$
is a derivation of $A$ as
$d$ of Eq.(1.4).
We further introduce 
$Q(x,y,z)\in \ {\rm End}\ A$
by
$$
Q(x,y,z):=
d_{0}(x,yz)+
d_{1}(z,xy)+
d_{2}(y,zx).
\eqno(1.33)$$
\par
\vskip 3mm
{\bf Proposition 1.5}
\par
\noindent
\vskip 3mm
(1)\quad
If $A$ is a regular triality algebra,
then
$$
(Q(x,y,z),Q(y,z,x),Q(z,x,y))\in
s\circ Lrt({\it A}),i.e.,$$
$$
Q(x,y,z)(uv)=
(Q(y,z,x)u)v+
u(Q(z,x,y)v).\eqno(1.34)$$
Also,we have
$$
Q(x,y,z)+
Q(y,z,x)+
Q(z,x,y)=
D(x,yz)+D(y,zx)+
D(z,xy).\eqno(1.35)
$$
(2)\quad
Moreover,
if $A$  is pre-normal triality algebra,
then
$$
Q(x,y,z)w=Q(w,y,z)x.\eqno(1.36)
$$
Further,
if $A$ is involutive with the validity of Eq.(1.23) in addition, it
satisfies also
$$
\overline{Q(x,y,z)}=
Q({\bar x},{\bar z},{\bar y}).
\eqno(1.37)
$$
\par
\vskip 3mm
{\bf Proof}
\par
\vskip 3mm
By Eq.(1.19),
we calculate
$$
\begin{array}{l}
d_{0}(x,yz)(uv)=
(d_{1}(x,yz)u)v+
u(d_{2}(x,yz)u)\\
d_{1}(z,xy)(uv)=
(d_{2}(z,xy)u)v+
u(d_{0}(z,xy)v)\\
d_{2}(y,zx)(uv)=
(d_{0}(y,zx)u)v+
u(d_{1}(y,zx)v).
\end{array}
$$
Adding all of these,
we obtain
Eq.(1.34).
Similarly for
Eq.(1.35).
Since Eq.(1.23)
gives
$\overline{d_{j}(x,y)}=
d_{3-j}(\overline{x},\overline{y}),$
Eq.(1.37) follows immediately from Eq.(1.33).
\par
Finally in order to prove
Eq.(1.36),
we calculate
$$
Q(x,y,z)w-Q(w,y,z)x=
\{d_{0}(x,yz)w-d_{0}(w,yz)x\}$$
$$
+\{d_{1}(z,xy)+d_{2}(y,zx)\}w-
\{d_{1}(z,wy)+d_{2}(y,zw)\}x,\eqno(1.38)
$$
and note
$$
\begin{array}{l}
d_{0}(x,yz)w-d_{0}(w,yz)x=-d_{0}(yz,x)w-d_{0}(w,yz)x\\
=d_{0}(x,w)(yz)
=\{d_{1}(x,w)y\}z+y\{d_{2}(x,w)z\}
\end{array}
$$
in view of Eqs.(1.21) and
(1.19) for
$j=0.$
Then,
Eq.(1.38)
becomes
$$
\begin{array}{l}
Q(x,y,z)w-Q(w,y,z)x\\
=\{d_{1}(x,w)y\}z+y\{d_{2}(x,w)z\}+\{d_{1}(z,xy)+d_{2}(y,zx)\}w\\
-\{d_{1}(z,wy)+d_{2}(y,zw)\}x\\
=\{(R(w)L(x)-R(x)L(w))y\}z+y\{(L(w)R(x)-L(z)R(w))z\}\\
+\{R(xy)L(z)-R(z)L(xy)+L(zx)R(y)-L(y)R(zx)\}w\\
-\{R(wy)L(z)-R(z)L(wy)+L(zw)R(y)-L(y)R(zw)\}x\\
=\{(xy)w\}z-\{(wy)x\}z+
y\{w(zx)\}-y\{x(zw)\}\\
+(zw)(xy)-\{(xy)w\}z+(zx)(wy)-y\{w(zx)\}\\
-(zx)(wy)+\{(wy)x\}z-(zw)(xy)+y\{x(zw)\}=0
\end{array}
$$
identically.
This completes the proof.$\square$
\par
We note that Eq.(1.34),(1.36)
and (1.37)
are consitent with the ansatz of
$Q(x,y,z)=0,$
and we further define the following.
\par
\vskip 3mm
{\bf Def.1.6}
\par
\vskip 3mm
We call a pre-normal triality algebra be a normal triality
algebra if it satisfies
$Q(x,y,z)=0$
in addition.
The conjugate algebra
$A^{*}$ of a normal
triality algebra $A$
satisfying
Eq.(1.23) is called a normal Lie-related 
triality algebra
(normal Lrt. algebra).
More explicitly,
it is defined by
\par
\noindent
(i)
$$
d_{1}(x,y)=l(\bar{y})l(x)-l(\bar{x})l(y),\eqno(1.39a)
$$
$$
d_{2}(x,y)=
r(\bar{y})r(x)-r(\bar{x})r(y)\eqno(1.39b)
$$
(ii)
$$\overline{d_{j}(x,y)}(u\star v)=
(d_{j+1}(x,y)u)\star v+
u\star (d_{j+2}(x,y)v)\eqno(1.39c)
$$
(iii)
$$
d_{0}(x,y)z+d_{0}(y,z)x+
d_{0}(z,x)y=0\eqno(1.39d)
$$
(iv)
$$
[d_{j}(u,v),d_{k}(x,y)]=
d_{k}(d_{j-k}(u,v)x,y)+
d_{k}(x,d_{j-k}(u,v)y)\eqno(1.39e)
$$
(v)
$$
Q(x,y,z)=d_{0}(x,\overline{y\star z})+
d_{1}(z,\overline{x\star y})+
d_{2}(y,\overline{z\star x})=0\eqno(1.39f)
$$
(vi)
$$
\overline{d_{j}(x,y)}=d_{3-j}(\overline{x},\overline{y}).\eqno(1.39g)
$$
\par
We note that Eqs.(1.39 a-f)
are simple rewriting of the corresponding relations for
the normal triality algebra $A$,
when we note,
for example,
Eqs.(1.15) for Eq.s.(1.39).
If $A^{*}$ is
unital with the unit element $e$,
then both conditions (B) and
(C)
are automatically satisfied,
because,
by
$xe=ex={\bar x}$
for any $x$,
and by
$be=0\Rightarrow b=0.$
Then,
we can omit Eqs.(1.39,d,e,and g)
since they are consequence of other postulates 
by Proposition 1.3.
Moreover,
if we set
$y=e$
or $z=e$
in Eq.(1.39f)
or alternately if we set
$u=e$
or $v=e$
in Eq.(1.39c),
then
$d_{0}(x,y)$
is determined 
to be
$$ 
d_{0}(x,y)=r(\overline{x}\star y-\overline{y}\star x)+
l(y)l(\overline{x})-
l(x)l(\overline{y})\eqno(1.40a)
$$
$$
=l(y\star  \overline{x}-x\star \overline{y})
+r(y)r(\overline{x})-
r(x)r(\overline{y}).\eqno(1.40b)
$$
We can then redifine the structurable algebra of Allison
[A.78]
to be a unital normal Lrt. algebra
(see [O.05]):
\par
\vskip 3mm
{\bf Def.1.7}
\par
\vskip 3mm
A pre-structurable algebra $A^{*}$
is a unital involutive algebra satisfying
\par
\noindent
(i)
$$
d_{1}(x,y)=l({\overline y})l(x)-l({\overline x})l(y),\eqno(1.41a)$$
$$
d_{2}(x,y)=r({\overline y})r(x)-r({\overline x})r(y),\eqno(1.41b)$$
$$
d_{0}(x,y)=r({\overline x}\star y-{\overline y}\star x)
+l(y)l({\overline x})
-l(x)l({\overline y})\eqno(1.41c)$$
$$
=l(y\star {\overline x}-x\star {\overline y})+r(y)r({\overline x})-
r(x)r({\overline y})$$
(ii)
$$
\overline{ d_{j}(x,y)}(u\star v)=
(d_{j+1}(x,y)u)\star v+
u\star(d_{j+2}(x,y)v).\eqno(1.41d)$$
Moreover if it satisfies the additional condition
\par
\noindent
(iii)
$$
Q(x,y,z)=
d_{0}(x,\overline{y\star z})+d_{1}(z,\overline{x\star y})
+d_{2}(y,\overline{z\star x})=0,\eqno(1.42)$$
then we call
${\it A}^{*}$ be a structurable algebra. ([K-O.14])
\par
\vskip 3mm
{\bf Remark 1.8}
\par
If $A$ is a normal triality algebra,
then
$D(x,y)$ defined by
Eq.(1.32) is a derivation satisfying
$$
D(x,yz)+
D(y,zx)+
D(z,xy)=0\eqno(1.43)$$
in view of Eq.(1.35).
In [Kam.95],
any algebra 
$A$ which posseses a derivation
$D(x,y)=-D(y,x)$
satisfying
Eq.(1.43)
has been called a generalized structurable algebra.
Therefore,
any normal triality algebra is a generalized structurable algebra if $D(x,y)$
is {\it not}
trivial.
Note that there exists a triality algebra
with
$D(x,y)=0$
identically
(see Eq.(2.20)).
\par
Many interesting algebra such as Malcev,
structurable,
admissible cubic algebra
([E-O.00])
and pseudo-composition algebra
[M-O.93]
are known to be generalized structurable
algebras.
(see [Kam.95],[O.05]).
\par
\vskip 3mm
{\bf Remark 1.9}
\par
\vskip 3mm
We can generalize the idea to super-algebra ([K-O.00]).
Let
$A$ be
$Z_{2}$-graded as
$$
A=A_{\overline 0}\oplus
A_{\overline 1}.\eqno(1.44)
$$
We write for simplicity
$$
(-1)^{x}=(-1)^{grad\ x},\eqno(1.45a)
$$
where
$$
{\rm grad}
\ x=
\left\{
\begin{array}{lll}
0,&{\rm if} &x\in A_{\overline 0}\\
1,&{\rm if} &x\in {\it A}_{\overline 1}.
\end{array}
\right.
\eqno(1.45b)
$$
Then,
we replace the definition for
$d_{j}(x,y)^{'}$s as
$$
d_{1}(x,y)=
(-1)^{xy}R(y)L(x)-R(x)L(y)\eqno(1.46a)$$
$$
d_{2}(x,y)=
(-1)^{xy}L(y)R(x)-L(x)R(y)\eqno(1.46b)$$
while the triality relation
Eq.(1.19)
must be replaced by
$$
d_{j}(x,y)(uv)=
(d_{j+1}(x,y)u)v+
(-1)^{(x+y)u}u(d_{j+2}(x,y)v)\eqno(1.47)$$
etc.
Then,
all statements so for given in
this section will proceed accordingly.
\par
\vskip 3mm
{\bf 2.\ Examples of Normal Triality Algebras}
\par
\vskip 3mm
{\bf Example 2.1,(Lie and Jordan algebra)}
\par
\vskip 3mm
Both Lie and Jordan algebras are normal triality algebras.
Writing the bi-linear product of these algebras as
$xy$,
we have
$$
xy=\varepsilon yx\eqno(2.1)$$
for
$\varepsilon=+1$
or $-1$,
respectively for Jordan or Lie algebra,
so that 
$$
L(x)=\varepsilon R(x).$$
Setting then
$$
d(x,y):=
d_{0}(x,y)=
d_{1}(x,y)=
d_{2}(x,y)=
-\varepsilon[L(x),L(y)],\eqno(2.2)$$
it is a inner  derivation of these algebra,
satisfying ([Kam.95])
$$
Q(x,y,z)=d(x,yz)+d(y,zx)+d(z,xy)=0.$$
Moreover Eq.(1.21)
is a cosequence of the Jacobi identity for Lie,
while it is
trivially satisfied for the case of the
Jordan algebra.
\par
Moreover,
they are involutive with the involution
$$
{\overline x}=+\varepsilon x,$$
so that they are also normal Lrt algebra with
$x\star y=\overline{xy}=\varepsilon xy.$
\par
\vskip 3mm
{\bf Example 2.2,(Symmetric Composition Algebras)}
\par
\vskip 3mm
Let $A$ be an algebra with symmetric bi-linear non-degenerate form
$<\cdot|\cdot>$
over the field $F$ of charachteristic $\not=2.$
Suppose that we have 
$$
x(yx)=(xy)x=<x|x>y,\eqno(2.3)$$
for $x,y\in A$.
Then,
{\it A} is known as a symmetric composition algebra,
since then it satisfies also
$$
<xy|xy>=<x|x><y|y>,\ 
<xy|z>=<x|yz>.\eqno(2.4)$$
Conversely the validity of Eq.(2.4) gives Eq.(2.3)
([O-O.81]).
Moreover,
a symmetric composition algebra is either
a para-Hurwitz
algebra or a eight-dimensional pseudo-octonion algebra.
([O-O.81], [O.95])
\par
Here,
the para-Hurwitz algebra is the conjugate algebra of the
Hurwitz
(i.e. unital composition)
algebra.
Any symmetric composition algebra satisfy
the triality relation for the choice of
$$
d_{0}(x,y)=2\{[L(x),L(y)]-R([x,y])\}\eqno(2.5a)$$
or equivalently by
$$
d_{0}(x,y)z=
4\{<x|z>y-<y|z>x\},\eqno(2.5b)$$
as has been noted in
([KMRT.98]
and [E.97]),
and it is a normal triality algebra ([O.05]).
\par
We also note that the para-Hurwitz algebra has
the para-unit $e$ but the pseudo-octonion
algebra possesses neither unit nor para-unit.
\par
\vskip 3mm
{\bf Example 2.3,(Tensor product)}
\par
\vskip 3mm
Let $A_{1}$ and $A_{2}$ be
two independent symmetric composition algebras.
Then,
their tensor product
$A_{1}\otimes A_{2}$
is normal triality algebra with
(see [O.05])
$$
D_{j}(x_{1}\otimes x_{2},y_{1}\otimes y_{2}):=
d^{(1)}_{j}(x_{1},y_{1})\otimes
<x_{2}|y_{2}>_{2}\mbox{id}+
<x_{1}|y_{1}>_{1}\mbox{id}  
\otimes d^{(2)}_{j}(x_{2},y_{2}),\eqno(2.6)$$
for $x_{1},y_{1}\in A_{1}$
and $x_{2},y_{2}\in A_{2}.$
\par
As we will show in the next section,
this case is relevant for a construction of the so-called
Freudenthal's magic square.
\par
\vskip 3mm
{\bf Example 2.4}
\par
\vskip 3mm
Let $A$ be a normal triality algebra with a order $3$
automorphism
$\Phi(i.,e,\Phi^{3}=1).$
Suppose that it also satisfies
$$
\Phi d_{0}(x,y)\Phi^{-1}=
d_{0}(\Phi x,\Phi y),\eqno(2.7)
$$
which holds automatically if the condition (B) or (C) is valid.
If we introduce then a new bi-linear product in the
same
vector space
$A$
by
$$
x\circ y=(\Phi x)(\Phi^{2}y)\eqno(2.8)$$
then the resulting new algebra
$A^{(\circ)}$
is a normal triality algebra
([E-O.07]),
so that a symmetric composition algebra
$A$ is transformed into another
symmetric composition algebra
$A^{(\circ)}$([E.97]).
\par
As an example,
consider the
$so(3)$ Le algebra:
$$
e_{i}e_{j}=\Sigma^{3}_{k=1}\epsilon_{ijk} e_{k}\quad
(i,j=1,2,3)$$
for a Levi-Civita symbol $\epsilon_{ijk}$.
Since it is a Lie algebra, it is a normal symmetric triality algebra by
Example 2.1.
Moreover,
$\Phi \in \mbox{End}(so(3))$
defined by
$$
\Phi:e_{1}\rightarrow e_{2}\rightarrow e_{3}\rightarrow e_{1}$$
is its order
$3$ automorphysm.
We then calculate the new product 
to satisfy
\begin{description}
\item[(1)]
$e_{1}\circ e_{1}=e_{1},\quad
e_{2}\circ e_{2}=e_{2},\quad
e_{3}\circ e_{3}=e_{3}$
\item[(2)]
$
e_{1}\circ e_{2}=-e_{3},\quad
e_{2}\circ e_{3}=-e_{1},\quad
e_{3}\circ e_{1}=-e_{2}
\hskip 60mm
{\rm(2.9)}$
\item[(3)]
$
e_{2}\circ e_{1}=e_{1}\circ e_{3}=
e_{3}\circ e_{2}=0$
\end{description}
\par
\noindent
as in [O.05].
This algebra has some interesting property.
We intoduce the bi-linear symmetric non-degenerate
form
$<\circ|\circ>$ by
$$
<e_{i}|e_{j}>=\delta_{ij}\quad
(i,j=1,2,3).\eqno(2.10)$$
Then,
it is a normal triality algebra with
$d_{0}(x,y)$
given by
$$
d_{0}(x,y)z=
<x|z>y-<y|z>x.\eqno(2.11)
$$
 Moreover,
we have
$$
(x\circ x)\circ(x\circ x)=<x|x\circ x>x\eqno(2.12)$$
so that the 3rd
bi-linear product defined by
$$
x\cdot y=\frac{1}{2}(x\circ y+y\circ x)$$
gives a 3-dimensional 
admissible-cubic
 algebra
([E-O.06]).
Moreover for
$x=\lambda_{1}e_{1}+\lambda_{2}e_{2}+\lambda_{3}e_{3},\ 
(\lambda_{j}\in F),$
we set
$$
t(x)=\lambda_{1}+\lambda_{2}+\lambda_{3},\quad
q(x)=
\lambda_{1}\lambda_{2}+\lambda_{1}\lambda_{3}+
\lambda_{2}\lambda_{3}$$
with
$f=e_{1}+e_{2}+e_{3}.$
Then,
they satisfy 
quadratic relation of ([O.06])
\begin{description}
\item
[(i)]
$$
f\circ f=0$$
\item[(ii)]
$$
x\circ x-t(x)x+q(x)f=0.\eqno(2.13)
$$
\end{description}
\par 
\vskip 3mm
{\bf Example 2.5}
\par
\vskip 3mm
Let
$$
{\it A}=\mbox{span}
<e,f,x_{\mu},x^{\mu},
\ (\mu=1,2,\cdots ,N)>$$
with the multiplication table of
\begin{description}
\item[(1)] $ee=e\ ,ff=e,\ ef=fe=-f$
\item[(2)] $ex_{\mu}=x_{\mu}e=x_{\mu},\ x^{\mu}e=ex^{\mu}=x^{\mu}$
\item[(3)] $fx_{\mu}=-x_{\mu}f=x_{\mu},\ fx^{\mu}=-x^{\mu}f=-x^{\mu}$
\item[(4)] $x_{\mu}x_{\nu}=0=x^{\mu}x^{\nu}$
\item[(5)] $x^{\mu}x_{\nu}=-2\delta^{\mu}_{\nu}(f+e)$
\item[(6)] $x_{\nu}x^{\mu}=2\delta^{\mu}_{\nu}(f-e)$
\end{description}
\par
\noindent
for $\mu, \nu=1,2,\cdots N.$
Then $A$ is a normal triality algebra.
Note that
{\it A} possesses a few involution maps:
\par
\vskip 3mm
{\bf Involution 1} 
\par
\vskip 3mm
${\overline f}=-f,$
but ${\overline x}=x,$
for
$x=e, x^{\mu},$
and $x_{\mu}.$
\par
\vskip 3mm
{\bf Involution 2}
\par
\vskip 3mm
$\overline{ x^{\mu}}=x_{\mu},\ \overline {x_{\mu}}=x^{\mu},$
but ${\overline x}=x,$
for $x=e,$
and $f$.
\par
\vskip 3mm
{\bf Involution 3}
\par
\vskip 3mm
${\overline e}=e,$
but ${\overline x}=-x$
for $x=f,\ x_{\mu}$
and
$x^{\mu}.$
\par
\vskip 3mm
The case of the involution 1 is of interest,
since then it satisfies
$ex=xe={\overline x}$
so that $e$ is the para-unit of {\it A}.Then,
its conjugate algebra
${\it A}^{*}$ is structurable.
\par
In section 4,
we will show that
this algebra is intimately
related to the
$A_{4}$
or $S_{4}$
symmetry of the Lie algebra
$sl(N),\ (N\geq 4)$.
\par
\vskip 3mm
{\bf Example 2.6 (Structurable Algebra)}
\par
\vskip 3mm
It is known ([A-F.93])
that any unital involutive alternative or Jordan algebra is structurable.
Especially,
any unital composition algebra as well as any unital involutive
associative algebra is structurable.
Moreover some class of Zorn's
vector matrix
algebras are also structurable.
Let $B$ be a involutive algebra over a field
$F$ with bi-linear product 
$xy$ and with a bi-linear form
$(\circ|\circ),$
and consider a vector space of form
$$
{\it A}=\left(\begin{array}{ll}
F&{\it B}\\
{\it B}&F
\end{array}
\right).\eqno(2.14)$$
\par
Designating a generic element of
{\it A} as
$$
X=
\left(\begin{array}{ll}
\alpha& x\\
y&\beta
\end{array}
\right),
\ (x,y\in {\it B},\ \alpha,\beta\in F)\eqno(2.15)
$$
we introduce a bi-linear product in
{\it A} by
$$
X_{1}\star X_{2}=
$$
$$
\left(\begin{array}{ll}
\alpha_{1}&x_{1}\\
y_{1}&\beta_{1}
\end{array}
\right)
\star
\left(\begin{array}{ll}
\alpha_{2}&x_{2}\\
y_{2}&\beta_{2}\end{array}
\right)=
\left(
\begin{array}{ll}
\alpha_{1}\alpha_{2}+(x_{1}|y_{2}),&
\alpha_{1}x_{2}+\beta_{2}x_{1}+
ky_{1}y_{2}\\
\alpha_{2}y_{1}+\beta_{1}y_{2}+kx_{1}x_{2},&
\beta_{1}\beta_{2}+(y_{1}|x_{2})
\end{array}
\right)\eqno(2.16)
$$
for a constant 
$k\in F$ and for
variables
$\alpha_{j},\beta_{j}\in F$
and
$x_{j},y_{j}\in {\it B}(j=1,2).$
Then,
$$
X\rightarrow \overline{X}=
\left(\begin{array}{ll}
\beta&{\bar x}\\
{\bar y}&\alpha
\end{array}
\right)
\eqno(2.17)$$
is a involution map of
$A$,
provided that
$(\circ|\circ)$
satisfies
$$
({\bar x}|y)=
({\bar y}|x),\ 
(=\mbox{symmetric in}  \ x \ 
\mbox{and} 
\ y).\eqno(2.18)$$
If $B$ is a commutative  cubic-admissible algebra over the field $F$
of characteristic
$\not=2$,
and $\not=3,$
satisfying
$$
x^{2}x^{2}=<x|x^{2}>x,\mbox{with}\ 
(x|y)=3<x|y>,\eqno(2.19)$$
then
$A$ is known to be structurable for
the choice
$k=2$
([O.05]).
\par
As an example,
consider the case of $\mbox{Dim} B=1$
with
$B=Fb,$
where $b\in B$
satisfies
$$
bb=b,\ <b|b>=1.$$
\par
If we set now
$$
e=\left(
\begin{array}{ll}
1&0\\
0&1
\end{array}
\right),
f=
\left(
\begin{array}{ll}
1&0\\
0&-1
\end{array}
\right),
g=
\left(
\begin{array}{ll}
0&0\\
b&0
\end{array}
\right),
h=
\left(
\begin{array}{ll}
0&b\\
0&0\end{array}
\right),$$
then
$e$ is the unit element of 
$A^{*}$
so that
$e\star x=x\star e=x,$
and the multiplication table 
is given by
$$
f\star f=e,\ 
f\star g=-g\star f=-g,\ f\star h=-h\star f=h,$$
$$
g\star g=2h,\ h\star h=2g,\ 
g\star h=\frac{3}{2}(e-f),\ 
h\star g=
\frac{3}{2}(e+f)\eqno(2.20)$$
as in [O,06].
A peculiar aspect of this algebra is
that we have
$D(x,y)=0$
identically.
Further,
$A^{*}$ admitts few involutions:
\begin{description}
\item[(1)] ${\bar f}=-f,$
but ${\bar x}=x$
for $x=e,g,$
and $h$, corresponding to Eq.(2.17).
\item[(2)] 
${\overline g}=h,{\overline h}=g,$
but ${\overline x}=x$
for $x=e$
and $f$.
\item[(3)] 
${\overline e}=e,$
but 
${\overline x}=-x$
for $x=f,g,$
and $h.$
\end{description}
\par
On the other side,
if {\it B}
is an anti-commutative algebra,
then
{\it A} is an alternative algebra,
provided that we have
$$
\begin{array}{l}
x(yz)=(x|y)z-(x|z)y,\\
(x|yz)=(y|zx)=(z|xy)
\end{array}
$$
with
${\overline x}=-x$
and
$k=1.$
This case yields the octonion algebra
as well as a unconventional six-dimensional degenerate composition
algebra associated with a five-dimentional Malcev
algebra 
[K-O,14],
although we will not go into
its details here.
\par
\vskip 3mm
{\bf 3 Lie Algebra satisfying Triality}
\par
\vskip 3mm
Let $A$ be a pre-normal triality algebra
as in Def.1.2,
and consider linear maps:
$$
\rho_{j}:A \rightarrow V,\mbox{and}
\ T_{j}:A\otimes A\rightarrow V
\eqno(3.1)
$$
for $j=0,1,2,$
where $V$ is an unspecified algebra with
skew symmetric
bi-linear product
$[\circ,\circ].$
We set now
$$
T(A,A)=
\mbox{span}<T_{j}(x,y),
\forall j=0,1,2,
\forall x,y\in A>\eqno(3.2)
$$
and
$$
L(A)=
\rho_{0}(A)\oplus
\rho_{1}(A)\oplus
\rho_{2}(A)\oplus
T(A,A).\eqno(3.3)
$$
Following
[A-F,93],
let
$(i,j,k)$
be a cyclic permutation of indices
$(0,1,2)$,
and assume the following anti-communtative
multiplication relations:
\par
\noindent
(1)
$$
[\rho_{i}(x),\rho_{i}(y)]=
-[\rho_{i}(y),\rho_{i}(x)]=
\gamma_{j}\gamma_{k}^{-1}T_{3-i}(x,y)\eqno(3.4a)
$$
(2)
$$
[\rho_{i}(x),\rho_{j}(y)]=
-[\rho_{j}(y),\rho_{i}(x)]=
-\gamma_{j}\gamma_{i}^{-1}\rho_{k}(xy)\eqno(3.4b)$$
(3)
$$
[T_{l}(x,y),\rho_{j}(z)]=
-[\rho_{j}(z),T_{l}(x,y)]=
\rho_{j}(d_{l+j}(x,y)z)\eqno(3.4c)
$$
(4)
$$
[T_{l}(u,v),T_{m}(x,y)]=
T_{m}(d_{l-m}(u,v)x,y)+
T_{m}(x,d_{l-m}(u,v)y)$$
$$
=-T_{l}(d_{m-l}(x,y)u,v)-
T_{l}(u,d_{m-l}(x,y)v)\eqno(3.4d)$$
for
$l,m=0,1,2.$
Here,
$\gamma_{j}\in F$
are some non-zero constants.
We introduce the Jacobian in $L(A)$
by
$$
J(X,Y,Z)=
[[X,Y],Z]+
[[Y,Z],X]+
[[Z,X],Y]\eqno(3.5)
$$
for
$X,Y,Z\in L(A).$
\par
We first prove.
\par
\vskip 3mm
{\bf Lemma 3.1}
\par
\vskip 3mm
$T(A,A)$
and 
$T_{j}(A,A)$
for
$j=0,1,2$
are Lie algebras.
Also $T_{j}(A,A)$ is an ideal of $T(A,A)$.
\par
\vskip 3mm
{\bf Proof}
\par
\vskip 3mm
We calculate
now for any $j,k,l = 0,1,2,$
$$
\begin{array}{l}
[[T_{j}(u,v),T_{k}(x,y)],T_{l}(z,w)]=
-[T_{l}(z,w),[T_{j}(u,v),T_{k}(x,y)]]\\
=-[T_{l}(z,w),T_{k}(d_{j-k}(u,v)x,y)+
T_{k}(x,d_{j-k}(u,v)y)]\\
=-T_{k}(d_{l-k}(z,w)d_{j-k}(u,v)x,y)-
T_{k}(d_{j-k}(u,v)x,
d_{l-k}(u,v)y)\\
+T_{k}(d_{l-k}(z,w)x,d_{j-k}(u,v)y)-T_{k}(x,d_{l-k}(z,w)d_{j-k}(u,v)y)\\
\end{array}
$$
$$
\begin{array}{l}
[[T_{l}(z,w),T_{j}(u,v)],T_{k}(x,y)]=
-[[T_{j}(u,v),T_{l}(z,w)],T_{k}(x,y)]\\
=-[T_{l}(d_{j-l}(u,v)z,w)+
T_{l}(z,d_{j-l}(u,v)w),
T_{k}(x,y)]\\
=-T_{k}(d_{l-k}(d_{j-l}(u,v)z,w)x,y)-T_{k}(x,d_{l-k}(d_{j-l}(u,v)z,w)y)\\
-T_{k}(d_{l-k}(z,d_{j-l}(u,v)w)x,y)-
T_{k}(x,d_{l-k}(z,d_{j-l}(u,v)w)y)
\end{array}
$$
and
$$
\begin{array}{l}
[[T_{k}(x,y),T_{l}(z,w)],
T_{j}(u,v)]=
[T_{j}(u,v),[T_{l}(z,w),T_{k}(x,y)]]\\
=[T_{j}(u,v),T_{k}(d_{l-k}(z,w)x,y)+
T_{k}(x,d_{l-k}(z,w)y)]\\
=T_{k}(d_{j-k}(u,v)d_{l-k}(z,w)x,y)+
T_{k}(d_{l-k}(z,w)x,
d_{j-k}(u,v)y)\\
+T_{k}(d_{j-k}(u,v)x,
d_{l-k}(z,w)y)+
T_{k}(x,d_{j-k}(u,v)
d_{l-k}(z,w)y).
\end{array}
$$
Adding these three relations we find
$$
J(T_{j}(u,v),T_{k}(x,y),T_{l}(z,w))=
T_{k}(\lambda x,y)+
T_{k}(x,\lambda y)$$
where
$\lambda $
is given by
$$
\lambda=
[d_{j-k}(u,v),d_{l-k}(z,w)]-
d_{l-k}(d_{j-l}(u,v)z,w)-
d_{l-k}(z,d_{j-l}(u,v)w)=0$$
by the triality Lie relation
Eq.(1.20).$\square$
\par 
We next set
$$
J(x,y,z):=
J(\rho_{0}(x),\rho_{1}(y),\rho_{2}(z)).\eqno(3.6)
$$
\par
\vskip 3mm
{\bf Lemma 3.2}
\par
\vskip 3mm
We have
$$
J(x,y,z)=T_{0}(x,yz)+
T_{1}(z,xy)+
T_{2}(y,zx)
\eqno(3.7)$$
which satisfies
$$
[J(x,y,z),\rho_{i}(w)]=
\rho_{i}(Q(z,x,y)w)\eqno(3.8a)$$
$$
[J(x,y,z),T_{l}(u,v)]=
T_{l}(Q_{l}(z,x,y)u,v)+
T_{l}(u,Q_{l}(z,x,y)v)\eqno(3.8b)
$$
where we have set
$$
Q_{l}(z,x,y)=
d_{-l}(z,xy)+
d_{1-l}(y,zx)+
d_{2-l}(x,yz).\eqno(3.8c)
$$
Note the 
$Q_{0}(z,x,y)=
Q(z,x,y),\ 
Q_{1}(z,x,y)=Q(y,z,x)$ etc.
\par
\vskip 3mm
{\bf Proof}
\par
\vskip 3mm
These are straightforward results of
Eqs.(3.4).$\square$
\par
\vskip 3mm
{\bf Proposition 3.3}
\par
\vskip 3mm
Let
$A$ be a prenormal triality algebra.
Then we have
$J(X,Y,Z)=0$
for
$X,Y$
and $Z$ being any one of forms
$\rho_{i}(z)$
or
$T_{j}(x,y)$
except for
$J(\rho_{i}(x),\rho_{j}(y),\rho_{k}(z))$
or
$J(\rho_{0}(x),\rho_{1}(y),\rho_{2}(z)).$
\par
\vskip 3mm
{\bf Proof}
\par
\vskip 3mm
(1)\ We calculate
$$
\begin{array}{l}
[[\rho_{i}(x),\rho_{i}(y)],\rho_{i}(z)]=
[\gamma_{j}\gamma^{-1}_{k}T_{3-i}(x,y),\rho_{i}(z)]\\
=\gamma_{j}\gamma^{-1}_{k}\rho_{i}(d_{3}(x,y)z)=
\gamma_{j}\gamma^{-1}_{k}\rho_{i}(d_{0}(x,y)z)
\end{array}
$$
so that we have
$$J(\rho_{i}(x),\rho_{i}(y),\rho_{i}(z))=
\gamma_{j}\gamma^{-1}_{k}\rho_{i}(w)$$
with
$$
w=d_{0}(x,y)z+
d_{0}(y,z)x+
d_{0}(z,x)y=0$$
by Eq.(1.21).
Thus,
we have
$J(\rho_{i}(x),\rho_{i}(y),\rho_{i}(z))=0.$
\par
\vskip 3mm
(2)\ 
We similarly compute
$$
\begin{array}{l}
[[\rho_{i}(x),\rho_{i}(y)],
\rho_{j}(z)]=
[\gamma_{j}\gamma^{-1}_{k}T_{3-i}(x,y),\rho_{j}(z)]\\
=\gamma_{j}\gamma^{-1}_{k}\rho_{j}(d_{3-i+j}(x,y))=
\gamma_{j}\gamma^{-1}_{k}\rho_{j}(d_{1}(x,y)z)
\end{array}
$$
when we note
$j-i=1$(mod $3$)
since $(i,j,k)$ is a cyclic pertation of $(0,1,2).$
Further we note
$$
\begin{array}{l}
[[\rho_{i}(y),\rho_{j}(z)],
\rho_{i}(x)]=
[-\gamma_{i}\gamma^{-1}_{i}\rho_{k}(yz),\rho_{i}(x)]\\
=(-\gamma_{j}\gamma^{-1}_{i})
(-\gamma_{i}\gamma^{-1}_{k})
\rho_{j}((yz)x)=
\gamma_{j}\gamma^{-1}_{k}\rho_{j}((yz)x)
\end{array}
$$
so that
$$
J(\rho_{i}(x),\rho_{i}(y),\rho_{j}(z))=
\gamma_{j}\gamma^{-1}_{k}\rho_{j}(w)$$
with
$$
w=
d_{1}(x,y)z+
(yz)x-(xz)y=
\{d_{1}(x,y)+
R(x)L(y)-
R(y)L(x)\}z=0$$
by
Eq.(1.18b).
This shows
$J(\rho_{i}(x),\rho_{i}(y),\rho_{j}(z))=0.$
\par
\vskip 3mm
(3)\ 
We analogously compute
$$
\begin{array}{l}
[[\rho_{i}(x),\rho_{i}(y)],\rho_{k}(z)]=
[\gamma_{j}\gamma^{-1}_{k}T_{3-i}(x,y),
\rho_{k}(z)]\\
=\gamma_{j}\gamma^{-1}_{k}\rho_{k}
(d_{3-i+k}(x,y)z)=
\gamma_{j}\gamma^{-1}_{k}(d_{2}(x,y)z)
\end{array}
$$
since
$k-i=2$(mod $3$),
while
$$
\begin{array}{l}
[[\rho_{i}(y),\rho_{k}(z)],\rho_{i}(x)]=
-[[\rho_{k}(z),\rho_{i}(y)],\rho_{i}(x)]\\
=\gamma_{i}\gamma^{-1}_{k}
[\rho_{j}(zy),\rho_{i}(x)]=
-\gamma_{i}\gamma^{-1}_{k}
[\rho_{i}(x),\rho_{j}(zy)]\\
=(-\gamma_{i}\gamma^{-1}_{k})
(-\gamma_{j}\gamma^{-1}_{i})
\rho_{k}(x(zy))=
\gamma_{j}\gamma^{-1}_{k}\rho_{k}(x(zy)).
\end{array}
$$
In this way, we obtain
$$
J(\rho_{i}(x),\rho_{i}(y),\rho_{k}(z))=
\gamma_{j}\gamma^{-1}_{k}\rho_{k}(w)$$
where
$$
w=d_{2}(x,y)z+
x(zy)-y(zx)=
\{d_{2}(x,y)+
L(x)R(y)-
L(y)R(x)\}z=0,
$$
by Eq.(1.18c).
\par
\vskip 3mm
(4)\ However,
$$
[[\rho_{i}(x),\rho_{j}(y)],\rho_{k}(z)]=
[-\gamma_{j}\gamma^{-1}_{i}\rho_{k}(xy),\rho_{k}(z)]=
-\gamma_{j}\gamma^{-1}_{i}\gamma_{i}\gamma^{-1}_{j}T_{3-k}(xy,z)$$
$$
=-T_{3-k}(xy,z)=T_{3-k}(z,xy)$$
so that
$$
J(\rho_{i}(x),\rho_{j}(y),\rho_{k}(z))=
T_{3-k}(z,xy)+
T_{3-i}(x,yz)+
T_{3-i}(y,zx)\eqno(3.6)'$$
which gives
Eq.(3.7)
for
$i=0,j=1$
and $k=2.$
\par
\vskip 3mm
(5)\ We similarly compute
$$
\begin{array}{l}
[[\rho_{i}(x),\rho_{i}(y)],
T_{l}(u,v)]
=[\gamma_{j}\gamma^{-1}_{k}T_{3-i}(x,y),
T_{l}(u,v)]\\
=-\gamma_{j}\gamma^{-1}_{k}
\{T_{3-i}(d_{l+i}(u,v)x,y)+
T_{3-i}(x,d_{l+i}(u,v)y)\}
\end{array}
$$
and
$$
\begin{array}{l}
[[\rho_{i}(y),T_{l}(u,v)],\rho_{i}(x)]=
[-\rho_{i}(d_{i+l}(u,v)y),\rho_{i}(x)]\\
=\gamma_{j}\gamma^{-1}_{k}T_{3-i}
(x,d_{i+l}(u,v)y).
\end{array}
$$
Then,
we see
$J(\rho_{i}(x),\rho_{i}(y),
T_{l}(u,v))=0.$
\par
\vskip 3mm
(6)\ Moreover,
We note
$$
\begin{array}{l}
[[\rho_{i}(x),\rho_{j}(y)],T_{l}(u,v)]=
[-\gamma_{j}\gamma^{-1}_{i}\rho_{k}(xy),
T_{l}(u,v)]=
\gamma_{j}\gamma^{-1}_{i}
\rho_{k}(d_{l+k}(u,v)(xy)),\\

[[\rho_{j}(y),T_{l}(u,v)],
\rho_{i}(x)]=
[-\rho_{j}(d_{j+l}(u,v)y),
\rho_{i}(x)]=
-\gamma_{j}\gamma^{-1}_{i}\rho_{k}
(x\{d_{j+l}(u,v)y\}),

\end{array}
$$
and
$$
[[T_{l}(u,v),\rho_{i}(x)],\rho_{j}(y)]=
[\rho_{i}(d_{l+i}(u,v)x),
\rho_{j}(y)]=
-\gamma_{j}\gamma^{-1}_{i}\rho_{k}
(\{d_{l+i}(u,v)x\}y).
$$
Thus we obtain
$$
J(\rho_{i}(x),\rho_{j}(y),T_{l}(u,v))=
\gamma_{j}\gamma^{-1}_{i}\rho_{k}(w),$$
with
$$
w=d_{l+k}(u,v)(xy)-
x\{d_{j+l}(u,v)y\}-
\{d_{i+l}(u,v)x\}y.$$
But then
$w=0$ by the triality relation
Eq.(1.19).
\par
\vskip 3mm
(7)\ We similarly find
$$
J(\rho_{k}(x),T_{l}(u,v),
T_{m}(x,y))=
\rho_{k}(\lambda z)$$
with
$$
\begin{array}{ll}
\lambda&=[d_{k+m}(x,y),d_{k+l}(u,v)]+
d_{k+m}(d_{l-m}(u,v)x,y)+d_{k+m}(x, d_{l-m}(u,v)y)\\
&=0
\end{array}
$$
by Eq.(1.20).
\par
\vskip 3mm
(8)\ We have already noted in Lemma 3.1
that we have for any $j,k,l =0,1,2,$
$$
J(T_{j}(u,v),
T_{k}(x,y),
T_{l}(z,w))=0.\square
$$
\par
In this connection,
we consider
\par
\vskip 3mm
{\bf Condition (D)}
\par
\vskip 3mm
Suppose that we have
$\rho_{i}(x)=0$
for some
$x\in A$ and for some value of
$i=0,1,2.$
We then have
$x=0.$
\par
\vskip 3mm
{\bf  Corollary 3.4}
\par
\vskip 3mm
Let $A$ be a pre-normal triality algebra.
If we have
$$
J(x,y,z)=
T_{0}(x,yz)+
T_{1}(z,xy)+
T_{2}(y,zx)=0,\eqno(3.9)$$
then
$L(A)$ is a Lie algebra.
Moreover,
if the condition
$(D)$ holds,
then
$A$ is a normal triality algebra.
Conversly,
if $L(A)$ is a Lie algebra and if the condition
$(D)$ holds,
then $A$ is a normal triality
algebra with the validity of
Eq.(3.9).
\par
\vskip 3mm
{\bf Proof}
\par
\vskip 3mm
This follows from Lemma 3.2 as well as the proof given in
Proposition 3.3. $\square$
\par
If we do {\it not} assume the validity of Eq.(3.9),
we set
$$
J=\mbox{span}<J(x,y,z),x,y,z\in A>.\eqno(3.10)$$
If $A$ is a normal triality algebra,
then Lemma 3.2 implies that
$J(x,y,z)$
are center elements of
$A$,
since
$Q(x,y,z)=0.$
Then,
we find
\par
\vskip 3mm
{\bf Theorem 3.5}
\par
\vskip 3mm
Let $A$ be a normal triality algebra.
Then, the quotient algebra
${\tilde L}=L/J$ is a Lie algebra.
\par
Hereafter in this section,
we assume $A$ to be a normal triality algebra 
unless it is stated otherwise.
Then,
in view of Theorem 3.5 we can 
effectively assume the validity of
Eq.(3.9).
As a matter of fact,
if we identify
$T_{j}(x,y)$
with the triple
$$
T_{j}(x,y)=
(d_{j}(x,y),d_{j+1}(x,y),d_{j+2}(x,y))\quad
(j=0,1,2)\eqno(3.11)
$$
(see i.e. [A-F,93],
and [E.04]),
then we find
$$
T_{0}(x,yz)+
T_{1}(z,xy)+
T_{2}(y,zx)=
(Q(x,y,z),Q(y,z,x),Q(z,x,y))=0.$$
Moreover,
they will yield
$T_{0}(x,y)=
T_{1}(x,y)=
T_{2}(x,y)$
if we have
$d_{0}(x,y)=d_{1}(x,y)=
d_{2}(x,y)$
as in the case of Lie and Jordan algebra
(see Example 2.1).
This can be also justified without
assuming Eq.(3.11) as follows:
If
$d_{0}(x,y)=d_{1}(x,y)=d_{2}(x,y),$
then we see from
Eq.(3.4) that the differences
$T_{i}(x,y)-T_{j}(x,y)$
for
$i\not= j$ are center element of
$L(A)$ so that we can effectively set
$T_{i}(x,y)=T_{j}(x,y).$
This fact will be assumed and 
used in the next section for
$S_{4}$-symmetry of the Lie algebra $so(N).$
\par
We will assume also  for simplicity
the validity of
Eq.(3.9) or Eq.(3.11) hereafter unless it is stated otherwise.
\par
For the case of $A^{*}$ being a structurable algebra, we need
simply replace
$\rho_{k}(xy)$
in Eq.(3.4b)
by
$$
\rho_{k}(xy)\rightarrow \rho_{k}(\overline{x\star y})=
\rho_{k}({\overline y}\star {\overline x})\eqno(3.12a)$$
and Eq.(3.9) by
$$
J(x,y,z)=
T_{0}(x,\overline{y\star z})+
T_{1}(z,\overline{x\star y})+
T_{2}(y,\overline{z\star x})=0\eqno(3.12b)$$
according to Eq.(1.11) for the Lie algebra
$L(A).$
\par
Now,a special choice of
$\gamma_{0}=\gamma_{1}=\gamma_{2}=1$
for constants
$\gamma_{j}$ in Eqs.(3.4)
is of a particular interest,
since the Lie algebra
$L(A)$
will admit then an alternative group
(or equivalently
tetrahedral
group $T_{4}$)
$A_{4}$
as automorphysm.
\par
First ,
$L(A)$ is clearly invariant under actions of a cyclic group $Z_{3}$
generated by $\phi\in \mbox{End}\ L(A)$
given by
$$
\rho_{i}(x)\rightarrow \rho_{i+1}(x),\ T_{i}(x,y)\rightarrow
T_{i-1}(x,y).
\eqno(3.13)$$
\par
Next,
let
$\tau_{\mu}\in \mbox{End}\ L(A)$
for
$\mu=1,2,3$
be defined by
$$
\tau_{1}:\rho_{1}(x)\rightarrow \rho_{1}(x),\ 
\rho_{2}(x)\rightarrow -\rho_{2}(x),\ 
\rho_{3}(x)\rightarrow -\rho_{3}(x),$$
$$
\tau_{2}:\rho_{1}(x)\rightarrow -\rho_{1}(x),\ 
\rho_{2}(x)\rightarrow \rho_{2}(x),\ 
\rho_{3}(x)\rightarrow -\rho_{3}(x)
\eqno
(3.14)$$
$$
\tau_{3}:\rho_{1}(x)\rightarrow-\rho_{1}(x),\ 
\rho_{2}(x)\rightarrow-\rho_{2}(x),\ 
\rho_{3}(x)\rightarrow \rho_{3}(x)
$$
while $T_{j}(x,y)$ for $j=0,1,2$ remains unchanged by actions of 
$\tau_{\mu}$.
Then,
$L(A)$ is also invariant under $\tau_{\mu}.$
Moreover,
we note
$$
\tau_{\mu}\tau_{\nu}=\tau_{\nu}\tau_{\mu},\ 
\tau_{\mu}\tau_{\mu}=1,\ 
\tau_{1}\tau_{2}\tau_{3}=1,
\ (\mu , \nu=1,2,3)\eqno(3.15)$$
so that
$(1,\tau_{1},\tau_{2},\tau_{3})$
is isomorphic to the Klein's 
4-group $K_{4}.$
\par
Further,
we see
$$
\phi\tau_{\mu}\phi^{-1}=\tau_{\mu+1}\ (\mu=1,2,3)\eqno(3.16)$$
with
$\tau_{4}=\tau_{1}.$
Since
$Z_{3}$ and
$K_{4}$ generate the alternative group
${\it A}_{4}$
(an equivalently the tetrahedral group $T_{4}$),
the Lie algebra
$L(A)$
is invariant under
$A_{4}$.
\par
If $A$ is involutive with the involution map
$x\rightarrow{\overline x}$
in addition,
then
$\tau\in \mbox{End} \ L(A)$ given by
$$
\tau:\rho_{1}(x)\leftrightarrow -\rho_{2}({\overline x}),\ 
\rho_{3}(x)\rightarrow -\rho_{3}({\overline x}),
\eqno(3.17)$$
$$
T_{1}(x,y)\leftrightarrow T_{2}({\overline x},{\overline y}),\ 
T_{3}(x,y)\rightarrow
T_{3}({\overline x},{\overline y})
$$
also defines an automorphsm of $L(A)$
satisfying
$$
\tau^{2}=1,\ 
\tau\tau_{1}\tau^{-1}=\tau_{2},\ 
\tau\tau_{3}\tau^{-1}=\tau_{3},\ 
\phi\tau\phi=\tau.\eqno(3.18)$$
Then,
$\tau$
and
$A_{4}$
generate the symmetric group  $S_{4}$ with 
identifications of
$$
\tau_{1}=(2,3)(1,4),\ 
\tau_{2}=(3,1)(2,4),\ 
\tau_{3}=(1,2)(3,4),\ 
\phi=(1,2,3),\ 
\tau=(1,2)\eqno(3.19)$$
in the standard notation for symmetric group.
\par
Regarding
$L(A)$ as a $A_{4}$-module,
the triple
$(\rho_{0}(x),\rho_{1}(x),\rho_{2}(x))$
for any 
$x\in A$ realizes then a $3$-dimensional 
irreducible
module of
$A_{4}$.
For $T(A,A),$
we assume for simplicity,
that the underlying field
$F$
is of charachtericsitc
$\not=2,$
and $\not= 3$.
If $F$ contains $\omega\in F$
satisfying $\omega^{3}=1$
but $\omega\not= 1,$
then
$T(A,A)$
is a direct sum of three inequivalent
one-dimensional modules given by
$$
\varphi_{n}(x,y)=
T_{0}(x,y)+
\omega^{n}T_{1}(x,y)+
\omega^{2n}T_{2}(x,y)$$
for
$n=0,1,2.$
However,
if $F$ does {\it not} contains such
$\omega\in F$,
then,
$$\varphi_{0}(x,y)=
T_{0}(x,y)+
T_{1}(x,y)+T_{2}(x,y)$$
is the trivial module of
$A_{4}$ and
$(\theta_{1}(x,y),\theta_{2}(x,y))$
defined by
$$
\begin{array}{l}
\theta_{1}(x,y)=
T_{1}(x,y)+
T_{2}(x,y)-2T_{0}(x,y),\\
\theta_{2}(x,y)=
T_{1}(x,y)-T_{2}(x,y)\end{array}
$$
represents two-dimensional irreducible
module of $A_{4}$.
\par
The case of the
$S_{4}$-symmetry is 
slightly more involved,
since we have to take account of the action of
$\tau=(1,2)$
in addition.
In that case,
depending upon
${\overline x}=x$
or
${\overline x}=-x,$
the triple
$(\rho_{0}(x),\rho_{1}(x),
\rho_{2}(x)),$
represents two inequivalent
$3$-dimentional modules of
$S_{4}$,
while for
$T_{j}(A,A),$
we have to consider
$4$ cases of 
${\overline x}=\pm x$
and
${\overline y}=\pm y$
or ${\overline y}=\mp y$
to find two inequivalent 
two-dimensional modules
$(\theta_{1}(x,y),
\theta_{2}(x,y))$
and
one-dimentional
modules $\varphi_{0}(x,y)$
of $S_{4}.$
\par
Returning to the
structure of
$L(A),$
we set
$$
L_{j}(A)=
\rho_{j}(A)\oplus
T_{3-j}(A,A),\ 
(j=0,1,2).\eqno(3.20)
$$
\par
We have then   
$$
L(A)=
L_{0}(A)+
L_{1}(A)+
L_{2}(A).\eqno(3.21)
$$
As we see from Eqs.(3.4),
$L_{j}(A)\ (j=0,1,2)$
are sub-Lie algebras of
$L(A),$
while
$T_{3-j}(A,A)$
is a sub-Lie algebra of
$L_{j}(A).$,
Moreover,
under action of
$Z_{3},$
we have
$$
\begin{array}{ll}
\phi:&
L_{0}(A)\rightarrow
L_{1}(A)\rightarrow 
L_{2}(A)\rightarrow 
L_{0}(A)
\hskip 75mm(3.22)\\
&
T_{0}(A,A)\rightarrow
T_{2}(A,A)\rightarrow
T_{1}(A,A)\rightarrow
T_{0}(A,A)
\end{array}
$$
while they transform among themselves
under action of the
Klein's $4$-group $K_{4}.$
\par
It may be instructive to depict
$L({\it A})$
as in Fig.1, exhibiting the triality.
\par
\noindent
\vskip 5mm
\begin{center}
\unitlength 0.1in
\begin{picture}(41.50,27.19)(38.10,-42.15)
%
\special{pn 8}%
\special{ar 6000 3920 1320 1320  3.1340170 3.1415927}%
%
\special{pn 8}%
\special{ar 6670 3750 1290 1290  3.1415927 6.2831853}%
\special{ar 6670 3750 1290 1290  0.0000000 0.0155026}%
%
\special{pn 8}%
\special{ar 7330 2640 1298 1298  1.0659350 4.2210668}%
%
\special{pn 8}%
\special{ar 6030 2630 1314 1314  5.2541089 6.2831853}%
\special{ar 6030 2630 1314 1314  0.0000000 2.0861220}%
\put(50.3000,-26.2000){\makebox(0,0){$L(A)=$}}%
\put(66.1000,-22.2000){\makebox(0,0){$\rho_{0}(A)$}}%
\put(64.2000,-30.5000){\makebox(0,0)[lb]{$T(A,A)$}}%
\put(58.0000,-35.9000){\makebox(0,0){$\rho_{1}(A)$}}%
\put(72.7000,-36.0000){\makebox(0,0){$\rho_{2}(A)$}}%
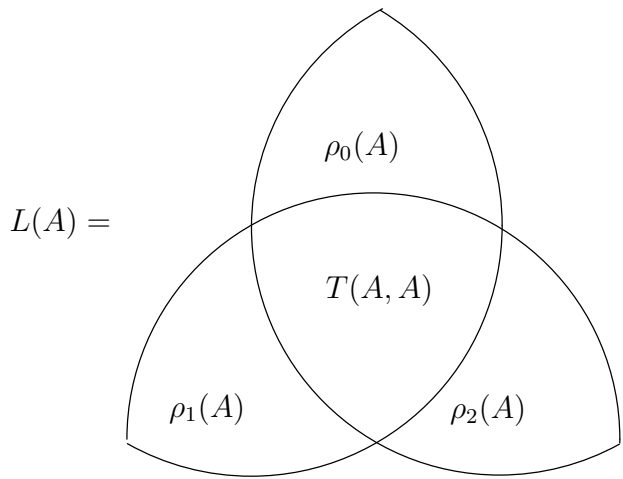
\put(66.9000,-43.0000){\makebox(0,0){Fig.1\ \it{Graphical Representation}\ of the Lie Algebra $L(A).$}}%
\end{picture}%
\end{center}
\par
\vskip 3mm
As illustration,
let us examine specific cases of Examples given in section 2,
assuming the underlying field $F$ to be algebraically
closed and of charateristic
$\not= 2,\not=3$
for simplicity.
\par
\vskip 3mm
{\bf Example 3.6(Lie algebra $G_{2}$)}
\par
\vskip 3mm
The $4$-dimensional structurable algebra
${\it A}^{*}=<e,f,g,h>$
given by
Eq.(2.20)
leads to
$$
L(A)=G_{2},\ L_{j}(A)=A_{1}\oplus A_{1}
(j=0,1,2)\ 
\mbox{and\ }
T_{j}(A,A)=
gl(1)\oplus
gl(1).\ (j=0,1,2)$$
as in [O.06].
\par
\vskip 3mm
{\bf Example 3.7(magic Square)}
\par
\vskip 3mm
Let $A=A_{1}\otimes A_{2}$
be the tensor product algebra of two
independest symmetric composition algebra as in Example 2.3.
Then,
$A$ is also a normal triality algebra,
and we can construct Lie algebras by Theorem 3.5.
Following [E.04]
and [E.06],
this leads to the
Freudenthal's magic square for the
Lie algebra
$L(A)$ as in Fig.2
(see also
[Ba-S.03]):
\begin{center}
\begin{tabular}{|c|c|c|c|c|} \hline
$
Dim\ A_{1} \setminus Dim\ A_{2}$ &1 &2 &4 &8\\ \hline
1&$A_{1}$ &$A_{2}$                    &$C_{3}$ &$F_{4}$\\ \hline
2&$A_{2}$ &$A_{2}\oplus A_{2}$ &$A_{5}$ &$E_{6}$\\ \hline
4&$C_{3}$ &$A_{5}$                  &$D_{6}$ &$E_{7}$\\ \hline
8&$F_{4}$ &$E_{6}$                    &$E_{7}$ &$E_{8}$\\ \hline
\end{tabular}
\end{center}
\vskip 2mm
\begin{center}
Fig.2:Magic Square
\end{center}
\par
\noindent
if $A_{1}$ is a para-octonion or pseudo-octonion algebra,
and if we choose $\mbox{Dim}\ A_{2}=1,$
then the resulting Lie algebras
are
$$
L(A)=F_{4},\ 
L_{j}(A)=B_{4},\ 
T_{j}(A,A)=D_{4}\ (j=0,1,2),$$
corresponding to the classical triality
case of
$A^{*}_{1}$
being octonion algebra.
\par
For other case of $\mbox{Dim}\ A_{1}=\mbox{Dim}\ A_{2}=8,$
where
$A_{1}$
and $A_{2}$
are either para-octonion or
pseudo-octonion algebra,
we obtain
$$
L(A)=E_{8},\ L_{j}(A)=D_{8},\ 
T_{j}(A,A)=D_{4}\oplus D_{4}.
$$
\par
\vskip 3mm
{\bf Example 3.8(Zorn's Vector Matrix Algebra)}
\par
\vskip 3mm
Let us consider
Example 2.6 again where
the algebra $B$ is now the $27$-dimensional cubic-admissible
algebra associated with the Albert algebra.
In that case,
it is known
([Kan 73] and [Kam 89])
that $L(A)$
is also the Lie algebra
$E_{8}$.
However,
we have
$$
L_{j}(A)=
E_{7}\oplus A_{1}\ \mbox{and}\ T_{j}(A,A)=E_{6}\oplus gl(1)\oplus gl(1)$$
in contrast to the previous case of
example 3.7.
\par
\vskip 3mm
{\bf Remark 3.9}
\par
\vskip 3mm
In ending this section,
we note that
any finite dimensional
normal triality algebra satisfying the condition
$(D)$ may be identified with
some symmetric space.
For example,
Eq.(3.4)
implies
$$
\begin{array}{l}
[\rho_{0}(x),\rho_{0}(y)]=
T_{0}(x,y),\\

[T_{0}(x,y),\rho_{0}(z)]=
\rho_{0}(d_{0}(x,y)z),\\

[T_{0}(u,v),T_{0}(x,y)]=
T_{0}(d_{0}(u,v)x,y)+
T_{0}(x,d_{0}(u,v)y)
\end{array}
$$
for a Lie algebra
$L_{0}(A)$,
so that we may identify
$\rho_{0}(A)$
with the symmetric space
$$
L_{0}(A)/T_{0}(A,A).$$
Moreover,
if the condition
$(D)$ for
$j=0$
is satisfied,
then
$\rho_{0}(x)=0$
for some
$x\in A$
implies
$x=0,$
so that
$A\rightarrow \rho_{0}(A)$
is one-to-one map.
Hence,
we can identify $A$ with the symmetric space.
\par
\vskip 3mm
{\bf 4.Tetrahedral Lie Algebras}
\par 
\vskip 3mm
In the previous section,
we have seen that we can construct a 
$A_{4}$-invariant Lie algebra out of a normal triality
algebra.
We will show in this section that
the converse statement holds valid also.
\par
Let $V$ be an algebra over a filed $F$ of charachteristic
$\not=2$,
endowed 
with a group homomorphism
$$
A_{4}\rightarrow \mbox{Auto}(V).\eqno(4.1)$$
Let $\phi$ and
$\tau_{\mu}(\mu=1,2,3)\in A_{4}$
as in Eq.(3.19),
satisfying relations
Eqs.(3.15),
 with
$\phi^{3}=1$.
Then,
$V$ can be decompsed by
actions of the Klein's $4$-group
$K_{4}=\{1,\tau_{1},\tau_{2},\tau_{3}\}$
into a direct sum
$$
V=t\oplus g_{1}\oplus g_{2}\oplus g_{3}\eqno(4.2)$$
where
$$
t=\{x\in V;\ \tau_{1}(x)=\tau_{2}(x)=\tau_{3}(x)=x\}
\eqno
(4.3a)
$$
$$
g_{1}=\{x\in V;\ \tau_{1}(x)=x,\ \tau_{2}(x)=\tau_{3}(x)=-x\}\eqno
(4.3b)
$$
$$
g_{2}=\{x\in V;\ \tau_{2}(x)=x,\ \tau_{1}(x)=\tau_{3}(x)=-x\}\eqno
(4.3c)
$$
$$
g_{3}=\{x\in V;\ \tau_{3}(x)=x,\ \tau_{1}(x)=\tau_{2}(x)=-x\}.\eqno
(4.3d)
$$
We then have 
\par
\vskip 3mm
{\bf Lemma 4.1}
\par
\vskip 3mm
\begin{description}
\item[(1)]
$$
\phi (g_{i})=g_{i+1}
(\mbox{with}\  g_{4}=g_{1})
\eqno
(4.4a)
$$
$$
\phi(t)=t
$$
\item[(2)]
$$
tt\subset t,\ \mbox{so that}\ 
t\ \mbox{is a subalgebra of}\ 
V
\eqno
(4,4b)
$$
\item[(3)]
$$ 
tg_{i}\subset g_{i},\ 
\mbox{and}\  
g_{i}t\subset g_{i}
\eqno
(4.4c)
$$
\item[(4)]
$$
g_{i}g_{i}\subset t
\eqno(4.4d)$$
\item[(5)]
If $(i,j,k)$
is a cyclic permutation of indices
$
(1,2,3),$ 
then
$$
g_{i}g_{j}\subset g_{k},\ 
\mbox{and}\ 
g_{j}g_{i}\subset g_{k}.
\eqno
(4.4e)
$$
\end{description}
\par
\vskip 3mm
{\bf Proof}
\par
\vskip 3mm
Noting
$\phi\tau_{\mu}=\tau_{\mu+1}\phi$
(with
$\tau_{4}=\tau_{1}$)
by Eq.(3.16),
we obtain
$$
\phi (g_{i})\subseteq g_{i+1},\ 
\mbox{and}\ 
\phi (t)\subseteq t.$$
For example,
if $x\in g_{1},$
then we calculate
$$
\tau_{2}\phi x=\phi \tau_{1}x=\phi x,\  
\tau_{3}\phi x=\phi \tau_{2} x=-\phi x,\  
\tau_{1}\phi x=\phi \tau_{3}x=
-\phi x$$
which gives
$\phi (g_{1})\subseteq g_{2}$.
Then we calculate
$$
g_{i}=
\phi^{3}(g_{i})
\subseteq
\phi^{2}(g_{i+1})
\subseteq
\phi (g_{i+2})
\subseteq 
g_{i},$$
which yields
$\phi (g_{i+2})=g_{i},$
\  
i.e. Eq.(4.4a).
\par
The rest of relations in Eq.(4.4)
can be similarly
verified,
when we note
$$
\tau_{\mu}(xy)=
(\tau_{\mu}x)(\tau_{\mu}y)$$
for $x,y\in V.\ \square$
\par
\vskip 3mm
{\bf Remark 4.2}
\par
\vskip 3mm
Setting
$$
V_{j}=t\oplus g_{j}\ (j=1,2,3),\eqno(4.5)
$$
we have
$$
V=V_{1}+V_{2}+V_{3}\eqno(4.6)
$$
and we may depict the situation as in Fig.3
\par
\vskip 5mm
\begin{center}
\unitlength 0.1in
\begin{picture}(32.80,26.39)(46.80,-41.35)
%
\special{pn 8}%
\special{ar 6000 3920 1320 1320  3.1340170 3.1415927}%
%
\special{pn 8}%
\special{ar 6670 3750 1290 1290  3.1415927 6.2831853}%
\special{ar 6670 3750 1290 1290  0.0000000 0.0155026}%
%
\special{pn 8}%
\special{ar 7330 2640 1298 1298  1.0659350 4.2210668}%
%
\special{pn 8}%
\special{ar 6030 2630 1314 1314  5.2541089 6.2831853}%
\special{ar 6030 2630 1314 1314  0.0000000 2.0861220}%
\put(66.8000,-22.6000){\makebox(0,0){$g_{1}$}}%
\put(66.9000,-30.7000){\makebox(0,0){$t$}}%
\put(59.6000,-35.3000){\makebox(0,0){$g_{2}$}}%
\put(74.7000,-35.3000){\makebox(0,0){$g_{3}$}}%
\put(49.4000,-26.1000){\makebox(0,0){$V=$}}%
\put(67.2000,-42.2000){\makebox(0,0){Fig.3\ \it{Graphical Representation of $V$.}}}%
\end{picture}%
\end{center}
\vskip 10mm
Note that $V_{j}$ $(j=1,2,3)$ 
are sub-algebras of $V$.
\par
{\bf Example 4.3}
\par
\vskip 3mm
Let $V$ be the Cayley algebra with the basis
$<e_{0},e_{1},e_{2},\cdots ,e_{7}>$
with the unit element $e=e_{0}$,
satisfying the multiplication table
of
$$
e_{i}e_{j}=
-\delta_{ij}e+
\sum^{7}_{k=1}f_{ijk}e_{k}$$
for
$i, j=1,2,\cdots,7,$
when $f_{ijk}$
is the totally anti-symmetric constants with
values
$1,0-1.$
Moreover,
$f_{ijk}=1$
are possible only for
$i,j,k=123,516,624,435,174,376,275$
with their cyclic
permutations.
We introduce a self-dual tensor 
$f_{\mu\nu}$
for
$\mu,\nu=1,2,3,4$
satisfying
$$
f_{\mu\nu}=-f_{\nu\mu}
=
\ ^{*}f_{\mu\nu}={1\over 2}
\sum^{4}_{\alpha,\beta=1}
\varepsilon_{\mu\nu\alpha\beta}f_{\alpha \beta}$$
(see [O,95]) by
$$
e_{1}=f_{23}=f_{14},\ 
e_{2}=f_{31}=f_{24},\ 
e_{3}=f_{12}=f_{34}
$$
where
$\varepsilon_{\mu\nu\alpha\beta}$
is the $4$-dimensional Levi-Civita symbol with
$\varepsilon_{1234}=1.$
\par
Moreover,setting
$$
a_{1}=e_{4},\ 
a_{2}=e_{5},\ 
a_{3}=e_{6},\ 
a_{4}=e_{7},$$
then
$a_{\mu}$
and $f_{\mu\nu}$ for
$\mu, \nu=1,2,3,4$
satisfy
$$
\begin{array}{rll}
a_{\mu}a_{\nu}
&=
&
-f_{\mu\nu}-\delta_{\mu\nu}e,
\\
f_{\mu\nu}a_{\lambda}
&=
&
-a_{\lambda}f_{\mu\nu}=
-\delta_{\mu\lambda}a_{\nu}+
\delta_{\nu\lambda}a_{\mu}-
\sum^{4}_{\alpha=1}\varepsilon_{\mu\nu\lambda \alpha}
e_{\alpha},
\\
f_{\mu \nu}f_{\alpha\beta}
&=
&
-\delta_{\nu\alpha}f_{\mu\beta}+\delta_{\nu\beta}f_{\mu\alpha}-
\delta_{\mu\beta}f_{\nu\alpha}+\delta_{\mu\alpha}f_{\nu\beta}
\\
& &
-(\delta_{\mu\alpha}\delta_{\nu\beta}-\delta_{\mu\beta}\delta_{\nu\alpha}+
\varepsilon_{\mu\nu\alpha\beta})e,
\end{array}
$$
which are clearly invariant under any even-permutations
of indices $1,2,3,$ and
$4$,
i.e.
under the alternative group
$A_{4}.$
We then find
$$
\begin{array}{rl}
g_{1}
&=\{e_{2},e_{4}-e_{5}-e_{6}+e_{7}\},
\\
g_{2}
&=\{e_{3},e_{4}+e_{5}-e_{6}-e_{7}\}
\\
g_{3}
&=\{e_{1},-e_{4}+e_{5}+e_{6}-e_{7}\},
\\
t
&=\{e,e_{4}+e_{5}+e_{6}+e_{7}\}
\end{array}
$$
by Eqs.(4.3).
Note that
$V_{j}=t\oplus g_{j}\ (j=1,2,3)$
are then quaternion
sub-algebras of the Cayley algebra.
\par
Actually,
the Cayley algebra is
invariant under
$S_{4},$
if we define
$\tau=(1,2)$
by
$$
\tau:e_{1}\rightarrow -e_{1},\ 
 e_{2}\rightarrow -e_{3}\rightarrow e_{2},$$
and
$$
e_{\mu}\rightarrow {1\over 2}(e_{4}+e_{5}+e_{6}+e_{7})-\widetilde{ e}_{\mu}, 
{\rm for } \  \mu=4,5,6,7,$$
where $\widetilde{e}_{4}=e_{5}, \widetilde{e}_{5}=e_{4}, \widetilde{e}_{6}=e_{6}, \ and \ \widetilde{e}_{7}=e_{7}. $
\par
Further,
any split Cayley algebra is also invariant under
$S_{4}.$
This fact has been used
in [E-O.08]
to show that all exceptional Lie algebras are
$S_{4}$-invariant.
\par
However,
the most interesting case is
obtained,
when $V$ is a Lie algebra,
as we see from the following Theorem
([E-O.07]).
\par
\vskip 3mm
{\bf Theorem 4.4}
\par
\vskip 3mm
Let $L$ be a Lie algebra over the field $F$ of charachteristic $\not=2,$
which is invariant under the action of the alternative group $A_{4}.$
Then,
there exists a normal triality algebra $A$
such that $L$ is written as a direct sum of
$$
L=\rho_{0}(A)\oplus\rho_{1}(A)\oplus \rho_{2}(A)\oplus
t\eqno(4.7)
$$
of some vector spaces
$\rho_{j}(A)$ and a sub-Lie algebra
$t$ of $L$.
Moreover,
there exits a sub-Lie algebra
$T(A,A)$
of $t$ such that
$$
\tilde{L}=
\rho_{0}(A)\oplus \rho_{1}(A)\oplus \rho_{2}(A)\oplus T(A,A)
\eqno(4.8)$$
is a $A_{4}$-invariant ideal of $L$,
which coincides with the Lie algebra
constructed in the previous section in terms of the normal triality algebra
$A$,
satisfying Eq.(3.4)
for 
$\gamma_{0}=\gamma_{1}=\gamma_{2}=1$
as well as Eq.(3.9),i.e,
$$
T_{0}(x,yz)+
T_{1}(z,xy)+T_{2}(y,zx)=0.\eqno(4.9)
$$
Further,
if $L$ is invariant under a larger group
$S_{4}$,
then $A$ is involutive with a involutive map
$x\rightarrow {\overline x}.$
\par
\vskip 3mm
{\bf Proof}
\par
\vskip 3mm
We identify $A$ with $g_{3}$
in Eq.(4.3d),i.e,
$$
A=\{x\in L;\ \tau_{3}(x)=x,\ 
\tau_{1}(x)=\tau_{2}(x)=-x\}\eqno(4.10)$$
and write
$$
\rho_{3}(x)=\rho_{0}(x)=x,\ \mbox{if}\ x\in A\eqno(4.11a)
$$
and set
$$
\rho_{1}(x)=\phi x,\ \rho_{2}(x)=\phi ^{2}x,\ \mbox{for}\ x\in A
\eqno(4.11b)
$$
so that we have
$$
\phi \rho_{j}(x)=
\rho_{j+1}(x)\eqno(4.12)$$
for any
$x\in A$
and for any
$j=1,2,3.$ 
\par
We note that the condition
$(D)$ of the previous section is automotically
satisfied  that if
$\rho_{j}(x)=0$
for some 
$x\in A$ and for some
$j=0,1,2,$
then
$x=0$ in view of
Eqs.(4.11).
\par
Next,
since
$[g_{1},g_{2}]\subseteq g_{3}$
by Eq.(4.4e)
of Lemma 4.1,
we can introduce a bi-linear product
$xy$
in $A$ by
$$
[\rho_{1}(x),\rho_{2}(y)]:=
-\rho_{3}(xy).\eqno(4.13)$$
Applying
$\phi \in Z_{3}$
to this relation,
and noting
Eq.(4.12),
this yield
$$
[\rho_{i}(x),\rho_{j}(y)]:=-\rho_{k}(xy)\eqno(4.14)$$
for any cyclic permutation
$(i,j,k)$
of indices
$(1,2,3).$
This reproduces
Eq.(3.4b)
for
$\gamma_{0}=\gamma_{1}=\gamma_{2}=1.$
Similarly,
$[g_{j},g_{j}]\subseteq t$
by Lemma 4.1 and
we define
$T_{j}(x,y)\in t$
by
$$
[\rho_{j}(x),\rho_{j}(y)]:=
T_{3-j}(x,y).\eqno(4.15)
$$
Applying
$\phi$ to this relation,
it gives
$$
\phi T_{3-j}(x,y)=
T_{3-(j+1)}(x,y)\eqno(4.16)
$$
since
$$
\begin{array}{ll}
\phi T_{3-j}(x,y)&
=\phi [\rho_{j}(x),\rho_{j}(y)]=
[\phi \rho_{j}(x),\phi \rho_{j}(y)]=
[\rho_{j+1}(x),\rho_{j+1}(y)]\\
&
=T_{3-(j+1)}(x,y).
\end{array}
$$
Analagously,
$[t,g_{k}]\subseteq g_{k}$
implies that we can define
$$
t_{j,k}:
A\otimes A\rightarrow \mbox{End}A,\ (j,k=0,1,2)
$$
by
$$
[T_{3-j}(x,y),\rho_{k}(z)]=
:\rho_{k}(t_{j,k}(x,y)z).\eqno(4.17)
$$
Operating
$\phi$ to this relation,
and noting
Eq.(4.16),
we calculate
$$
[T_{3-(j+1)}(x,y),
\rho_{k+1}(z)]=
\rho_{k+1}(t_{j,k}(xy)z)$$
or
$$
\rho_{k+1}(t_{k+1,j+1}(x,y)z)=
\rho_{k+1}(t_{j,k}(x,y)z),$$
which implies
$$
t_{k+1,j+1}(x,y)z=
t_{j,k}(x,y)z
\eqno(4.18)
$$
because of the condition $(D)$.
Then,
$t_{j,k}(x,y)$
depends upon
$j$ and $k$ only in the combination of
their difference
$k-j$,
and we can set
$$
t_{j,k}(x,y)=d_{k-j}(x,y)$$
for some
$d_{i}(x,y)\in \mbox{End}A.$
Therefore,
Eq.(4.17)
becomes
$$
[T_{j}(x,y),\rho_{k}(z)]=
\rho_{k}(d_{j+k}(x,y)z)\eqno(4.19)
$$
which is Eq.(3.4).
Here,
we have changed
$j\rightarrow 3-j.$
\par
\vskip 3mm
Finally,
since $L$ is a Lie algebra,
Eq.(3.4d) follows from
Eqs.(3.4a)
and (3.4c),
satisfying all relations in Eqs.(3.4).
As the result,
Corollary 3.4 implies $A$ to be a normal triality algebra with
the validity of
Eq.(4.9).
\par
In order to show that
$\tilde{L}$
is a ideal of $L$,
we first note
$$
[\rho_{j}(A),t]\subseteq \rho_{j}(A)\eqno(4.20)$$
by Lemma 4.1.
Moreover,
we calcultate
$$
\begin{array}{l}
[T_{j}(x,y),t]=[[\rho_{j}(x),\rho_{j}(y)],t]
\\
=-[[\rho_{j}(y),t],\rho_{j}(x)]-
[[t,\rho_{j}(x)],
\rho_{j}(y)]
\subseteq
[\rho_{j}(A),\rho_{j}(A)]
\subseteq 
T_{j}(A,A)
\end{array}
$$
so that we have
$$
[T_{j}(A,A),t]\subseteq T_{j}(A,A).\eqno(4.21)$$
\par
If $L$ is invariant under
$S_{4},$
we define
${\overline x}$ for any
$x\in A$ by
$$
\tau\rho_{0}(x)=
-\rho_{0}({\overline x}),\ 
(\mbox{i.e.}
\tau x=-{\bar x})\eqno(4.22)$$
for the transposition $\tau=(1,2)\in S_{4}.$
We can then prove that
$x\rightarrow
{\overline x}$
is a involution of $A$.$\square$
\par
\vskip 3mm
{\bf Remark 4.5}
\par
\vskip 3mm
If $L$ is simple, and if $\tilde{L}$ is {\it not}
trivial,
then
$\tilde{L}=L.$
Suppose that $L$ is {\it not}
simple,
and
$S_{4}$-invariant.
Then,
both
$t$ and
$T(A,A)$
is $S_{3}$-invariant
since
$\phi\tau_{\mu}\phi^{-1}=
\tau_{\mu+1},$
so that
$L/\tilde{L}
=t/T(A,A)$
is now $S_{3}$-invariant.
A co-ordinatization of any Lie algebra which
is invariant under
$S_{3}$,
or more genelly
$di$-cyclic group has been given
in [E-O.09], [E-O.11].
\par
\vskip 3mm
{\bf Remark 4.6}
\par
\vskip 3mm
It has been noted in [E-O.08] that any
simple Lie algebra over the algebraicaly
closed field of charachristic
zero is $S_{4}$-invariant
so that all these algebras can be constructed by some normal
triality algebras.
We will study some cases below.
\par
\vskip 3mm
{\bf Example 4.7(a)}
\par
\vskip 3mm
The 
$so(3)$
Lie algebra defined by
$$
[e_{i},e_{j}]=
\sum^{3}_{k=1}\varepsilon_{ijk}e_{k},
(i, j=1.,2,3)$$
is $S_{4}$-invariant.
First,
$\phi=(1,2,3)$
and
$\tau=(1,2)$
are given by
$$
\phi :e_{1}\rightarrow e_{2}\rightarrow e_{3}\rightarrow e_{1},\ 
\tau:
e_{1}\leftrightarrow e_{2},\ e_{3}\rightarrow-e_{3}$$
while the Klein's $4$-group
$K_{4}$ acts as 
$$
\begin{array}{l}
\tau_{1}:
e_{1}\rightarrow e_{1},\ e_{2}\rightarrow-e_{2},\ e_{3}\rightarrow-e_{3}
\\
\tau_{2}:
e_{2}\rightarrow e_{2},\ 
e_{1}\rightarrow -e_{1},\ 
e_{3}\rightarrow -e_{3}
\\
\tau_{3}:
e_{3}\rightarrow e_{3},\ 
e_{1}\rightarrow -e_{1},\ 
e_{2}\rightarrow -e_{2}.
\end{array}
$$
Then,
$g_{i}=Fe_{i}$
with $T_{j}(A,A)=0.$
Hence,
the resulting normal triality algebra or
structurable algebra
$A$ is isomorphic to the
field $F$ itself.
Similary,
we note that the
quaternion algebra is also
$S_{4}$-invariant.
\par
\vskip 3mm
{\bf Example 4.7(b)\ ($so(N)$ algebra for $N\geq 4 $)}
\par
\vskip 3mm
The $so(N)$ Lie algebra is defined by
$J_{\mu \nu}=-J_{\nu \mu}$
satisfying
$$
[J_{\mu \nu},J_{\alpha\beta}]=
\delta_{\mu \alpha}J_{\nu\beta}-\delta_{\nu\alpha}J_{\mu\beta}-
\delta_{\mu\beta}J_{\nu\alpha}+
\delta_{\nu\beta}J_{\mu\alpha}$$
for $\mu,\nu,\alpha,\beta=1,2,\cdots ,N.$
It is clearly invariant under the
symmetric group
$S_{N}$
permuting
$N$ indices $1,2,\cdots,N.$
For $N\geq 4$,
it is then invariant also under
its sub-group $S_{4}$
which permutes $4$ indices
$1,2,3$
and $4$,
but leaves other indices $5,6,\cdots ,N$
being unchanged.
Then,
the decomposition Eq.(4.7)
of $L=so(N)$
by the Klein's $4$-group is readily
computed to yield
$$
\begin{array}{rl}
t:
&
\left\{
\begin{array}{l}
(1)\ J_{1j}+J_{2j}+
J_{3j}+J_{4j}\ (j\geq 5)\\
(2)\ J_{ij},\ (i,j\geq 5)
\end{array}
\right.
\\
g_{3}:
&
\left\{
\begin{array}{l}
(1)\ J_{13}+J_{24}\\
(2)\ J_{14}+J_{23}\\
(3)\ J_{1j}+J_{2j}-J_{3j}-J_{4j},\ (j\geq 5)
\end{array}
\right.
\\
g_{1}:
&
\left\{
\begin{array}{l}
(1)\ J_{12}-J_{34}\\
(2)\ J_{13}-J_{24}\\
(3)\ J_{1j}-J_{2j}-J_{3j}+J_{4j},\ (j\geq 5)
\end{array}
\right.
\\
g_{2}:
&
\left\{
\begin{array}{l}
(1)\ J_{12}+J_{34}\\
(2)\ J_{14}-J_{23}\\
(3)\ J_{1j}-J_{2j}+
J_{3j}-J_{4j}.\ (j\geq 5)
\end{array}
\right.
\end{array}
$$
Assuming that the field
$F$ is of charachteristic $\not= 2,$
we set
$$
\begin{array}{rl}
e
&
={1\over 2}(J_{13}+J_{24}-J_{14}-J_{23}),\\
f_{0}
&=
{1\over 2}(J_{13}+J_{24}+J_{14}+J_{23}),\\
f_{j-4}
&=
{1\over 4}(J_{1j}+J_{2j}-J_{3j}-J_{4j}),\ (j\geq 5).
\end{array}
$$
Then, by Eq,(4.13), we calculate
$$
A=g_{3}=\mbox{span}
<e,f_{\mu},\ 
(\mu=0,1,2,\cdots,N-4)> \eqno(4.23)
$$
to be a unital commutative algebra
with the multiptication table of
$$
ef_{\mu}=f_{\mu}e=f_{\mu},\ f_{\mu}f_{\nu}=\delta_{\mu\nu}e\eqno(4.24)
$$ 
for $\mu,\nu=0,1,2,\cdots,N$.
If we further introduce a symmetric bi-linear non-degenerate form
$<\cdot|\cdot>$
in $A$ by
$$
<f_{\mu}|f_{\nu}>=\delta_{\mu\nu},\ 
<f_{\mu}|e>=
<e|f_{\mu}>=0,\ <e|e>=1,$$
$A$ is a quadratic algebra satisfying
$$
x^{2}-2<x|e>x+
<x|x>e=0\eqno(4.25)$$
for any $x\in A.$
Especially,
$A$ is Jordan algebra.
Therefore,
$A$ is also a structurable algebra
with ${\overline x}=x.$
Note that Eq.(4.21)
will give,
contralily
${\overline e}=e$
but
${\overline f}_{\mu}=-f_{\mu},$
for
$\mu=0,1,\cdots,N-4.$
\par
In this case,
we have
$L(A)=so(N),\ L_{j}(A)=s0(N-2)\oplus gl{(1)}$
and
$T_{j}(A,A)=so(N-3)$
as well as
$T_{1}(x,y)=T_{2}(x,y)=T_{0}(x,y)$
in accordance with
$d_{1}(x,y)=d_{2}(x,y)=d_{0}(x,y)$
(see discussion given after Eq.(3.11)).
We also note that
$L_{j}(A)$
is still
$S_{N-4}$-invariant,
permuting indices
$5,6,-,N.$
\par
\vskip 3mm
{\bf Remark 4.8(Some Lie superalgebras)}

\par
\vskip 3mm
Some Lie superalgebras are also
$S_{4}$-invariant,
and we can apply the same tecnique to show the
triality
(see Remark 1.9). 
Consider for example of the Lie superalgebra $osp(N,2)$ for $N\geq 4 $.
They can be invariant under the
$S_{4}$ symmetry by extending the action of
$S_{4}$
of its even-part
$L_{{\overline 0}}=so(N)\ (N\geq 4).$
Then the resulting normal
triality super-algebra $A$ has its even part
$A_{{\overline 0}}$
given by Eq.(4.23),
while its odd part
$A_{{\overline 1}}$
is
$$
A_{{\overline 1}}=
\mbox{span}<\xi_{1},\xi_{2}>$$
satisfying
$$
\begin{array}{ll}
\xi_{1}\xi_{2}=-\xi_{2}\xi_{1}=e,&
\xi_{1}\xi_{1}=\xi_{2}\xi_{2}=0, \ 
e\xi_{\alpha}=\xi_{\alpha} e =\xi_{\alpha}\ (\alpha =1,2)\\
f_{\mu}\xi_{\alpha}=\xi_{\alpha}f_{\mu}=0,&
(\mu=0,1,2,\cdots,N-4,\mbox{and}\ \alpha=1,2,).
\end{array}
$$
The cases of Lie superalgebras
$G(3)$ and $F(4)$
have been disscussed in [E-O,08].
\par
\vskip 3mm
{\bf Example 4.9($sl(N)$\ Lie algebra for $N\geq 4$)}
\par
\vskip 3mm
The $sl(N)$ Lie algebra is specified by the
commutation relation
$$
[X^{\mu}_{\nu},X^{\alpha}_{\beta}]=
\delta^{\mu}_{\beta}X^{\alpha}_{\nu}-
\delta^{\alpha}_{\nu}X^{\mu}_{\beta},\ 
\sum^{N}_{\mu=1}X^{\mu}_{\mu}=0$$
for $\mu,\nu,\alpha,\beta=1,2,\cdots,N,$
which is invariant under $S_{N}$ again.
Assuming
$N\geq 4$,
and restricting ourselves to
its sub-group $S_{4}$ as in Example
4.7b,
we find the resulting normal triality algebra to be given by
$$
A=g_{3}=\mbox{span}
<e,f,x^{\mu},x_{\mu},
\ \mu=0,1,2,\cdots,N-4> \eqno(4.26)
$$
where we have set
$$
\begin{array}{rl}
e&
={1\over 2}\{(X^{3}_{1}-X^{1}_{3}+X^{4}_{2}-X^{2}_{4})
+(X^{2}_{3}-X^{3}_{2}+X^{1}_{4}-X^{4}_{1})\}
\\
f&
={1\over 2}\{(X^{1}_{1}+X^{2}_{2}-
X^{3}_{3}-X^{4}_{4})-
(X^{1}_{2}+X^{2}_{1}-X^{3}_{4}-X^{4}_{3})\}
\\
x_{j-4}&
=X^{1}_{j}+X^{2}_{j}-X^{3}_{j}-X^{4}_{j},\ (j\geq 5)
\\
x^{j-4}
&
=X^{j}_{1}+
X^{j}_{2}-X^{j}_{3}-X^{j}_{4}\ (j\geq 5)
\end{array}
$$
while
$x_{0}$
and 
$x^{0}$
are defined by
$$
\begin{array}{rl}
x_{0}&
=g-k,\ \mbox{and} \ \ x^{0}=g+k,
\\
g
&={1\over 2}\{(X^{1}_{1}+
X^{2}_{2}-X^{3}_{3}-X^{4}_{4})
+
(X^{1}_{2}+X^{2}_{1}-X^{3}_{4}-X^{4}_{3})\}\\
k&={1\over 2}\{
(X^{3}_{1}-X^{1}_{3}+X^{4}_{2}-X^{2}_{4})
-(X^{2}_{3}-X^{3}_{2}+X^{1}_{4}-X^{4}_{1})\}.
\end{array}
$$
They satisfy by Eq.(4.13)the multiplication table of
\begin{description}
\item[(1)]
$ef=fe=-f,$
\ but\ $ex=xe=x\ \mbox{for}\ x=x_{\mu}$
and
$
x^{\mu},$
\item[(2)]
$ff=e,$
\item[(3)]
$fx_{\mu}=-x_{\mu}f=x_{\mu},\ fx^{\mu}=-x^{\mu}f=-x^{\mu},$
\item[(4)]
$x_{\mu}x_{\nu}=0=x^{\mu}x^{\nu},$
\item[(5)]
$x_{\mu}x^{v}=2\delta^{\nu}_{\mu}(f-e),\ x^{\nu}x_{\mu}=-2\delta^{\nu}_{\mu}(f+e)$
\end{description}
for
$\mu,\nu =0,1,2,\cdots,N-4,$
reproducing the result of Example 2.5.
\par
For $L(A)=sl(N),$
we have $L_{i}(A)=sl(N-2)\oplus gl(1),$
and
$T_{j}(A,A)=sl(N-3).$
\par
Also in the Example (2.5),
we have seen that this algebra has more than
one involution and the involution 1 corresponds
to the case of $A^{*}$ being structurable.
Note that the involution 3
 is the one obtained  by
Eq.(4.21),
i.e,
$\tau\rho_{0}(x)=-\rho_{0}({\overline x}).$
\par
\vskip 3mm
{\bf Example 4.10(Lie algebra $E_{8}$)}
\par
\vskip 3mm
The exceptional Lie algebra
$E_{8}$ as well as Lie superalgebras
$G(3)$ and
$F(4)$
are $S_{4}$-invariant,
and are discussed in
[E-O.08].
We will not go into details.
\par
\vskip 3mm
{\bf Remark 4.11(Tetrahodron Algebra)}
\par
\vskip 3mm
The tetrahedron Lie algebra
$\boxtimes$ of Hartwig and Terlliger
[H-T,07] is generated by
$$
\{X_{ij},|i,j\in I,\ i\not=j\},I=\{0,1,2,3\}$$
with
\begin{description}
\item[(1)]
For distinct
$ 
i,j\in I,\ X_{ij}+X_{ji}=0$
\item[(2)]
For mutually district 
$k,i,j\in I,$
$$[X_{ki},X_{ij}]=2(X_{ki}+X_{ij})$$
\item[(3)]
For mutully district
$h,i,j,k\in I,$
$$
[X_{ki},[X_{ki},[X_{ki},X_{jh}]]]=
4[X_{ki},X_{jh}].$$
\end{description}
\par
\noindent
It is clearly
$S_{4}$-invariant,
and we have the decomposition
$$
\boxtimes =\Omega\oplus \Omega^{'}\oplus \Omega^{''},$$
where $\Omega\ (resp\ \Omega^{'} )\ and \ (resp \ \Omega^{''}) $
is a sub-algebra of
$\boxtimes$
generated by
$X_{12},X_{23}\ (\mbox{resp.}\ X_{23},X_{01}$)
and (resp.$X_{31},X_{02}$).
All
$\Omega,\Omega^{'}$
and 
$\Omega^{''}$
are Onsager Lie algebras.
\par
A remarkable fact is that it is isomorphic to three point
$sl(2)$ loop algebra by
$$
\Phi:\boxtimes \rightarrow sl(2)\otimes F(t,
{1\over t},{1\over 1-t})$$
for a indefinite variable $t$.
For details,
see
[H-T.07]
and [E.07]. 
However, the relationship between these facts and Theorem 4.4 is
not transparent. 
\par
In ending this section,
let us return to a further
study of the normal triality algebra
$A$
associated with the Lie algebra
$\tilde{L}$
in Theorem 4.4.
We introduce a bi-linear form
$<\cdot|\cdot>$
in $A$ by
$$
<x|y>:=
\sum^{3}_{j=1}\ 
\mbox{Tr}\ (\mbox{ad}\ \rho_{j}(x)
\mbox{ad}\ \rho_{j}(y))
\eqno(4.27)$$
where
"ad"
implies the adjoint operation.
We then find
\par
\vskip 3mm
{\bf Lemma 4.12}
\par
\vskip 3mm
\noindent
(1)
$$
<y|x>=<x|y>\eqno(4.28a)
$$
(2)
$$
<xy|z>=
<x|yz>\eqno(4.28b)$$
(3)
$$
<x|d_{j}(z,w)y>=
-<d_{j}(z,w)x|y>=
<z|d_{3-j}(x,y)w>
\eqno(4.28c)
$$
for any
$x,y,z\in A$
and
$j=1,2,3.$
\par
\vskip 3mm
{\bf Proof}
\par
\vskip 3mm
Eq.(4.28a) is a immediate consequence of
Eq(4.27).
Let $(i,j,k)$
be any cyclic permutation of
indices
$(1,2,3).$
Then,
the trace identity
$$
\begin{array}{l}
\mbox{Tr}([\mbox{ad}\rho_{i}(x),
\mbox{ad}\rho_{j}(y)]
\mbox{ad}\rho_{k}(z)\\
=
\mbox{Tr}(
\mbox{ad}\rho_{i}(x)
[\mbox{ad}\rho_{j}(y),\mbox{ad}\rho_{k}(z)])
\end{array}
$$
gives
$$
\mbox{Tr}(\mbox{ad}\rho_{k}(xy)
\mbox{ad}\rho_{k}(z))=
\mbox{Tr}
(\mbox{ad}\rho_{i}(x)\mbox{ad}\rho_{i}(yz))$$
by
Eq.(4.14).
Summing over $k$,
this yields Eqs.(4.28b).
\par
Next,
Eqs.(4.28c)
for $j=1$
or
$2$ are consequences of 
Eqs.(4.28b),
and (1.18).
For example,
we calculate
$$
\begin{array}{l}
<x|d_{1}(z,w)y>=
<x|(R(w)L(z)-R(z)L(w))y>=<x|(zy)w-(wy)z>\\
=<wx|zy>-<zx|wy>
\end{array}
$$
which is anti-symmetric in 
$x\leftrightarrow y$,
giving
$$
<x|d_{1}(z,w)y>=
-<y|d_{1}(z,w)x>=
-<x|d_{1}(z,w)y>.$$
Also,
we note
$$
\begin{array}{l}
<z|d_{2}(x,y)w>=
<z|(L(y)R(x)-L(x)R(y))w>\\
=<z|y(wx)-x(wy)>
=<zy|wx>-<zx|wy>
\end{array}
$$
so that we have
$<x|d_{1}(z,w)y>=
<z|d_{2}(x,y)w>.$
The case of 
$j=2$
for Eqs.(4.28c)
can be similarly proved.
In order to show its validity for the
case of
$j=0,$
we compute now
$$
\sum^{3}_{j=1}
\mbox{Tr}(\mbox{ad}T_{j}(x,y)\mbox{ad}T_{j}(z,w))
$$
$$
=\sum^{3}_{j=1}
\mbox{Tr}[(\mbox{ad}\rho_{3-j}(x),
\mbox{ad}\rho_{3-j}(y)]
\mbox{ad}T_{j}(z,w))
$$
$$
=\sum^{3}_{j=1}
\mbox{Tr}(\mbox{ad}\rho_{3-j}(x)
[\mbox{ad}\rho_{3-j}(y),
\mbox{ad}T_{j}(z,w)])
$$
$$
=-\sum^{3}_{j=1}
\mbox{Tr}(\mbox{ad}\rho_{3-j}(x)
\mbox{ad}\rho_{3-j}(d_{0}(z,w)y))
$$
by Eqs.(3.4a)
and (3.4b)
so that we obtain
$$
\sum^{3}_{j=1}
\mbox{Tr}(
\mbox{ad}T_{j}(x,y)
\mbox{ad}T_{j}(z,w))=
-<x|d_{0}(z,w)y>.\eqno(4.29)
$$
However,
 the left side of this relation is
symmetric for
$x\leftrightarrow z$
and $y\leftrightarrow w$
but anti-symmetric in
$x\leftrightarrow y.$
Those give Eqs.(4.28c)
for
$j=0$.
$\square$
\par
We will assume hereafter that
$<\cdot|\cdot>$
is non-trivial,
i.e.
it is not identically zero.
\par
If we set
$$
A_{0}=\{x|<x|A>=0,x\in A\},
\eqno(4.30)$$
Then $A_{0}$ is a ideal of $A$ in view of Eqs.(4.28b).
\par
\vskip 3mm
{\bf Lemma 4.13}
\par
\vskip 3mm
We have
\par
(1)\quad $d_{j}(A,A)A_{0}\subseteq A_{0}$
\par
and
\par
(1)\quad$d_{j}(A_{0},A)A\subseteq A_{0}$
\par
for $j=1,2,3.$
\par
\vskip 3mm
{\bf Proof}
\par
\vskip 3mm
Since
$$
<d_{j}(A,A)A_{0}|A>=
-<A_{0}|d_{j}(A,A)A>=
0$$
by Eqs.(4.28c),
this proves
$d_{j}(A,A)A_{0}\subseteq A_{0}.$
Similarly,
if $a\in A_{0},$
we calculate
$$
<z|d_{j}(a,x)w>=
<a|d_{j}(z,w)x>=0$$
again by the second relation of
Eq.(4.28c)
to give
$$
d_{j}(A_{0},A)A\subseteq A_{0}. \ \ \square$$
\par
\newpage
{\bf Proposition 4.14}
\par
\vskip 3mm
Suppose that
$<\cdot|\cdot>$
 is non-trivial.
If $\tilde{L}$ is a simple Lie algebra,
then
$<\cdot|\cdot>$ is non-degenerate.
\par
\vskip 3mm
{\bf Proof}
\par
\vskip 3mm
Suppose that
$<\cdot|\cdot>$ is degenerate.
Then,
$A_{0}$ is a proper ideal of $A$.
Further,
$$
L_{0}=
\rho_{0}(A_{0})\oplus
\rho_{1}(A_{0})
\oplus
\rho_{2}(A_{0})
\oplus
T(A_{0},A)$$
can be readily verified to be a proper ideal of
$\tilde{L}$,
if we note
Eqs.(3.4)
and Lemma 4.13.
Therefore,
$\tilde{L}$
is {\it not} simple, so that
$<\cdot|\cdot>$ must be non-degenerate.
$\square$
\par
\vskip 3mm
{\bf Remark 4.15}
\par
\vskip 3mm
If the conjugate algebra
$A^{\star}$ of $A$ is unital and involutive with the unit
 element $e$.
Then,
$e$ is the para-unit of $A$.
Then,
Eqs.(4.28b)
yields
$$
<\bar{x}|\bar{y}>=
<x|y>,\eqno(4.31a)$$
$$
<\bar{x}|y\star z>=
<\bar{y}|z\star x>=
<\bar{z}|x\star y>.\eqno(4.31b)$$
Relations of Lemma 4.13 as well as
Eqs.(4.31)
for
$A^{\star}$ have been noted in
[O.05]
for case of Examples (2.2) and
(2.6) 
\par
\vskip5mm
{\bf 5.\ Pre-structurable Algebra}
\par
\vskip 3mm
Although we have already defined a pre-structurable algebra by
Def.1.6,
we will here introduce the following slightly more 
generalization for
a later purpose.
\par
\vskip 3mm
{\bf Def.5.1}
\par
\vskip 3mm
Let $A^{*}$ be a involutive algebra with the bi-linear product
$x\star y$ and the involution 
$x\rightarrow {\overline x}.$
Suppose that it satisfies the
triality relation
$$
\overline{ d_{j}(x,y)}
(u\star v)=
(d_{j+1}(x,y)u)\star v+
u\star(d_{j+2}(x,y)v)\eqno(5.1)
$$
for
$d_{j}(x,y)\in
\mbox{End}\ A^{*}\ (j=0,1,2)$
given by
$$
d_{1}(x,y)
=l({\overline y})l(x)-
l({\overline x})l(y),
\eqno
(5.2a)
$$
$$
d_{2}(x,y)
=r({\overline y})r(x)-r({\overline x})r(y)
\eqno(5.2b)$$
$$
d_{0}(x,y)
=r({\overline x}\star y-{\overline y}\star x)+
l(y)l({\overline x})-
l(x)l({\overline y})
$$
$$
=l(y\star {\overline x}-x\star {\overline y})+
r(y)r({\overline x})-
r(x)r({\overline y}).\eqno(5.3)
$$
We call then
$A^{*}$ be an almost pre-structurable algebra.
Note that if 
$A^{*}$ is unital in addition,
then
$A^{*}$ is pre-sturcturable.
Moreover, setting
$$
Q(x,y,z)=
d_{0}(x,\overline{y\star z})+
d_{1}(z,\overline{x\star y})+
d_{2}(y,\overline{z\star x})\eqno(5.4)
$$
as before,
we call a pre-sturcturable algebra be
structurable when we have
$Q(x,y,z)=0$
furthermore.
\par
\vskip 3mm
{\bf Def.5.2}
\par
\vskip 3mm
Let $A^{*}$ be a involutive algebra.
We introduce multiplication operators
by
(see [A-F,93])
$$
A(x,y,z)w
:=\{(w\star x)\star {\overline y}\}\star z-
w\star\{x\star( {\overline y}\star z)\}\eqno
(5.5a)
$$
$$
B(x,y,z)w
:=
\{(w\star x)\star{\overline y}\}\star z-
w\star\{(x\star{\overline y})\star z\}\eqno
(5.5b)
$$
$$
C(x,y,z)w
:=
\{x\star({\overline y}\star {\overline w})\}\star z-
(x\star{\overline y})\star
({\overline w}\star z)\eqno
(5.5c)
$$
$$
C^{'}(x,y,z)w
=C(x,y,z){\overline w}
=\{x\star({\overline y}\star w)\}\star z-
(x\star {\overline y})\star(w\star z)\eqno
(5.5d)$$
for $x,y,z,w\in A^{*}.$
\par
\vskip 3mm
{\bf Lemma 5.3}
\par
\vskip 3mm
If $A^{*}$ is an almost pre-structurable algebra,
we have
\par
\noindent
(1)
$$
A(x,y,z)-A(y,x,z)=
A(z,x,y)-A(z,y,x)\eqno(5.6a)$$
and
\par
\noindent
(2)
$$
B(z,x,y)-B(z,y,x)=
C^{'}(y,x,z)-
C^{'}(x,y,z).\eqno(5.6b)$$
Conversly,
if an involutive algebra
$A^{*}$ with
$d_{j}(x,y)$'s
being given by Eqs.(5.2) and (5.3) 
satisfies Eq.(5.6a)
or (5.6b),
respectively,
then the triality relation Eq.(5.1)
holds respectively for
$j=1$ and $2$ or
for
$j=0.$
\par
\vskip 3mm
{\bf Proof}
\par
\vskip 3mm
We may easily verify that
\par
\noindent
(1)\quad
Eq.(5.1) for
$j=1$
with
$d_{0}(x,y)=r({\overline x}\star y-{\overline y}\star x)+
l(y)l({\overline x})
-l(x)l({\overline y})$
is rewritten as
$$
\{A(w,x,y)-
A(w,y,x)\}z=
\{A(x,y,w)-A(y,x,w)\}z.$$
Similarly,
Eq.(5.1) for
$j=2$
with
$d_{0}(x,y)=
l(y\star{\overline x}-
x\star{\overline y})+
r(y)r({\overline x})-
r(x)r({\overline y})$
is equivalent
to the same relation,
if we take the involution of the
relation.
\par
\noindent
(2)\quad
Eq.(5.1)
for $j=0$
is similarly shown
to be equivalent
to the validity of
Eq.(5.6b).
$\square$
\par
\vskip 3mm
{\bf Proposition 5.4(see [A-F,93])}
\par
\vskip 3mm
Let $A^{*}$ be now a pre-structurable algebra.
We then also have
\begin{description}
\item[(1)]
$$ 
B(x,y,z)-B(y,z,x)=
B(z,x,y)-B(z,y,x),\eqno(B)
$$
\item[(2)]
$$ [x-{\overline x},y,z]=
-[y,x-{\overline x},z]=
[y,z,x-{\overline x}],\eqno(sk)
$$
\item[(3)]
$$
 [x,{\overline y},z]-
[y,{\overline x},z]=
[z,{\overline x},y]-
[z,{\overline y},x]=
[z,x,{\overline y}]
-[z,y,{\overline x}]
\eqno(A.1)
$$
\end{description}
where
$[x,y,z]$
is the associater of $A^{*}$ defined by
$$
[x,y,z]=
(x\star y)\star z-
x\star(y\star z).\eqno(5.7)
$$
\par
\vskip 3mm
{\bf Proof}
\par
\vskip 3mm
If we note
$A(x,y,z)e=[x,{\overline y},z]$
for the unit element $e$ of
$A^{*}$,
then Eq.(5.6a) becomes
$$
[x,{\overline y},z]-
[y,{\overline x},z]=
[z,{\overline x},y]-
[z,{\overline y},x]\eqno(5.6)'
$$
which is a part of Eq.(A1)
consistent with Eq.(5.3).
Also from Eq.(5.5),
we see
$$
A(x,y,z)w=B(x, y,z)w
-w\star 
[x,{\overline y},z]
\eqno(5.7)'
$$
so that
Eq.(5.6a) together with
Eq.(5.6)' and
(5.7)'
gives Eq.(B).
\par
If we next set
$y=e$ in Eq.(B),
it yields
$$
[w,x-{\overline x},z]=
[w,z,{\overline x}-x].$$
Taking the involution of this relation,
and changing the notation suitably we obtain
Eq.(sk).
Other relations
can be similarly 
proved.
$\square$
\par
\vskip 3mm
{\bf Lemma 5.5}
\par
\vskip 3mm
Let $A^{*}$ be a pre-structurable algebra.
Then
$$
D(x,y)=
d_{0}(x,y)+
d_{1}(x,y)+ 
d_{2}(x,y)\eqno(5.8a)$$
is a derivation of $A^{*}$,
satisfying
$$
\overline{D(x,y)}=
D({\overline x},{\overline y})=D(x,y).\eqno(5.8b)$$
\par
\vskip 3mm
{\bf Proof}
\par
\vskip 3mm
Form Eqs.(5.2),
we see that
$d_{j}(x,y)$'s satisfy
$$
\overline{d_{j}(x,y)}=
d_{3-j}({\overline x },{\overline y})\eqno(5.9a)$$
and
$$
d_{0}(x,y)z+
d_{0}(y,z)x+
d_{0}(z,x)y=0.\eqno(5.9b)$$
Then,
$\overline{D(x,y)}=
D({\overline x},{\overline y})$
immediately follows from
Eq.(5.8a)
and
(5.9a).
\par
We note
$$
\begin{array}{ll}
D(x,y)z
&
=z\star ({\overline x}\star y-{\overline y}\star x)+
y\star ({\overline x}\star z)-
z\star ({\overline y}\star x)\\
&
+{\overline y}\star(z\star x)-
{\overline x}\star (y\star z)+
(z\star x)\star {\overline y}-
(z\star y)\star {\overline x}.
\end{array}
$$
by Eqs.(5.2).
We then calculate
$$
\{D(x,y)-D({\overline x},{\overline y})\}z=
[z,x,{\overline y}]-
[z,y,{\overline x}]+
[z,{\overline y},x]-[z,{\overline x},y]=0$$
by Eq.(A1) so that we have
$D(x,y)=D({\overline x},{\overline y}).$
Finally, summing over 
$j=0,1,2$
in Eq.(5.1),
we obtain
$$
\overline{D(x,y)}(u\star v)=
(D(x,y)u)\star v)+
u\star(D(x,y)v)$$
which shows
$D(x,y)$
to be a derivation of $A^{*}$ in view of Eq.(5.8b).
$\square$
\par
We next consider two sets of
$$
S=\{x|{\overline x}=x,\ x\in A^{*}\}
$$
$$
H=\{x|{\overline x}=-x,\ x\in A^{*}\}.\eqno(5.10)$$
Then,
if the underlying field $F$ is of characteristic $\not=2$,
Eq.(sk)
indicates that $H$ is a generalized alternative nucleus of $A^{*}$.
As the consequence,$H$ is a Malcev algebra with respect to the
commutor product
$[x,y]^{*}=x\star y-y\star x$(see [P-S.04]).
\par
\vskip 3mm
{\bf Theorem 5.6} ([K-O.14])
\par
\vskip 3mm
Let $A^{*}$ be a pre-structurable algebra.
We then have
\par
\begin{description}
\item[(1)]
$Q(x,y,z)w$ is totally symmetric in $x,y,z,w\in A^{*}$.
\item[(2)]
$Q(x,y,z)w=0$ identically,
if at least one of
$x,y,z$ and $w$ is the unit element $e$ of $A^{*}$.
\item[(3)]
Supose that the underlying field $F$ is of charachteristic $\not=2.$
Then,
$Q(x,y,z)w=0$
identically again,
provided that
at least one of $x,y,z,$
and $w$ is an element of $H$.
\item
[(4)]
$\overline{Q(x,y,z)}=Q({\overline x},
{\overline y},{\overline z})=Q(x,y,z)$
is a derivation of $A^{*}.$
\item
[(5)]
\par
$3Q(x,y,z)=
D(x,{\overline y}\star {\overline z})+
	D(y,{\overline z}\star {\overline x})+
D(z,{\overline x}\star {\overline y})$
\item
[(6)]
$
[Q(x,y,z),Q(u,v,w)]=
Q(Q(x,y,z)u,v,w)+
Q(u,Q(x,y,z)v,w)+
Q(u,v,Q(x,y,z)w).$
\end{description}
\par
For the proof of this Theorem,
we start from the
following Lemma.
\par
\vskip 3mm
{\bf Lemma 5.7}
\par
\vskip 3mm
We have $$Q(x,y,z)e=0.$$
\par
\vskip 3mm
{\bf Proof}
\par
\vskip 3mm
We calculate
$$
Q(x,y,z)=
r({\overline x}\star ({\overline z}\star {\overline y}))-
r((y\star z)\star x)+
l({\overline z}\star {\overline y})l({\overline x})-
l(x)l(y\star z)$$
$$+
l(x\star y)l(z)-
l({\overline z})l({\overline y}\star {\overline x})+
r(z\star x)r (y)-
r({\overline y})r({\overline x}\star {\overline z}),
\eqno
(5.11)
$$
from Eq.(5.2) and
(5.4),
so that we obtain
$$
\begin{array}{ll}
Q(x,y,z)e&
={\overline x}\star ({\overline z}\star {\overline y})-
(y\star z)\star x+
({\overline z}\star {\overline y})\star {\overline x}-x\star(y\star z)\\
&+(x\star y)\star z-{\overline z}\star({\overline y}\star{\overline x})
+y\star(z\star x)-
({\overline x}\star {\overline z})\star {\overline y}\\
&=-[{\overline x},{\overline z},{\overline y}]-
[y,z,{\overline x}]+
[{\overline z},{\overline y},{\overline x}]+
[x,y,z]=0
\end{array}
$$
by Eq.(A.1).
$\square$
\par
\vskip 3mm
{\bf Lemma 5.8}
\par
\noindent
(1)
$$
\overline{Q(x,y,z)}=Q({\overline x},{\overline z},{\overline y})\eqno(5.12a)$$
(2)
$$
\overline{Q(x,y,z)}(u\star v)=
\{Q(y,z,x)u\}\star v+
u\star\{Q(z,x,y)v\}.\eqno(5.12b)
$$
\par
\vskip 3mm
{\bf Proof}
\par
\vskip 3mm
Eq.(5.12a) is nothing but Eq.(1.37),
while we note
$$
\begin{array}{l}
\overline{d_{0}(x,\overline{y\star z})}
(u\star v)=
\{d_{1}(x,\overline{y\star z})u\}\star v+
u\star \{d_{2}(x,\overline{y\star z})v\}\\
\overline{d_{1}(z,\overline{x\star y})}
(u\star v)=\{d_{2}(z,\overline{x\star y})u\}\star v+
u\star\{d_{0}(z,\overline{x\star y})v\}\\
\overline{d_{2}(y,\overline{z\star x})}(u\star v)=
\{d_{0}(y,\overline{z\star x})u\}\star v+
u\star\{d_{1}(y,\overline{z\star x})v\}.
\end{array}
$$
Adding all these relations,
we obtain Eq.(5.12b).
$\square$
\par
We next first set
$u=e$ and
$v=w,$
and also
$v=e$
and $u=w$
in Eq.(5.12b) to find
$$
\overline{Q(x,y,z)}w=
Q(z,x,y)w=
Q(y,z,x)w$$
where we used
$Q(x,y,z)e=0$
by Lemma 5.7.
This implies the validity of 
$$
\overline{Q(x,y,z)}=
Q(z,x,y)=Q(y,z,x).\eqno(5.12c)$$
Especially,
letting further
$x\rightarrow y\rightarrow z\rightarrow x,$
this leads to
$$
Q(z,x,y)=Q(y,z,x)=Q(x,y,z)\eqno(5.12d)$$
to be cyclically invariant,
and then
$$
\overline{Q(x,y,z)}=Q(x,y,z)\eqno(5.12e)$$
by Eq.(5.12c) again.
\par
Moreover,
since $A^{*}$ is unital,
its conjugate algebra $A$ satisfies
$ex=xe=\overline{x}$
so that the conditions
$(B)$ and
$(C)$ of section 1
are automatically satisfied.
Especially,
$A$ is a pre-normal triality algebra,
so that 
Eq.(1.36) holds with
$$
Q(x,y,z)w=Q(w,y,z)x,$$
by Proposition 1.5.
We then calculate
$$
\begin{array}{ll}
Q(x,y,z)w&
=Q(w,y,z)x=Q(y,z,w)x=Q(x,z,w)y\\
&=Q(w,x,z)y=Q(y,x,z)w
\end{array}
$$
which yields
$Q(x,y,z)=Q(y,x,z).$
Together with
Eq.(5.12d),
these imply that $Q(x,y,z)w$
is totally
symmetric in
$x,y,z$
and $w$.
Then Lemma 5.7 shows that
$Q(x,y,z)w=0$
if at least one of
$x,y,z$ and
$w$ coincides with the unit element $e$.
Moreover Eqs.(5.12a,b,c),
and
(5.12d) imply also
$\overline{Q(x,y,z)}=Q(\overline{x},\overline{y},\overline{z})=Q(x,y,z)$
to be a derivation of $A^{*}$.
\par
In order to prove the statement (3) of Theorem 5.6,
we write generic element of
$S$ and $H$ as
$x_{0}$ and
$x_{1}$
respectively,
so that
$\overline{x_{0}}=x_{0}$
and $\overline{x_{1}}=-x_{1}.$
Then,
$Q(\overline{x},\overline{y},\overline{z})=
Q(x,y,z)$
yield immediately
$Q(x_{1},y_{1},z_{1})=0=Q(x_{0},y_{0},z_{1}),$
provided that the
field $F$ is of charachteristic $\not= 2.$
Moreover,
we note
$$
Q(x_{1},y_{1},z_{0})w_{1}=
Q(x_{1},y_{1},w_{1})z_{0}=0,$$
and
$$
Q(x_{1},y_{1},z_{0})w_{0}=Q(x_{1},w_{0},z_{0})y_{1}=0
$$
so that we have also
$Q(x_{1},y_{1},z_{0})=0.$
This proves the statement (3) of the Theorem.
\par
\vskip 3mm
Also, Eq.(1.35)
together with
Eq.(5.12d)
yields immediately the relation of
$$
3Q(x,y,z)=
D(x,\overline{y\star z})+
D(y,\overline{z\star x})+
D(z,\overline{x\star y}).$$
Therefore,
it remains only to prove the final statement (6).
To show it,
we note the following:
\par
\vskip 3mm
{\bf Lemma 5.9}
\par
\vskip 3mm
Let $D$ be a derivation of
$A^{*}$ satisfying
$\overline{D}=D,$
then we have
$$
[D,Q(u,v,w)]=
Q(Du,v,w)+
Q(u,Dv,w)+
Q(u,v,Dw).\eqno(5.13)
$$
\par
\vskip 3mm
{\bf Proof}
\par
\vskip 3mm
Since $D$ is a derivation of $A^{*}$,
we have
$$
D(u\star v)=
(Du)\star v+
u\star (Dv)$$
which is equivalent to the validity of
$$
[D,l(u)]=
l(Du),\quad
[D,r(v)]=
r(Dv).$$
Moreover,
$\overline{D}=D$
implies
$\overline{Dx}=
D\overline{x}.$
Then,
these are sufficient
to prove Eq.(5.13). $\square$
\par
Since $D=Q(x,y,z)$ 
satisfies the condition of Lemma 5.9,
these give
$$
[Q(x,y,z),Q(u,v,w)]=
Q(Q(x,y,z)u,v,w)+
Q(u,Q(x,y,z)v,w)+
Q(u,v,Q(x,y,z)w).
$$
These results complete the proof of Theorem 5.6.
\par
\vskip 3mm
{\bf Remark 5.10}
\par
\vskip 3mm
If we choose
$D=D(x,y)$
in Lemma 5.9,
we have
$$
[D(x,y),Q(u,v,w)]=
Q(D(x,y)u,v,w)+
Q(u,D(x,y)v,w)+
Q(u,v,D(x,y)w).\eqno(5.14a)
$$
Moreover,
we have also
$$
[Q(u,v,w),D(x,y)]=
D(Q(u,v,w)x,y)+
D(x,Q(u,v,w)y)\eqno(5.14b)$$
by the following reason.
Since $A$ is a pre-normal triality algebra,
we have Eq.(1.20),i.e.
$$
[d_{j}(u,v),d_{k}(x,y)]=
d_{k}(d_{j-k}(u,v)x,y)+
d_{k}(x,d_{j-k}(u,v)y).$$
 Letting $v\rightarrow \overline{v}\star \overline{w},$
and then letting
$u\rightarrow v\rightarrow v\rightarrow w\rightarrow u,$
these give
$$
[Q(u,v,w),d_{k}(x,y)]=
d_{k}(Q(u,v,w)x,y)+d_{k}(x,Q(u,v,w)y)\eqno(5.15)$$
when we further sum over
$j\equiv 0,1,2$ and note that
$Q(u,v,w)$
is totaly symmetric in $u,v,w.$
Finally,
summing over $k$,it gives Eq.(5.14b).
\par
\vskip 3mm
{\bf Proposition 5.11}
\par
\vskip 3mm
Let $A^{*}$ be a pre-structurable algebra over the field $F$ of charachterictic
$\not= 2,
\not= 3.$
If $A^{*}$ is power-associative,
then $A^{*}$
is structurable.
\par
\vskip 3mm
{\bf Proof}
\par
\vskip 3mm
For any $a\in A^{*}$ satisfying $\overline{a}=a,
(i.e, a\in S).$
we calculate
$$
Q(a,a,a)a
=[a,a\star a^{2}]^{\star}+
3\{a^{2}\star a^{2}-a\star(a^{2}\star a)\}
\eqno
(5.16a)
$$
$$
=[a^{2}\star a,a]^{\star}+
3\{a^{2}\star a^{2}-
(a\star a^{2})\star a\}\eqno
(5.16b)
$$
where we have set
$a^{2}=a\star a$ and
$[x,y]^{\star}\equiv x\star y-y\star x.$
Note that
Eq.(5.16a) follows immediately
from Eq.(5.11)
by setting
$x=y=z=a,$
while Eq.(5.16b)
results from taking the involution of
Eq.(5.16a).
Then,
if $A^{*}$ is power-associative,
Eq.(5.16)
implies
$Q(a,a,a)a=0.$
Therefore,
linearizing the relation,
we obtain
$Q(x,y,z)w=0$
for $x,y,z,w\in S,$
since we are assuming
the field $F$ to be of charachteristic
$\not= 2,\not=3$.
Together with Theorem 5.6,
this shows
$A^{*}$ to be structurable.
$\square$
\par
\vskip 3mm
{\bf Proposition 5.12([O.05])}
\par
\vskip 3mm
If a pre-sturucturable algebra
$A^{*}$ possesses a symmetric bi-linear non-degenerate form
$<\cdot|\cdot>$
satisfying
$$
<\overline{x}|y\star z>=
<\overline{y}|z\star x>=
<\overline{z}|x\star y>,$$
then $A^{*}$ is structurable.
\par
\vskip 3mm
{\bf Proposition 5.13}
\par
\vskip 3mm
Let $A^{*}$ be a pre-structurable algebra and set
$A_{0}=\{x|x\in A^{*},\mbox{and}\
Q(u,v,w)x=0\ \mbox{for any}\ 
u,v,w\in A^{*}\},$
then
$A_{0}$ is a structurable algebra.
Moreover $A_{0}$ contains a structurable sub-algebra generated by 
the unit element $e$ and members of $H$ assuming $2\not= 0$.
\par
\vskip 3mm
{\bf Proof}
\par
\vskip 3mm
First, we show that
$A_{0}$ is a sub-algebra of $A^{*}$ since we calculate
$$
Q(u,v,w)(xy)=
(Q(u,v,w)x)y+
x(Q(u,v,w)y)=0$$
for any
$x,y\in A_{0}$  to get $xy \in A_{0}$,
by the derivation property of
$Q(v,u,w).$
Moreover,
$e\in A_{0}$ also by Theorem 1.4.
Further,
if $x\in A_{0}$,then $\overline{x}\in A_{0}$ also since
$$
0=\overline{Q(u,v,w)x}=
\overline{Q(u,v,w)}{\overline x}=
Q(u,v,w)\overline{x}.$$
Then, these imply $d_{j}(x,y)\in End \ A_{0}$, 
for
$x,y\in A_{0},$
so that
$A_{0}$ is pre-structurable.
Since
$Q(u,v,w)=0$
restricted to $A_{0},$
this proves $A_{0}$ to be structurable.
The fact that
$A_{0}$ contains a structurable sub-algebra generated by 
$e$ and
$H$ follows from Theorem 5.6.
$\square$
\par
We can prove the converse statement
of Theorem 5.6
\par
\vskip 3mm
{\bf Theorem 5.14}
\par
\vskip 3mm
Let $A^{*}$ be a unital involutive algebra satisfying
\begin{description}\item[(i)]
$Q(x,y,z)w$ is totally symmetric in $x,y,z,w\in A^{*}.$
\item[(ii)]
$Q(x,y,z)=0$
identically whenever at least one of $x,y$ and $z$ is a element of $H.$
\item[(iii)]
The validity of Eq.(sk).
\end{description}
\par
Then
$A^{*}$ is pre-structurable.
\par
In order to prove this Theorem,
we note the following
\par
\vskip 3mm
{\bf Lemma 5.15}
\par
\vskip 3mm
Under the conditions for
$A^{*}$ given in Theorem 5.14,
we have
$$
\overline{Q(x,y,z)}=
B(x,y,z)-C(y,x,z)-
C(z,x,y)-C^{'}(z,y,x).\eqno(5.17)$$
\par
\vskip 3mm
{\bf Proof}
\par
\vskip 3mm
Eq.(5.11) leads to
$$
\begin{array}{ll}
Q(x,y,z)w&
=w\star \{\overline{x}\star(\overline{z}\star\overline{y})\}-
\{w\star(\overline{x}\star \overline{z})\}\star \overline{y}\\
&
-w\star\{(y\star z)\star x\}+
(w\star y)\star(z\star x)\\
&
-\overline{z}\star\{(\overline{y}\star\overline{x})\star w\}+
(\overline{z}\star\overline{y})\star(\overline{x}\star w)\\
&
-x\star\{(y\star z)\star w\}+
(x\star y)\star(z\star w).
\end{array}
$$
Taking
the involution of this relation,
we have
$$
\overline{Q(x,y,z)w}=
\{B(z,\overline{x},\overline{w})-
C(\overline{x},z,\overline{w})-
C(\overline{w},z,\overline{x})-
C^{'}(\overline{w},\overline{x},z)\}y.
\eqno(5.18)$$
The left-hand side is rewritten as
$$
\overline{Q(x,y,z)w}=
\overline{Q(x,\overline{y},z)w}=
\overline{Q(x,w,z)\overline{y}}=
\overline{Q(x,w,z)}y$$
since
$Q(x,y-\overline{y},z)=0.$
Therefore,
Eq.(5.18)
is rewritten as
$$
\overline{Q(x,w,z)}=
B(z,\overline{x},\overline{w})-
C(\overline{x},z,\overline{w})-
C(\overline{w},z,\overline{x})-
C^{'}(\overline{w},\overline{x},z).$$
Letting
$x\rightarrow \overline{x}$
and
$w\rightarrow\overline{w},$
and noting
$Q(\overline{x},\overline{w},z)=
Q(x,w,z)$
this yields
$$
\overline{Q(x,w,z)}=
B(z,x,w)-
C(x,z,w)-
C(w,z,x)-
C^{'}(w,x,z).$$
Changing 
$w\rightarrow z\rightarrow x\rightarrow y,$
this gives Eq.(5.17).$\square$
\par
Since
$Q(x,y,z)$
is totally symmetric in
$x,y$
and $z$,
Eq.(5.17)
immediately gives
\par
\noindent
(1)
$$
B(x,y,z)-B(x,z,y)=
C^{'}(z,y,x)-
C^{'}(y,z,x),\eqno(5.19)
$$
and
\par
\noindent
(2)
$$
B(x,y,z)-B(y,x,z)=
C(y,x,z)+
C(z,x,y)-
C(x,y,z)-
C(z,y,x)+
C^{'}(z,y,x)-C^{'}(z,x,y)\eqno(5.20)
$$
from $Q(x,y,z)=Q(x,z,y)$
for Eq.(5.19)
and
$Q(x,y,z)=Q(y,x,z)$
for
Eq.(5.20).
Moreover,
letting $x\leftrightarrow z$ 
in Eq.(5.19) and adding it
to Eq.(5.20),
we obtain
$$
\begin{array}{l}
\{
B(x,y,z)-B(y,x,z)+
B(z,y,x)-B(z,x,y)\}w\\
=-\{C(x,y,z)-C(y,x,z)+
C(z,y,x)-C(z,x,y)\}w\\
+\{C^{'}(x,y,z)-C^{'}(y,x,z)+
C^{'}(z,y,x)-C^{'}(z,x,y)\}w\\
=-\{C(x,y,z)-C(y,x,z)+
C(x,y,z)-C(z,x,y)\}(w-\overline{w})\hskip 38.3mm
(5.21)
\end{array}
$$
since
$C^{'}(x,y,z)w=
C(x,y,z)\overline{w}.$
\par
Further,
if Eq.(sk)
holds,
we have
(see,[A-F,93])
$$
C(x,y,z)(w-\overline{w})=
B(x,y,z)(w-\overline{w})\eqno(5.22)
$$
when we calculate
(with $s=w-\overline{w}$),
$$
\begin{array}{ll}
C(x,y,z)s&
=\{x\star(\overline{y}\star\overline{s})\}\star
z-(x\star {\overline y})\star ({\overline s} \star z)\\
&
=-\{x\star (\overline{y}\star s)\}\star z+
(x\star \overline{y})\star(s\star z)\\
&=\{[x,\overline{y},s]-
(x\star \overline{y})\star s\}\star z+
(x\star \overline{y})\star(s\star z)\\
&
=[x,\overline{y},s]\star z-
[x\star\overline{y},s,z]=
[s,x,\overline{y}]\star z
+[s,x\star \overline{y},z]\\
&=\{(s\star x)\star \overline{y}-
s\star(x\star\overline{y})\}\star z+
\{s\star(x\star \overline{y})\}\star z-
s\star\{(x\star \overline{y})\star z\}\\
&=\{(s\star x)\star\overline{y}\}\star z-
s\star\{(x\star\overline{y})\star z\}=
B(x,y,z)s.
\end{array}
$$
Then,Eq.(5.21)
is rewritten as 
$$
\{B(x,y,z)-B(y,x,z)+
B(z,y,x)-B(z,x,y)\}w$$
$$
=-\{B(x,y,z)-
B(y,x,z)+
B(z,y,x)-
B(z,x,y)\}(w-\overline{w}).\eqno(5.23)
$$
However,
Eq.(5.19)
for
$(x\leftrightarrow z),$
gives also
$$
\{B(z,y,x)-B(z,x,y)\}(w-\overline{w})=
\{C^{'}(x,y,z)-C^{'}(y,x,z)\}(w-\overline{w})$$
$$
=-\{B(x,y,z)-B(y,x,z)\}(w-\overline{w})$$
or
$$
\{B(x,y,z)-
B(y,x,z)+
B(z,y,x)-B(z,x,y)\}(w-\overline{w})=0$$
which yields Eq.(B),i.e.
$$
B(x,y,z)-B(y,x,z)+
B(z,y,x)-B(z,x,y)=0\eqno(B)
$$
in view of Eq.(5.23).
\par
We next set 
$x=e$
in Eq.(5.19) to obtaion
$$
\{B(e,y,z)-
B(e,z,y)\}w=
\{C^{'}(z,y,e)-C^{'}(y,z,e)\}w$$
or equivalently
$$
[w,\overline{y},z]-
[w,\overline{z},y]=
[y,\overline{z},w]-
[z,\overline{y},w]$$
which is one of Eq.(A.1),
if we change variables suitably.
Then together with
Eq.(5.9),
we find the validity of Eq.(5.6a)i.e.
$$
A(x,y,z)-A(y,z,x)=
A(z,x,y)-A(z,y,x).\eqno(A)$$
Since Eq.(5.6b)
is nothing but
Eq.(5.19),
then
Lemma 5.3
show that
the
$A^{*}$ is an almost pre-structurable algebra.
But $A^{*}$ is unital by  assumption
and these prove $A^{*}$ to be pre-structurable,
This completes the proof of Theorem 5.15.$\square$
\par
The special case of $Q(x,y,z)=0$
identically in Theorem 5.15 immediately
reproduces
(iii) of Theorem 5.5 of
[A-F,93]
by giving
$$
B(x,y,z)=
C(y,x,z)+C(z,x,y)+
C^{'}(z,y,x).\eqno(X)
$$
\par
\vskip 3mm
{\bf Theorem 5.16}
\par
\vskip 3mm
A necessary and sufficient condition that
a unitary involutive
algebra
$A^{*}$
being  structurable is the validily of
Eq.(sk)
and Eq.(X)
(or equivalently
$Q(x,y,z)=0).$
\par
\vskip 3mm
{\bf Remark 5.17}
\par
\vskip 3mm
Many interesting unital involution algebra containing Jordan and
alternative algebras are structurable.
It is rather hard
to find examples of a simple
pre-structurable but not
structurable algebra.
\par
\vskip 3mm
{\bf 6.\ Kantor Triple System and $A$-ternary Algebra}
\par
\vskip 3mm
Let $V$ be a vector space over a field $F$,
equipped with a tri-linear map
$$
V\otimes V\otimes V\rightarrow V
$$
$$
x\otimes y\otimes z\rightarrow xyz.
\eqno(6.1)
$$
If the triple product $xyz$ satisfies
$$
uv(xyz)=
(uvx)yz-
x(vuy)z+
xy(uvz)\eqno(6.2)
$$
for any
$u,v,x,y,z\in V,$
then
$(V,xyz)$
is called a generalized
Jordan triple system.
Moreover,
if it satisfies
$$
xyz=zyx,\eqno(6.3)$$
then
$(V,xyz)$
defines a Jordan triple system
[J.68].
It is often more convenient to introduce a
multiplication operator
$L(x,y)\in\mbox{End} \ V$
by
$$
L(x,y)z:=xyz.\eqno(6.4)$$
Then,
Eq.(6.2)
is equivalent to a Lie algebra relation of
$$
[L(u,v),L(x,y)]=
L(uvx,y)-
L(x,vuy).\eqno(6.5)$$
Moreover,
suppose that
$K(x,y)\in\mbox{End}\ V$
given by
$$
K(x,y)z=
xzy-yzx\eqno(6.6)$$
satisfies
$$
K(K(u,v)x,y)=
L(y,x)K(u,v)+
K(u,v)L(x,y).\eqno(6.7)$$
Then
$(V,xyz)$
is called a Kantor triple systems
([Kan.73]).
Note that the Jordan triple system
is a Kantor triple system
with
$K(x,y)=0.$
\par
Also
it is known
(see Eq.(7.6))
that the condition Eq.(6.7)
is equivalent to
$$
K(xyz,w)-
K(xyw,z)=
-K(x,K(z,w)y),\eqno(6.8)$$
if Eq.(6.5)
holds valid.
\par
A main purpose of this section is to  note that
a structurable algebra
is intimately related to the Kantor triple system as is indicated in the
following Theorem
(see [F.94],[K-O.10]):
\par
\vskip 3mm
{\bf Theorem 6.1}
\par
\vskip 3mm
Let $A^{*}$ be a structurable algeba over a field
$F$ of
charachteristic $\not= 2.$
If we define a triple product $xyz$
in the vector space of
$A^{*}$ by
$$
xyz:=
(z\star\overline{y})\star x-
(z\star\overline{x})\star y+
(x\star\overline{y})\star z,\eqno(6.9)
$$
then
$(A^{*},xyz)$
is a Kantor triple system
such that it satisfies
$$
eex=x,\ \mbox{and}\ exe+2xee=3x\eqno(6.10)$$
for the unit element
$e$ of $A^{*}$.
Conversely if $(A^{*},xyz)$ is
a Kantor triple system over a field $F$
of charachteristic $\not=2,\not=3,$
satisfying Eq.(6.10)
for a privileged
element
$e$
of $A^{*}$,
and if we introduce a mapping
$x\rightarrow \overline{x}$
and a bi-linear product
$x\star y$
in $A^{*}$
by
$$
\bar{x}:=
2x-xee,\eqno(6.11a)$$
$$
x\star y:=
\bar{x}ey-
\bar{x} \bar{y}e+
yex,\eqno(6.11b)$$
then
$(A^{*},x\star y)$
is a structurable algebra with the
unit element
$e$ and
the involution map
$x\rightarrow \overline{x}.$
\par
First,
we shall prove here a slightly 
weaker
theorem in the following.
\par
\vskip 3mm
{\bf Theorem 6.2([O.05])}
\par
\vskip 3mm
Let $A^{*}$ be normal Lie-related
triality algebra
(see Def.1.6).
Then,
the triple product in
$A^{*}$ defined by
$$
xyz:=
k\{l(x\star\bar{y}+
y\star\bar{x})-
d_{0}(x,y)-
d_{2}(\bar{x},\bar{y})\}z\eqno(6.12)
$$
for $k\in F,(k\not =0)$,
leads to a generalized Jordan triple system
$(A^{*},xyz).$
\par
For a proof of this Theorem,
we need the following Lemma.
\par
\vskip 3mm
{\bf Lemma 6.3}
\par
\vskip 3mm
Let $A^{*}$ be a pre-normal Lie related triality algebra.
If we set
$$
D_{0}(x,y):=
d_{0}(x,y)+
d_{2}(\bar{x},\bar{y}),\eqno(6.13a)$$
then it satisfies
$$
[D_{0}(u,v),
D_{0}(x,y)]=
D_{0}(D_{0}(u,v)x,y)+
D_{0}(x,D_{0}(u,v)y).\eqno(6.13b)$$
We also have
$$
[d_{3-j}(\bar{x},\bar{y})+
d_{j+2}(x,y),l(z)]=
l((d_{j+1}(x,y)+
d_{2-j}(\bar{x},\bar{y}))z),\eqno(6.14a)$$
and
$$
\{d_{3-j}(\bar{x},\bar{y})-d_{j+2}(x,y),
l(z)\}_{(+)}=
l((d_{j+1}(x,y)-
d_{2-j}(\bar{x},\bar{y}))z)\eqno(6.14b)$$
where we have set
$$
\begin{array}{l}
[X,Y]=XY-YX,\\
\{X,Y\}_{+}=
XY+YX\end{array}$$
for $X,Y,\in\mbox{End}\ A^{*}.$
\par
\vskip 3mm
{\bf Proof}
\par
\vskip 3mm
By Eq.(1.20),
we calculate
$$
\begin{array}{ll}
[D_{0}(u,v),D_{0}(x,y)]&
=[d_{0}(u,v)+d_{2}(\overline{u},\overline{v}),d_{0}(x,y)+
d_{2}(\overline{x},\overline{y})]\\
&=
d_{0}((d_{0}(u,v)+d_{2}(\overline{u},\overline{v}))x,y)+
d_{0}(x,(d_{0}(u,v)+
d_{2}(\overline{u},\overline{v}))y)\\
&+
d_{2}((d_{1}(u,v)+
d_{0}(\overline{u},\overline{v}))\overline{x},\overline{y})+
d_{2}(x,(d_{1}(u,v)+
d_{0}(\overline{u},\overline{v}))\overline{y}).
\end{array}
$$
Also,
we note
$$
\begin{array}{ll}
D_{0}(D_{0}(u,v)x,y)
&
=d_{0}(D_{0}(u,v)x,y)+
d_{2}(\overline{D_{0}(u,v)x},\overline{y})\\
&
=
d_{0}((d_{0}(u,v)+
d_{2}(\overline{u},\overline{v}))x,y)+
d_{2}((d_{0}(\overline{u},\overline{v})+
d_{1}(u,{v}))
\overline{x},\overline{y})
\end{array}
$$
and
$$
\begin{array}{ll}
D_{0}(x,D_{0}(u,v)y)&
=d_{0}(x,D_{0}(u,v)y)+
d_{2}(\overline{x},
\overline{D_{0}(u,v)y})\\
&=
d_{0}(x,(d_{0}(u,v)+
d_{2}(\overline{u},\overline{v}))y)+
d_{2}(\overline{x},(d_{0}(\overline{u},\overline{v})
+d_{1}(u,v))\overline{y})
\end{array}
$$
where we noted Eq.(1.23).
These prove Eqs.(6.13b).
\par
We next rewrite the triality relation
Eq.(1.41d)
as
$$
d_{3-j}(\overline{x},\overline{y})(z\star w)=
(d_{j+1}(x,y)z)\star w+
z\star (d_{j+2}(u,v)w)$$
which yield
$$
d_{3-j}(\overline{x},\overline{y})l(z)=
l(d_{j-1}(x,y)z)+
l(z)d_{j+2}(\overline{x},\overline{y}).\eqno(6.15)
$$
Letting
	$x\rightarrow \bar{x}$ and
$y\rightarrow\bar{y}$
with
$j\rightarrow 1-j,$
this is rewritten as
$$
d_{j+2}(x,y)l(z)=
l(d_{2-j}(\overline{x},\overline{y})z)+
l(z)d_{3-j}(x,y).$$
Adding or sub-tracting both relations,
we obtain
Eqs.(6.14).$\square$
\par
After these preparations,
we will now proceed to the proof of Theorem 6.2.
First,
choosing $j=0$
in Eq.(6.14a)
and letting
$x\rightarrow\overline{x}$
and
$y\rightarrow\overline{y},$
it yields
$$
[D_{0}(x,y),l(z)]=
l((d_{1}(\overline{x},\overline{y})+
d_{2}(x,y))z).\eqno(6.16)$$
For simplicity,
we set
$$
s=u\star\overline{v}+
v\star\overline{u},\ t=x\star\overline{y}+
y\star\overline{x}\eqno(6.17)$$
so that we can write
$$
L(u,v)=k\{l(s)-D_{0}(u,v)\}$$
$$
L(x,y)=k\{l(t)-D_{0}(x,y)\},\eqno(6.18)
$$
and calculate
$$
[L(u,v),L(x,y)]
=
k^{2}[l(s)-D_{0}(u,v),
l(t)-D_{0}(x,y)]
$$
$$
=
k^{2}\{[l(s),l(t)]-
[D_{0}(u,v),l(t)]+
[D_{0}(x,y),l(s)]+
[D_{0}(u,v),D_{0}(x,y)]\}
$$
$$
=
k^{2}\{[l(s),l(t)]-
l((d_{1}(\overline{u},\overline{v})+
d_{2}(u,v))t)+
l((d_{1}(\overline{x},\overline{y})+
d_{2}(x,y))s)$$
$$
+
D_{0}(D_{0}(u,v)x,y)+D_{0}(x,D_{0}(u,v)y)\}.
\eqno(6.19)
$$
Also
$$
L(L(u,v)x,y)$$
$$
=
k^{2}\{l(\ (l(s)-D_{0}(u,v))x\star \overline{y}+
y\star
(\overline{(l(s)-D_{0}(u,v))x)}\ )$$
$$
-D_{0}((l(s)-D_{0}(u,v))x,y)\}$$
$$
=
k^{2}\{l((s\star x)\star\overline{y}+
y\star(\overline{x}\star s))-
l(D_{0}(u,v)x\star\overline{y}+
y\star
\overline{D_{0}(u,v)x)}$$
$$
-
D_{0}(s\star x,y)+
D_{0}(D_{0}(u,v)x,y)\}
$$
and
$$
L(x,L(v,u)y)$$
$$
=k(l(x\star\overline{L(v,u)y}+
L(v,u)y\star\overline{x}\}
-D_{0}(x,L(v,u)y)\}$$
$$
=
k^{2}\{l\{x\star
\overline{((l(s)+ D_{0}(u,v))y)}
+(l(s)+
D_{0}(u,v))y\star
\overline{x}\}-
D_{0}(x,(l(s)+D_{0}(u,v))y)\}$$
$$=
k^{2}\{l(x\star(\overline{y}\star s)+
(s\star y)\star\overline{x})+
l(x\star\overline{D_{0}(u,v)y})+
l(D_{0}(u,v)y\star\overline{x})$$
$$
-D_{0}(x,s\star y)-
D_{0}(x,D_{0}(u,v)y)\}
$$
where 
we noted $\overline{s}=s$
an $L(v,u)=l(s)-D_{0}(v,u)=l(s)+
D_{0}(u,v).$
Then setting
$$
R=[L(u,v),L(x,y)]-
L(L(u,v)x,y)+
L(x,L(v,u)y),$$
we can rewrite it as
$$
R=k^{2}\{
[l(s),l(t)]+
l(w)+D_{0}(s\star x,y)-D_{0}(x,s\star y)\},\eqno(6.20)$$
with
$$
w
=-(d_{1}(\overline{u},\overline{v})+
d_{2}(u,v))t+
(d_{1}(\overline{x},\overline{y})+
d_{2}(x,y))s
$$
$$
-(s\star x)\star\overline{y}
-y\star(\overline{x}\star s)+
x\star(\overline{y}\star s)+
(s\star y)\star \overline{x}+J,\eqno
(6.21)
$$
where we have set further
$$
J=(D_{0}(u,v)x)\star \overline{y}+
y\star\overline{D_{0}(u,v)x}+
x\star\overline{D_{0}(u,v)y}+
(D_{0}(u,v)y)\star\overline{x}.$$
We calculate
$J$ to be
$$
\begin{array}{ll}
J&
=(d_{0}(u,v)+d_{2}(\overline{u},\overline{v}))x\star{y}+
y\star(d_{0}(\overline{u},\overline{v})+
d_{1}(u,v))\overline{x}\\
&+x\star(d_{0}(\overline{u},\overline{v})+
d_{1}(u,v))\overline{y}+
(d_{0}(u,v)+
d_{2}(\overline{u},\overline{v}))y\star\overline{x}\\
&=
\overline{d_{2}(u,v)}
(x\star\overline{y})+
\overline{d_{1}(\overline{u},\overline{v})}(x\star y)+
\overline{d_{2}(u,v)}(y\star x)+
\overline{d_{1}(\overline{u},\overline{v})}
(y\star \overline{x})\\
&=
(d_{1}(\overline{u},\overline{v})+
d_{2}(u,v))
(x\star\overline{y}+y\star\overline{x})=(d_{1}(\overline{u},\overline{v})+
d_{2}(u,v))t.
\end{array}
$$
by the triality relation Eq.(1.39c) and
Eq.(1.39g).
This
then leads to $w=0$
since we calculate
$$
\begin{array}{ll}
w&
=-(d_{1}(\overline{u},\overline{v})+
d_{2}(u,v))t+
(d_{1}(\overline{x},\overline{y})+
d_{2}(x,y))s\\
&+\{-r(\overline{y})r(x)-
l(y)l(\overline{x})+
l(x)l(\overline{y})+
r(\overline{x})r(y)\}s\\
&+(d_{1}(\overline{u},\overline{v})+
d_{2}(u,v))t=0.
\end{array}
$$
Moreover,
we note
$$
\begin{array}{ll}
D_{0}(s\star x,y)-D_{0}(x,s\star y)&\\
&=
d_{0}(s\star x,y)+
d_{2}(\overline{s\star x},\overline{y})-
d_{0}(x,s\star y)-d_{2}(\overline{x},\overline{s\star y})\\
&=
-d_{0}(y,\overline{\overline{x}\star s})-
d_{2}(\overline{x},\overline{\overline{y}\star s})-
d_{0}(x,\overline{\overline{y}\star s})
-d_{2}(\bar{y},
\overline{s\star x})\\
&=
-\{Q(y,\bar{x},s)
-d_{1}(s,\overline{y\star\bar{x}})\}-
\{Q(x,\bar{y},s)-d_{1}(s,\overline{x\star\overline{y}})\}\\
&=
d_{1}(s,x\star\bar{y})+
d_{1}(s,y\star \bar{x})=
d_{1}(s,x\star\bar{y}+
y\star\bar{x})=d_{1}(s,t)=\\
&=-[l(s),l(t)]
\end{array}
$$
where we used $Q(x,y,z)=0$
together with
$\overline{s}=s$
and
$\overline{t}=t.$
Therefore we have
$R=0$
by Eq.(6.20),
completing the proof of Theorem 6.2.$\square$
\par
\vskip 3mm
{\bf Remark 6.4}
\par
\vskip 3mm
If we rewrite Eq.(6.12) as
$$
xyz=k\{(x\star\bar{y}+
y\star\bar{x})\star z+
(z\star\bar{y})\star x-
(z\star\bar{x})\star y-
d_{0}(x,y)z\},$$
and
note Eq.(1.39d)i.e.
$$
d_{0}(x,y)z+
d_{0}(y,z)x+
d_{0}(z,x)y=0,$$
this gives
$$
K(x,y)=
k\{r(y)r(\bar{x})-
r(x)r(\bar{y})+
l(x\star\bar{y}-
y\star\bar{x})-
d_{0}(x,y)\}.\eqno(6.22)
$$
Hereafter in this section,
we assume $A^{*}$ to be structurable over the
field $F$ of
charachterictic
$\not= 2.$
Then choosing
$k={1\over 2}$
and noting
$$
d_{0}(x,y)=r(y)r(\bar{x})-
r(x)r(\overline{y})-
l(x\star\bar{y}-
y\star\bar{x}),$$
Eq.(6.22) gives
$$
K(x,y)=
l(x\star\bar{y}-
y\star\bar{x})\eqno(6.23)
$$
while Eq.(6.12)
reproduces Eq.(6.9).
We can then  prove the validity of Eq.(6.7)
as follows:
Setting now
$$
s=u\star\overline{v}-
v\star\overline{u}=
w-\overline{w},\ 
(w=u\star\overline{v}),\eqno(6.24)$$
we have
$
K(u,v)=l(s)$
so that Eq.(6.7)
becomes
$$
K(s\star x,y)z=
yx(s\star z)+
s\star(xyz)$$
or
$$
\begin{array}{ll}
\{
(s\star x)\star\bar{y}-
y\star\overline{(s\star x)}
\}\star z
&=\{(s\star z)\star\bar{x}\}\star y-
\{(s\star z)\star\bar{y}\}
\star x+
(y\star\bar{x})
\star(s\star z)\\
&+
s\star\{
(z\star \bar{y})\star x-
(z\star\bar{x})
\star y+
(x\star \bar{y})\star z\}
\end{array}
$$
by Eq.(6.9) and
(6.23).
This relation is rewriten as
$$
\{B(x,y,z)-B(z,x,y)+
B(z,y,x)-
C(y,x,z)\}s=0$$
which is satisfied in view of Eqs.(B) and
(5.22).
This proves that
$(A^{*},xyz)$ is a Kantor triple system.
\par
Moreover,Eq.(6.10)
holds valid by Eq.(6.9).
However,
the converse statement that any Kantor triple system
$(A^{*},xyz)$ satisfying Eq.(6.10)
will give a structurable algebra
requires
more calculations.
See [F.94]
or [K-O.10] for its proof.
\par
\vskip 3mm
{\bf Remark 6.5}
\par
\vskip 3mm
The structurable algebra has been originally defined by Allison ([A.78])
in terms of the Kantor triple system as in Theorem 6.1.
We redefined
it in this note
as in [K-O.14].
It is often more appropriate to
consider a triple
($A^{*},x\star y,xyz$)
by defining:
\par
\vskip 3mm
{\bf Def.6.6}
\par
\vskip 3mm
A triple ($A^{*},x\star y,xyz$)
with a bi-linear product $x\star y$ and a triple
product
$xyz$
in a vector space $A^{*}$ is called an Allison ternary algebra or
simply $A$-ternary algebra,
provided that we have
\begin{description}
\item[(1)]
($A^{*},x\star y$)
is a structurable algebra
\item[(2)]
($A^{*}, xyz$)
is a Kantor triple system
\item[(3)]
The triple product
$xyz$ is expressed in terms of the
bi-linear product by
$$
xyz=(x\star\bar{y})\star z+
(z\star\bar{y})\star x-
(z\star\bar{x})\star y.$$
\end{description}
\par
Note that both the bi-linear product 
$x\star y$ and
the tri-linear one $xyz$
are intimralty
related to each other as in Theorem 6.1.
Other example of such triple are
$(\alpha,\beta,\gamma)$
ternary algebra
([K-O.10])
based upon unital
$(\alpha,\beta,\gamma)$
ternary system
as well as some balanced
$(-1,-1)$FKTS
(see [E-K-O.03 and 05]).
\par
For the $A$-ternary algebra,
we note that we have
$$
\begin{array}{l}
K(x,y)=d_{2}(\bar{x},\bar{y})-d_{0}(x,y)=
l(x\star\bar{y}-y\star\bar{x})\\
L(x,y)+
L(y,x)=l(x\star \bar{y}+y\star\bar{x})\hskip 85mm
{\rm(6.25)}\\
L(x,y)-L(y,x)=
-d_{0}(x,y)-d_{2}(\bar{x},\bar{y})
\end{array}
$$
which determine
$K(x,y)$
and $L(x,y)$
in terms of 
$d_{j}(x,y)$
and
$l(x).$
\par
Conversly,
we can express
$$
\begin{array}{l}
l(x+\bar{x})=L(e,x)+L(x,e)\\
l(x-\bar{x})=K(x,e)\\
d_{0}(x,y)=
{1\over 2}\{
L(y,x)-L(x,y)-K(x,y)\}\\
d_{2}(x,y)=
{1\over 2}\{K(\bar{x},\bar{y})-
L(\bar{x},\bar{y})+
L(\bar{y},\bar{x})\}\\
d_{1}(x,y)-
d_{1}(\bar{x},\bar{y})=
-K(\bar{x},\bar{y})+
K(x,y)\hskip 80mm
{\rm(6.26)}\\
d_{1}(x,y)+
d_{1}(\bar{x},\bar{y})=
L(y,x)-
L(x,y)+
L(e,\bar{x}\star y)-
L(\bar{x}\star y,e).
\end{array}
$$
These relations will be used
to prove
some results in the next section.
\par
\vskip 3mm
{\bf 7.\ Lie algebras and superlagebras associated with
$(\varepsilon,\delta)$
Freudenthal-Kantor Triple System (FKTS)}
\par
\vskip 3mm
In this section,
we first note that the Kantor triple
system is a special case of a more general
$(\varepsilon,\delta)$
Freudenthal-Kantor triple system
[Y-O.84],
(see also
[Kam.87]
for many earlier references on the subject)
and second that we can construct Lie or Lie superalgebra out of 
these triple systems.
\par
Let $V$ be a vector space over a
field $F$ with a triple
product $xyz.$
Let the multiplication operators
$L(x,y),$
and $K(x,y)\in \mbox{End}\  V$
be given by
$$
L(x,y)z:=xyz,\ 
K(x,y)z:=xzy-\delta yzx.\eqno(7.1)$$
If they satisfy
$$
[L(u,v),L(x,y)]=
L(uvx,y)+
\varepsilon L(x,vuy)\eqno(7.2)$$
and
$$
K(K(u,v)x,y)=
L(y,x)K(u,v)-
\varepsilon K(u,v)L(x,y)\eqno(7.3)$$ 
for any $u,v,x,y,\in V,$
then,
we call the triple system
$(V,xyz)$ be
a
$(\varepsilon,\delta)$
Freudenthal-Kantor triple system
[Y-O.84],
where $\varepsilon$ and $\delta$ are constants with values either $1$
or $-1$.
Then,
comparing these
 with
Eq.(6.5),(6.6),
and (6.7),
we see that the Kantor triple system
is precisely
$(-1,1)$
FKTS.
Before going into further details,
we note that
Eq.(7.1) implies
$$
K(y,x)=
-\delta K(x,y)\eqno(7.4)$$
while Eq.(7.2)
is equivalent to the validity of
$$
uv(xyz)=
(uvx)yz+
\varepsilon x(vuy)z+
xy(uvz).\eqno(7.5)$$
\par
First,
we note that
Eq.(7.3) can be replaced by
$$
K(K(u,v)x,y)=
K(yxu,v)+
K(u,yxv)\eqno(7.6)
$$
under the validity of Eq.(7.2)
when we note
\par
\vskip 3mm
{\bf Lemma 7.1}
\par\vskip 3mm
If Eq.(7.2) holds,
we then have 
$$
L(y,x)K(u,v)-\varepsilon K(u,v)L(x,y)=
K(yxu,v)+
K(u,yxv).\eqno(7.7)$$
\par
\vskip 3mm
{\bf Proof}
\par
\vskip 3mm
We calculate
$$
\begin{array}{l}
L(y,x)K(u,v)z-
\varepsilon K(u,v)L(x,y)z\\
=yx(uzv-\delta vzu)-
\varepsilon\{u(xyz)v-\delta v(xyz)u\}.
\end{array}
$$
Moreover,
we note
$$
yx(uzv)=
(yxu)zv+
\varepsilon u(xyz)v+
uz(yxv)$$
by Eq.(7.5)
so that
$$
\begin{array}{l}
L(y,x)K(u,v)z-
\varepsilon K(u,v)L(x,y)z\\
=(yxu)zv+
\varepsilon u(xyz)v+uz(yxv)\\
-\delta(yxv)zu-\varepsilon \delta v(xyz)u-
\delta vz(yxu)\\
-\varepsilon u(xyz)v+\varepsilon\delta v(xyz)u\\
=(yxu)zv+
uz(yxv)- 
\delta(yxv)zu-
\delta vz(yxu)\\
=\{(yxu)zv-\delta vz(yxu)\}+
\{uz(yxv)-
\delta (yxv)zu)\}\\
=K(yxu,v)z+
K(u,yxv)z,
\end{array}
$$
which yields
Eq.(7.7).
$\square$
\par
\vskip 3mm
{\bf Proposition 7.2}
\par
\vskip 3mm
Let $(V,xyz)$
be a $(\varepsilon,\delta)$
FKTS.
We then have
$$
K(u,v)K(x,y)
$$
$$
=\varepsilon\delta L(K(u,v)y,x)-\varepsilon L(K(u,v)x,y)\eqno
(7.8a)$$
$$
=L(v,K(x,y)u)-\delta L(u,K(x,y)v)\eqno
(7.8b)
$$
for any
$u,v,x,y\in V.$
\par
\vskip 3mm
{\bf Proof }
\par
\vskip 3mm
First,
let us define
$$
K(V,V):=
\mbox{span}
\{K(x,y),\forall,
x,y\in V\}.\eqno(7.9)
$$
Then,
for any
$\sigma \in K(V,V),$
Eq.(7.3)
gives
$$
K(\sigma x,y)=
L(y,x)\sigma-\varepsilon\sigma L(x,y)\eqno(7.10)$$
so that
$$
(\sigma x)zy-\delta yz(\sigma x)=
yx(\sigma z)-\varepsilon\sigma(xyz)$$
which is rewritten as
$$
\sigma(xyz)=
-\varepsilon(\sigma x)zy+
\varepsilon yx(\sigma z)+
\varepsilon \delta yz(\sigma x)\eqno(7.11)$$
since 
$\varepsilon^{2}=1.$
We then calculate
$$
\begin{array}{ll}
\sigma K(x,z)y&
=\sigma(xyz)-\delta\sigma(zyx)\\
&=-\varepsilon(\sigma x)zy+
\varepsilon yx(\sigma z)+
\varepsilon\delta yz(\sigma x)+
\varepsilon\delta(\sigma z)xy-
\varepsilon\delta yz(\sigma x)-
\varepsilon\delta^{2}yx(\sigma z)\\
&=
-\varepsilon(\sigma x)zy+
\varepsilon\delta(\sigma z)xy=
-\varepsilon L(\sigma x,z)y+
\varepsilon\delta L(\sigma z,x)y
\end{array}
$$
which gives
$$
\sigma K(x,z)=
-\varepsilon L(\sigma x,z)+
\varepsilon\delta L(\sigma z,x).$$
Letting
$z\rightarrow y$
and choosing $\sigma=K(u,v),$
this leads to Eq.(7.8a).
\par
In order to prove Eq.(7.8b),
we note
$$
\begin{array}{l}
[L(u,v),L(x,y)]
=L(uvx,y)+\varepsilon L(x,vuy)\\
=-[L(x,y),L(u,v)]=
-L(xyu,v)-\varepsilon L(u,yxv)
\end{array}$$
so that
$$
L(uvx,y)+
\varepsilon L(x,vuy)+
L(xyu,v)+
\varepsilon L(u,yxv)=0.\eqno(7.12)$$
Letting 
$x\leftrightarrow u,$
it also gives
$$
L(xvu,y)+
\varepsilon L(u,vxy)+
L(uyx,v)+
\varepsilon L(x,yuv)=0.$$
Multiplying $\delta$
and subtracting it from
Eq.(7.12),
we find
$$
L(uvx-\delta xvu,y)+
\varepsilon L(x,vuy-\delta yuv)+
\varepsilon L(u,yxv-\delta vxy)
+L(xyu-\delta uyx,v)=0,$$
or
$$
L(K(u,x)v,y)+
\varepsilon L(x,K(v,y)u)+
\varepsilon L(u,K(y,v)x)+
L(K(x,u)y,v)=0.$$
Changing $x\leftrightarrow v$
and noting
$K(y,x)=-\delta K(x,y),$
this is rewritten as 
$$
L(K(u,v)x,y)-
\delta L(K(u,v)y,x)=
\varepsilon \delta L(u,K(x,y)v)-
\varepsilon L(v,K(x,y)u)$$
which proves Eq.(7.8b)$\square$
\par
We can then prove the following
(see [K-M-O.10]).
\par
\vskip 3mm
{\bf Corollary 7.3}
\par
\vskip 3mm
$K(x,y)$'s satisfy
\begin{description}
\item[(1)]
$$
K(z,w)K(x,y)K(u,v)+
K(u,v)K(x,y)K(z,w)$$
$$
=K(K(z,w)K(x,y)u,v)+
K(u,K(z,w)K(x,y)v)\eqno(7.13a)$$
$$
=\varepsilon\delta K(K(z,w)x,K(u,v)y)+
\varepsilon \delta K(K(u,v)x,K(z,w)y)
$$
\item[(2)]
$$
[[K(z,w),K(u,v)],K(x,y)]
=
$$
$$
-\varepsilon K(x,[K(z,w),K(u,v)]y)-
\varepsilon K([K(z,w),K(u,v)]x,y)\eqno
(7.13b)
$$
\end{description}
for any
$u,v,x,y,z,w\in V.$
Especially,
if we introduce two triple products
in $K(V,V)$
by
\begin{description}
\item[(a)]
$$\{K_{1},K_{2},K_{3}\}:=
K_{1}K_{2}K_{3}+
K_{3}K_{2}K_{1}\eqno(7.14a)$$
\item[(b)]
$$[K_{1},K_{2},K_{3}]:=
[[K_{1},K_{2}],K_{3}]\eqno(7.14b)
$$
\end{description}
then they define a Jordan triple system for
$\{K_{1},K_{2},K_{3}\}$
and
a Lie triple system for
$[K_{1},K_{2},K_{3}],$
respectivaly
for
$K_{1},K_{2},K_{3}\in K(V,V).
$
\par
\vskip 3mm
{\bf Remark 7.4}
\par
\vskip 3mm
Let $V=A^{*}$ be a structurable algebra over 
the field $F$ of
charachteristic $\not= 2$.
Then Eq.(6.23)
implies
$K(A^{*},A^{*})=l(H)$
since 
$x\star \overline{e}-
e\star \overline{x}=
(x-\overline{x})\in H$
where
$H=\{\overline{x}=-x,\ x\in A^{*}\}.$
Then corollary 7.3 implies that
$l(H)$ admitts
both Jordan triple system
and Lie triple system
for $K_{j}\in l(H)$
by Eqs.(7.14)
with
$(\varepsilon,\delta)=(-1,1).$
\par
Eqs.(7.8) enables us to 
prove the following theorem:
\par
\vskip 3mm
{\bf Theorem 7.5}
\par
\vskip 3mm
Let $T(\varepsilon,\delta):=(V,xyz)$
be a
$(\varepsilon,\delta)$FKTS.
Suppose that
$J\in \mbox{End} \ V$
satisfies
\begin{description}
\item[(1)]
$$
J(xyz)=
(Jx)(Jy)(Jz)\eqno
(7.15b)$$
\item[(2)]
$$J^{2}=-\varepsilon \delta \mbox{id}\eqno
(7.15b)$$
\end{description}
then a new triple product defined by
$$
[x,y,z]:=
x(Jy)z-\delta y(Jx)z+
\delta x(Jz)y-
y(Jz)x\eqno(7.10)$$
satisfies
\begin{description}
\item[(1)]
$$[x,y,z]:=
-\delta[y,x,z]\eqno
(7.17a)$$
\item[(2)]
$$[x,y,z]+
[y,z,x]+
[z,x,y]=0\eqno
(7.17b)$$
\item[(3)]
$$[u,v,[x,y,z]]=
[[u,v,x],y,z]+
[x,[u,v,y],z]+
[x,y,[u,v,z]].\eqno(7.17c)
$$
\end{description}
In other words,
$[x,y,z]$
is a Lie triple system for
$\delta=1$ and
an anti-Lie triple system for
$\delta=-1.$
\par
\vskip 2mm
For a proof of this Theorem,
the readers are  reffered to [K-O.00]
or [K-M-O.10].
We note that for the case of the Kantor triple system
$(\varepsilon=-1,\delta=+1),$
we can chose 
$J=id.$
However,
a more interesting case is to
consider a larger vector space:
$$
W=V\otimes V,\mbox{or}
\ W=\left(
\begin{array}{c}
V\\V
\end{array}
\right),\eqno(7.18)$$
and intorduce a triple product in $W$ by
$$
\left(
\begin{array}{c}
x_{1}\\y_{1}
\end{array}
\right)
\left(
\begin{array}{c}
x_{2}\\
y_{2}
\end{array}
\right)
\left(
\begin{array}{c}
x_{3}\\
y_{3}
\end{array}
\right)=
\left(
\begin{array}{c}
x_{1}x_{2}x_{3}\\
y_{1}y_{2}y_{3}
\end{array}
\right).
\eqno(7.19)$$
Then
$W=\left(
\begin{array}{c}
V\\V\end{array}\right)=
V\oplus V$
is also
a $(\varepsilon,\delta)$
FKTS,
if $V$ is a $(\varepsilon,\delta)$FKTS.
Moreover
$J\in\mbox{End}\ W$ 
given by
$$
J=\left(
\begin{array}{cc}
0,&1\\
-\varepsilon\delta,&0
\end{array}
\right),\eqno(7.20)$$
satisfies the conditions of
Eqs.(7.15)
in $W$.
\par
\noindent
Then,
the resulting Lie or anti-Lie triple system in
$W$ is rewritten as
$$
[\left(
\begin{array}{c}
u\\v\end{array}
\right),
\left(
\begin{array}{c}
x\\y
\end{array}
\right),
\left(
\begin{array}{c}
z\\
w\end{array}
\right)]
$$
$$
=\left(
\begin{array}{ll}
L(u,y)-\delta L(x,v),&
\delta K(u,x)\\
-\varepsilon K(v,y),&
\varepsilon L(y,u)-\varepsilon \delta L(v,x)
\end{array}
\right)
\left(
\begin{array}{c}
z\\
w\end{array}
\right).
\eqno(7.21)
$$
Further,
we define a multiplication operator
$L(X_{1},X_{2})\in
\mbox{End}\  W$
by 
$$
L(X_{1},X_{2})X_{3}:=
[X_{1},X_{2},X_{3}]\eqno(7.22)
$$
and set
$$
L(W,W)=
\mbox{span}
\{L(X_{1},X_{2}),
\forall X_{1},X_{2}\in W\}.\eqno(7.23)
$$
Then,
for any
$x,y,z,u,v$
and
$w\in V,\ D$
given by
$$
D=\left(
\begin{array}{ll}
L(x,y),&
\delta K(z,w)\\
-\varepsilon K(u,v),&
\varepsilon L(y,x)
\end{array}
\right)
\in
L(W,W)\eqno(7.24)
$$
is a derivation of the triple product
$[X_{1},X_{2},X_{3}],$
by the analogue of
Eq.(7.17c)
when we replace
$x\rightarrow X,\ y\rightarrow Y,\ z\rightarrow Z$
then,
i.e.
we have
$$
D[X_{1},X_{2},X_{3}]=
[DX_{1},X_{2},X_{3}]+
[X_{1},DX_{2},X_{3}]+
[X_{1},X_{2},DX_{3}].\eqno(7.25)
$$
Further,
we set
$$
L_{\overline{1}}=
W=\{X=
\left(
\begin{array}{c}
x\\y
\end{array}\right)
|x,y\in V\}
$$
$$
L_{\overline{0}}=
\{D|D\ 
\mbox{is a derivation of}\ 
[X_{1},X_{2},X_{3}]\}.
\eqno
(7.26)
$$
Then
$$L=L_{\overline{0}}\oplus L_{\overline{1}}\eqno(7.27)$$
is a Lie algebra for
$\delta=1,$
but a Lie superalgebra
for $\delta=-1$ with
$L_{\overline{0}}$
and
$L_{\overline{1}}$
being its even and odd part,
respectively.
Here,
we define the commutater
by (see [Y-O.84])
$$
[D_{1}\oplus X_{1},D_{2}\oplus X_{2}]=
([D_{1},D_{2}]+
[X_{1},X_{2}])
\oplus
(D_{1}X_{2}-D_{2}X_{1})
\eqno(7.28)
$$
for
$D_{1},D_{2}\in L_{\overline{0}}$
and
$X_{1},X_{2}\in W=L_{\overline{1}}$
by
$$
[D_{1},D_{2}]:=
D_{1}D_{2}-D_{2}D_{1}\in L_{\overline{0}}
\eqno(7.29)$$
and
$$
[X_{1},X_{2}]=
[\left(\begin{array}{c}
x_{1}\\
y_{1}
\end{array}
\right),
\left(
\begin{array}{c}
x_{2}\\
y_{2}\end{array}
\right)]=
L(X_{1},X_{2})=
$$
$$
\left(
\begin{array}{ll}
L(x_{1},y_{2})-\delta L(x_{2},y_{1}),&
\delta K(x_{1},x_{2})
\\
-\varepsilon K(y_{1},y_{2}),&
\varepsilon L(y_{2},x_{1})-
\varepsilon\delta L(y_{1},x_{2})
\end{array}
\right)
\in L_{\overline{0}}.
\eqno(7.29c)$$
However,
we will restrict ourselves for
a choice of
the derivation to be those given by
Eq.(7.24)
and
write Eq.(7.27)
as 
$$
L=L(W,W)\oplus W.\eqno(7.30)
$$
Then,
$L$ is $5$-graded Lie algebra or Lie superalgebra:
$$
L=L_{-2}\oplus L_{-1}\oplus
L_{0}\oplus L_{1}\oplus L_{2}\eqno(7.31)
$$
where we have set
$$
L_{-2}=\mbox{span}
\{
\left(
\begin{array}{cc}
0,&0\\
K(x,y),&
0\end{array}\right)
|x,y\in V\}
\eqno
(7.32a)
$$
$$
L_{-1}=
\mbox{span}
\{
\left(
\begin{array}{c}
0\\
x\end{array}\right)|
x\in V\}\eqno
(7.32b)
$$
$$
L_{0}=
\mbox{span}
\{
\left(
\begin{array}{cc}
L(x,y),&
0\\
0,&
\varepsilon L(y,x)
\end{array}
\right)
|x,y\in V\}
\eqno
(7.32c)
$$
$$
L_{1}=
\mbox{span}
\{
\left(
\begin{array}{c}
x\\
0\end{array}\right)|
x\in V\}\eqno
(7.32d)
$$
$$
L_{2}=\mbox{span}
\{
\left(\begin{array}{cc}
0,&K(x,y)\\
0,&0
\end{array}
\right)|x,y\in V\}.\eqno
(7.32e)
$$
Note
$$
L_{\overline{0}}=
L_{-2}\oplus L_{0}\oplus L_{2},\eqno
(7.33a)$$
$$
L_{\overline{1}}=
L_{-1}\oplus L_{1}.\eqno
(7.33b)
$$
If we introduce operators
$\theta$
and
$\sigma(\lambda)$
for
$\lambda\in F,\lambda\not=0$
in $\mbox{End}(L)$
by
$$
\theta\left(\begin{array}{c}
x\\y
\end{array}
\right)=
\left(
\begin{array}{c}
-\varepsilon y\\
\delta x
\end{array}
\right)
\eqno
(7.34a)
$$
$$
\sigma(\lambda)
\left(\begin{array}{c}
x\\
y\end{array}
\right)
=
\left(\begin{array}{c}
\lambda x\\
y/{\lambda}
\end{array}
\right),\eqno(7.34b)
$$
it is easy to see that they are automorphysm of
$[W,W,W]$,i.e.
we have for example
$$
\theta([X_{1},X_{2},X_{3}])
=[\theta X_{1},\theta X_{2},\theta X_{3}].$$
Extending
these actions to the whole of $L$ in a natural way,
we find also that they are
also automorphysm of the
Lie algebra
or Lie superalgebra
$L$.
Moreover,
they satisfy
\begin{description}
\item[(1)]
$\theta^{4}=id,\ \sigma(1)=\mbox{id}$
\item[(2)]
$
\theta^{2}=-\varepsilon\delta\ \mbox{id}$
for $L_{\overline{1}}$
and
$\theta^{2}=\mbox{id for}\ L_{\overline{0}}$
\item[(3)]
$
\sigma(\mu)\sigma(\nu)=\sigma(\mu\nu),\ 
(\mu,\nu\in F,
\mu\nu\not=0)$
\item[(4)]
$
\sigma(\lambda)\theta\sigma(\lambda)=\theta,$
for any
$\lambda\in F,\lambda\not= 0.$
\end{description}
\par
\vskip 3mm
Conversely,
for any
$5$-graded Lie algebra or Lie
super-algebra $L$ satisfying these conditions,
$A=L_{\overline{1}}$
with a triple product given by
$$
\{
X,Y,Z\}:
=
[[X,\theta Y],Z]$$
for any
$X,Y,Z\in A=L_{\overline{1}},$
essentially define a
$(\varepsilon,\delta)$FKTS
(see [E-K-O.13]).
\par
After these preparations,
we shall now restrict ourselves to the case of Kantor
triple system derived 
from a structurable algebra $A^{*}$.
Then,
we can construct
 Lie algebras in the two defferent 
ways:
one as in the present-section,
and the second one from the
structurable algebra as 
in section 3.
A question
will arise for
relationship between these two Lie algebras.
We will show next that
we can construct a Lie algebra given in section 3
from that of the present section for
a $A$-ternary algebra
$
(A^{*},x\star y,xyz$).
\par
\vskip 3mm
{\bf Theorem7.6}
\par
\vskip 3mm
Let ($A^{*},x\star y,xyz)$
be a $A$-ternary algebra.
Regarding
$(A^{*},x\star y)$ as a Kantor triple system,
we contruct a Lie algebra
$$
L=
L(W,W)\oplus W\eqno(7.35)
$$
as in Eqs(7.28),
and (7.29).
\par
For any non-zero constant
$\alpha,\beta, k\in F,$
we introduce raties
$\gamma_{1}/\gamma_{3}$
and $\gamma_{2}/\gamma_{3}$
for
$\gamma_{j}\in F$ by
\begin{description}
\item[(1)]
$$
\gamma_{2}/\gamma_{3}=
-2\alpha\beta\eqno
(7.36)$$
\item[(2)]
$$(\gamma_{3})^{2}/\gamma_{1}\gamma_{2}=
-k^{2},$$
\end{description}

and introduce
$\rho_{j}(x)$
and
$T_{j}(x,y)\ (j=0,1,2)$
by
$$
\rho_{1}(x)=
\left(
\begin{array}{c}
\alpha x\\
\beta x
\end{array}
\right),\  
\rho_{2}(x)=
\left(
\begin{array}{c}
k\alpha\overline{x}\\
-k\beta\overline{x}
\end{array}
\right),\eqno(7.37a)$$
$$
\rho_{0}(x)=
(\frac{k\gamma_{1}}{\gamma_{2}}
)
\left(
\begin{array}{ll}
\alpha\beta l(x+\overline{x}),&
\alpha^{2}l(x-\overline{x})\\
-\beta^{2}l(x-\overline{x}),&
-\alpha\beta l(x+\overline{x})
\end{array}
\right)
\eqno(7.37b)
$$
and
$$
T_{j}(x,y)=-
{\gamma_{3}\over \gamma_{2}}
\left(
\begin{array}{ll}
\alpha\beta(d_{j+1}(x,y)+d_{1-j}(\overline{x},\overline{y})),&
\alpha^{2}(d_{j+1}(x,y)-d_{1-j}
({\overline x},{\overline y}))\\
\beta^{2}(d_{j+1}(x,y)-
d_{1-j}(\overline{x},\overline{y})),&
\alpha\beta(d_{j+1}(x,y)+
d_{1-j}(
\overline{x},\overline{y}))
\end{array}
\right)
\eqno(7.37c)$$
for
$j=0,1,2.$
\par
We then can show that both
$\rho_{j}(x)$
and
$T_{j}(x,y)$
are elements of
$L(W,W)\oplus W,$
and satisfy the Lie algebra relations
given in section 3.
Moreover,
they satisfy
Eq.(3.12b),
i.e.
$$
T_{0}(x,
\overline{y\star z})+
T_{1}(z,\overline{x\star y})+
T_{2}(y,\overline{z\star x})=0.$$
\par
\vskip 2mm
Since the proof of this Theorem is lengthy,
we will not
go into it,
(see [K-O.13]).
Here,we simply mention that we
utilized
Eqs.(6.14),
(6.25)
and
(6.26)
for the purpose.
\par
We also note that the choice of
$\gamma_{0}=\gamma_{1}=\gamma_{2}=1$
requires
$k^{2}=-1$
and
$\alpha\beta=-{1\over 2}$
by Eq.(7.36).
Then the underlying field $F$ must contain 
the square root of
$-1.$
\par
\vskip 3mm
{\bf 8.\ $BC_{1}$-graded 
Lie Algebra of Type $B_{1}$}
\par
\vskip 3mm
As we noted in the previous sections,
any $A$-ternary algebra admitts two
constructions of Lie algebras:
The one given in section 3 exhibits
the triality,
but {\it not}
the $5$-graded structure,
while the other one based upon
the standard construction for
$(\varepsilon,\delta)$FKTS in section 7
manifests
the explicit $5$-graded nature
but {\it not}
the triality.
\par
Here in this section,
we will show that the
second method can be used to
prove that its associated
Lie algebra
$L$ is a 
$BC_{1}$-graded Lie algebra of type
$B_{1}$ ([Be-S.03]),
i.e.,
there exists a sub-Lie algebra  $sl(2)$ of
$L$
such that
$L$ regarded as the $sl(2)$
module is a direct sum
of trivial,
$3$-dimensional,
and
$5$-dimensional modules of 
$sl(2)$.
\par
Let 
$(A^{*},\ x\star y,\ xyz)$
be an 
$A$-ternary algebra with the
unit element
$e$ for the structurable algebra
$(A^{*},x\star y)$ so that it satisfies
$$
eex=x,\qquad 2xee+exe=3x.\eqno(8.1)$$
Following 
[K-O.10] or [E-K-O,13],
let $R,M\in \mbox{End}\ A^{*}$
be defined by
$$
Rx=xee,\qquad
Mx=exe\eqno(8.2)$$
so that 
Eq.(8.1)
gives
$$
M+2R=3id.\eqno(8.3)$$
We then have,
(assuming $2\not= 0,$
hereafter.)
\par
\vskip 3mm
{\bf Lemma 8.1}
\par
\vskip 3mm
\begin{description}
\item[(1)]
$$
L(xye,e)=L(e,yxe)\eqno(8.4a)
$$
\item[(2)]
$$
L(Rx,e)=
L(e,Mx)\eqno(8.4b)
$$
\item[(3)]
$$
L(Mx,e)=L(e,Rx)\eqno(8.4c)
$$
\item[(4)]
$$
M^{2}=R^{2}\eqno(8.4d)
$$
\item[(5)]
$$
K(x,e)e=
(R-1)x\eqno(8.4e)$$
\item[(6)]
$$
K(x,y)=
{1\over 2}K(K(x,y)e,e).
\eqno(8.4f)$$
\item[(7)]
$$
(R-1)(R-3)=0
\eqno(8.4g)
$$
\end{description}
\par
\vskip 3mm
{\bf Proof}
\par
\vskip 3mm
Setting $x=y=e$
we have
$$
[L(u,v),L(e,e)]=
L(uve,e)-
L(e,vue).$$
Moreover,
$eex=x$
implies
$$
L(e,e)=\mbox{id}\ (\equiv 1)$$
so that it yields
Eq.(8.4a)
by changing
$u\rightarrow x$
and
$v\rightarrow y.$
Then Eqs.(8.4b)
and (8.4c)
are special cases of
Eq.(8.4a)
for either
$x=e$
or $y=e$.
\par
In order to prove Eq.(8.4d),Eq.(6.2)
gives
$$
xe(eee)=
(xee)ee-
e(exe)e+
ee(xee)$$
which is rewritten as
$$
Rx=R^{2}x-M^{2}x+Rx$$
i.e.$R^{2}=M^{2}.$
Moreover,
Eq.(6.6)
leads to
$K(x,e)e=xee-eex=(R-1)x$
i.e.
Eq.(8.4e).
Similarly,
we find
$$
K(K(u,v)e,e)=
L(e,e)K(u,v)+
K(u,v)L(e,e)=
2K(u,v)$$
from Eq.(6.7)
to give Eq.(8.4d).
Finally,
from
Eqs.(8.4e)
and
(8.4d),
we calculate
$$
(R-1)x=
K(x,e)e=
{1\over 2}K(K(x,e)e,e)e=
{1\over 2}(R-1)
K(x,e)e=
{1\over 2}(R-1)^{2}x,$$
so that
$R-1=
{1\over 2}(R-1)^{2}$
or $(R-1)(R-3)=0$
as in Eq.(8.4e).
Note that this relation is consistent
with
$M^{2}=R^{2}$
in
Eq.(8.3).
$\square$
\par
In view of
Eq.(8.4g),
$(R-1)(R-3)=0$,
we can decompose
$A^{*}$ as in
$$A^{*}=V_{1}\oplus V_{3}\eqno(8.5)$$
where we have set
$$
V_{1}=\{x|Rx=x,\ x\in A^{*}\}\eqno(8.6a)$$
$$
V_{3}=\{x|Rx=3x,\ x\in A^{*}\}.\eqno(8.6b)
$$
For the details of this decomposition,
see also
([K-K.03]).
Moreover by Eq.(6.11a),
we have
$$
\overline{x}=(2-R)x,\eqno(8.7)$$
so that we can rewrite
Eqs.(8.6)
also as
$$
V_{1}=\{x|\overline{x}=
x,x\in A^{*}\},\eqno(8.8a)
$$
$$
V_{3}=
\{x|\overline{x}=
-x,x\in A^{*}\}.
$$
We next set
$$
h=
\left(
\begin{array}{cc}
L(e,e),&0\\
0,&-L(e,e)
\end{array}
\right)=
\left(\begin{array}{cc}
1&0\\
0&-1\end{array}
\right)
\eqno(8.9a)
$$
$$
f=\left(
\begin{array}{c}
e\\o
\end{array}
\right),\qquad
g=
\left(\begin{array}{c}
0\\e
\end{array}\right).\eqno(8.9b)
$$
We see then
$f,g\in W$
and
$h\in L(W,W)$
so that 
they are elements
of the
Lie algebra
$L(W,W)\oplus W.$
Moreover,
they satisfy the 
$sl(2)$
Lie relations of
$$
[h,f]=f,\qquad
[h,g]=-g,\qquad
[f,g]=h\eqno(8.10)$$
since we calculate
$$
[h,f]=
\left(\begin{array}{cc}
1&0\\
0&-1
\end{array}
\right)
\left(\begin{array}{c}
e\\
0
\end{array}\right)
=
\left(\begin{array}{c}e\\0\end{array}
\right)=f
$$
$$
[h,g]=
\left(\begin{array}{cc}
1&0\\
0&-1\end{array}\right)
\left(\begin{array}{c}
0\\e\end{array}
\right)=
\left(
\begin{array}{c}
0\\
-e\end{array}\right)=
-g
$$
and
$$
[f,g]=
[\left(
\begin{array}{c}
e\\0
\end{array}
\right),
\left(\begin{array}{c}
0\\e\end{array}\right)]$$
$$
=
\left(
\begin{array}{ll}
L(e,e)-L(e,0),&
K(e,0)\\
K(0,e),&
-L(e,e)+
L(0,e)
\end{array}
\right)
=
\left(\begin{array}{cc}
1&0\\
0&-1
\end{array}
\right)=h,$$
by Eqs.(7.28)
and (7.29).
\par
We note
 moreover that we have
$$
K(x,e)=0,\qquad
\mbox{if}\ x\in V_{1},\eqno(8.11)
$$
since we calculate
$$
K(x,e)=l(x\star\overline{e}-
e\star\overline{x})
=l(x-\overline{x}).
$$
Then,
Eq.(8.4)
implies
$$
K(A^{*},A^{*})=
K(V_{3},e).\eqno(8.12)
$$
We can now construct
$5$-dimentional modules
$M_{5}$
and
$3$-dimensional
$M_{3}$
of the
$sl(2)$ by
.$$
M_{5}=
$$
$$
span <
\left(
\begin{array}{ll}
0,&0\\
K(x,e),&0
\end{array}\right),
\left(\begin{array}{c}
0\\x\end{array}
\right),
\left(\begin{array}{ll}
L(x,e),&
0\\
0,&-L(e,x)\end{array}
\right),
\left(\begin{array}{c}
x\\0
\end{array}
\right),
\left(\begin{array}{ll}
0,&K(x,e)\\
0,&0\end{array}
\right)>
\eqno(8.13)
$$
for $x\in V_{3}$
satisfying
$\overline{x}=-x,$
and
$$
M_{3}=$$
$$
span <
\left(\begin{array}{c}
0\\x
\end{array}
\right),
\left(
\begin{array}{ll}
L(x,e),& 0\\
0,& -L(e,x)\end{array}
\right),
\left(\begin{array}{c}
x\\0
\end{array}
\right)>
\eqno(8.14)
$$
for $x\in V_{1},$
satisfying
$\overline{x}=x.$ (Note $K(x,e)=0 \ for \ x\in V_{1}.)$
\par
We further note that
\begin{description}
\item[(1)]
If
$\overline{x}=x,$
then $Rx=x,\ Mx=x,$
and $L(x,e)=L(e,x)$
\item[(2)]
If
$\overline{x}=-x,$
then $Rx=3x,\ Mx=-3x,$
and
$L(x,e)=-L(e,x)$
\end{description}
by
Eqs.(8.4b)
and (8.4c),
assuming the underlying field
$F$ to be of
charachteristic
$\not= 3$ in addition,
Especially,
we need not consider
elements of
$L$ of form
$$
\left(\begin{array}{cc}
L(e,x),&0\\
0,&-L(x,e)
\end{array}\right).$$
Finally,
the trivial modules can be constructed as follows.
\par
For any
$x,y\in A^{*},$
we set
$$
u={1\over 2}(R-1)(xye)\in V_{3}$$
$$
v=-{1\over 2}(R-3)(xye)\in V_{1}.\eqno(8.15)$$
We then have
$$
X=\left(
\begin{array}{cc}
L(u,e),&0\\
0,&-L(e,u)
\end{array}
\right)
\in M_{5}
\eqno(8.16a)$$
and
$$
Y=
\left(
\begin{array}{cc}
L(v,e),&0\\
0,&-L(e,v)
\end{array}
\right)
\in M_{3}.\eqno(8.16b)$$
Then
$$
\xi:=
\left(\begin{array}{cc}
L(x,y),&0\\
0,&-L(y,x)
\end{array}
\right)
-{1\over 3}X-Y\eqno(8.17)
$$
can be
verified to be elements of the trivial modules of
the $sl(2).$
Then rewriting
$$
\left(\begin{array}{cc}
L(x,y),&0
\\
0,&-L(y,x)\end{array}
\right)=
\xi+
{1\over 3}X+Y\in
M_{1}\oplus M_{3}\oplus M_{5}$$
together
with
$K(A^{*},A^{*})=
K(V_{3},e),$
we have 
$$
M_{1}:=
\{
\left(
\begin{array}{cc}
\phi&0\\
0&\phi^{'}
\end{array}
\right)
|\phi(e)=\phi^{'}(e)=0,\ \phi ,\phi^{'} \in L(A^{*},A^{*})\},
$$ 
and we obtain
$$L=L(W,W)\oplus W=M_{1}\oplus M_{3}\oplus M_{5}.$$ 
\par
Therefore,
we have found:
\par
\vskip 3mm
{\bf Proposition 8.2}
\par
\vskip 3mm
Let
$(A^{*},x\star y,xyz)$
be an $A$-ternary algebra
over the field $F$ of 
charachteristic
$\not= 2,\ \not= 3.$
Then,
its associated Lie algebra is a 
$BC_{1}$-graded Lie algebra of type $B_{1}.$
\par
\vskip 3mm
If we assume for simplicity the
field $F$ to be an algebraically
closed field of
charachteristic zero,
then we can show conversely that any
$BC_{1}$-graded
Lie algebra of
type $B_{1}$ can be 
constructed from some
$A$-ternary algebra.
For this purpose,
it is convenient to use the
terminology familiar in
the angular momentum
algebra in Quantum Mechanics
(e.g.[C-D-L.77])
by setting
$$
J_{3}=h,\ J_{+}=\sqrt{2}f,\ J_{-}=\sqrt{2}g
\eqno(8.18)
$$
which satisfy
$$
[J_{3},J_{\pm}]=
\pm J_{\pm},\ 
[J_{+},J_{-}]=
2J_{3}.\eqno(8.19)
$$
We write the generic irreducible
state as
$$
\Phi(j,m;\alpha)=|j,m;\alpha>
\eqno(8.20)
$$
for $j=0,1,2,$
corresponding to the
trivial,
$3$-dimensional,
and $5$-dimensional modules,
while the sub-quantum number
$m$ can assume
$2j+1$
values of
$j,j-1,\cdots ,-(j-1), -j.$
Also,
$\alpha$
in Eq.(8.20)
simply designates
other labels.
For example,
we may identify
$\alpha=x$,
satisfying
${\overline x}=-x$
or ${\overline x}=+x$
for modules $M_{5}$
and 
$M_{3}$
in Eqs.(8.13)
and (8.14).
We note
the commutation relations of
$$
[J_{3},\Phi(j,m;\alpha)]=
m\Phi(j,m;\alpha),
$$
$$
[J_{\pm},\Phi(j,m;\alpha)]=
\pm\sqrt{(j
\mp m)(j\pm m+1)}
\Phi(j,m\pm 1;\alpha).
\eqno(8.21)
$$
\par
Now, the
$BC_{1}$-graded Lie algebra
of type $B_{1}$
is then $5$-graded as in
$$
L=g_{-2}\oplus g_{-1}\oplus
g_{0}\oplus g_{1}\oplus
g_{2}
\eqno(8.22)
$$
when we set
$$
g_{m}=
\{x|J_{3}x=mx,\quad
(m=0,\pm 1,\pm 2),\ x\in L\}.
\eqno(8.23)
$$
Moreover,
$\theta$ given by
$$
\theta:
\Phi(j,m;\alpha)\rightarrow
(-1)^{j-m}
\Phi(j,-m,\alpha)
\eqno(8.24)
$$
is an automorphism
of $L$ of
order $2$,
letting
$g_{m}\leftrightarrow g_{-m}.$
Especially,
we note
$$
\theta:h\rightarrow -h,\quad f\leftrightarrow g.
\eqno(8.25)
$$
Then,
by Theorem 4.1 of
[E-K-O.13],
$A=g_{1}$
becomes a Kantor triple system
with respect to the triple product
$$
xyz=[[x,\theta(y)],z]
\eqno(8.26)
$$
for $x,y,z\in g_{1}.$
Noting
$f={1\over{\sqrt{2}}} J_{+}\in g_{1},$
we calculate then
$$
ffx=xff=fxf=x,\ 
\mbox{for}\ x=\Phi(1,1,;\alpha)\in g_{1}$$
and
$$
ffx=x,\ xff=-fxf=3x\ 
\mbox{for}\ x=\Phi(2,1;\alpha)\in g_{1},$$
so that they satisfy
the condition
of Eq.(6.10)
for
$e=f.$
Therefore,
by Theorem 6.1,
$g_{1}=A$
becomes a
$A$-ternary algebra.
\par
\vskip 3mm
{\bf Remark 8.3}
\par
\vskip 3mm
We can relax the condition in Proposition 8.2
and others 
as in
[E-K-O.13],
although we will not go into its detail.
Also,
an analogous theorem on
 Lie superalgebra associated with
$(-1,-1)$
Freudenthal-Kantor triple system is
given there.
Similarly,
some class of
$(1,1)$FKTS lead to
$BC_{1}$-graded Lie algebra of type
$C_{1}$ as well to a ternary system called a
$J$-ternary algebra
(see [E-O.11],
and
[A-B-G.02]).
\par
\vskip 3mm
{\bf 9.\ Final Comments}
\par
\vskip 3mm
In the previous sections,
we have studied relationship between the structurable
algebra and the Kantor triple system.
Since other
$(\varepsilon,\delta)$ FKTS do not appear 
to have a direct connection to 
the triality relation,
we did not discuss other
$(\varepsilon,\delta)$ FKTS.
Here,
we simply mention that some
$(-1,-1)$ FKTS have been used 
to construct exceptional Lie 
superalgebra
$D(2,1;\alpha),\ G(3),$
and
$F(4)$
(see [K-O.03],
[E-K-O.03]
and [E-K-O.05]).
Also,
some connections exist between
$(1,1)$
and $(-1,1)$FKTS
([E-K-O.13]).
\par
\vskip 2mm
The triality relations discussed in this note are
of a local type.
If $\sigma_{j}\in Epi(A)$
for
$j=0,1,2$
satisfy
$$
\sigma_{j}(xy)=
(\sigma_{j+1}x)
(\sigma_{j+2}y),\eqno(9.1)$$
in contrast to Eq.(1.1),
it is called a global triality relation,
and
$$
G=
\{
\sigma_{j}|
\sigma_{j}(xy)=
(\sigma_{j+1}x)
(\sigma_{j+2}y),\ 
\forall j=0,1,2,\ \forall,x,y\in A\}\eqno(9.2)$$
is a group 
(in general a Lie group)
instead of a
Lie algebra,
which may be called  triality group.
Here, the indices over $j$
are defined modulo $3$.
Its general structure is
harder to analyze,
and has not been
studied much.
Here, 
we will give a example based
upon the symmetric composition algebra
(see Example 2.2)
satisfying
$$
x(yx)=
(xy)x=
<x|x>y\eqno(9.3)$$
where
$<\cdot|\cdot>$ is a symmetric bi-linear 
non-degenerate form in $A$.
For any two elements $a,b\in A$
satisfying
$<a|a>=<b|b>=1,$
we set
$a=a_{1},$ and
$b=a_{2}$
and 
define 
$a_{3}$
by
$
a_{3}=a_{1}a_{2}.$
We then find
$$
a_{j}a_{j+1}=
a_{j+2},\qquad
<a_{j}|a_{j}>=1\eqno(9.4)$$
for $j=1,2,3,$
where the indices over $j$ is defined 
again modulo $3$,
i.e.
$a_{j\pm 3}=a_{j}.$
Introducing the multiplication operators
$L(x)$
and $R(x)$
again by
$$
L(x)y=xy,\qquad
R(x)y=yx\eqno(9.5)$$
we set
$$
\sigma_{j}(a)=
R(a_{j+1})R(a_{j+2})\eqno(9.6a)$$
$$
\theta_{j}(a)=
L(a_{j+2})L(a_{j+1}),\eqno(9.6b)$$
for
$j=1,2,3$.
We can then show the validity of
\begin{description}
\item[(i)]
$$
\sigma_{j}(a)(xy)=
(\sigma_{j+1}(a)x)
(\sigma_{j+2}(a)y)\eqno(9.7a)$$
\item[(ii)]
$$
\theta_{j}(a)(xy)=
(\theta_{j+1}(a)x)
(\theta_{j+2}(a)y)\eqno(9.7b)$$
\item[(iii)]
$$
\sigma_{j}(a)\theta_{j}(a)=
\theta_{j}(a)\sigma_{j}(a)=1\eqno(9.7c)$$
\item[(iv)]
$$
\theta_{j}(a)\theta_{j+1}(a)
\theta_{j+2}(a)=
\sigma_{j+2}(a)\sigma_{j+1}(a)\sigma_{j}(a)=1\eqno(9.7d)$$
\item[(v)]
$$
<\sigma_{j}(a)x|y>=
<x|\theta_{j}(a)y>
\eqno(9.7e)
$$
\item[(vi)]
$$
<\sigma_{j}(a)x|\sigma_{j}(a)y>=
<\theta_{j}(a)x|
\theta_{j}(a)y>=
<x|y>.
\eqno(9.7f)
$$
\end{description}
\par
Especially,
both 
$\sigma_{j}(a)$
and $\theta_{j}(a)$
satisfy the global triality relation.
The details will be given elsewhere.
\par
\vskip 3mm
{\bf References}
\par
\rm
\vskip 3mm
\noindent
[A.78]:
Allison,B.N.;
"A class of non-associative algebras with involution
containing
a class of Jordan algebra"\ 
Math,Ann 
{\bf 237},
(1978)
133-156
\par
\noindent
[A-B-G.02]:
Allison,B.N.,Benkart,G.,Gao,Y.;
"Lie algebras graded by
the root 
system
\par
\noindent
$BC_{r}, r\geq 2$"\ 
Memoirs of Amer.math.
Soci.
vol.{\bf 78}
(2002)
(Princeton,N.J.,
American 
Math.Soc.)
\par
\noindent
[A-F.93]:
Allison,B.N.,Faulkner;J.R.;
"Non-associative coefficient algebras for
Steinberg
unitary Lie algebras"\ 
J.Algebras {\bf 161}
(1993)133-158
\par
\noindent
[Ba-S.03]:Barton,C.H.,Sudbury,A.;
"Magic Squares
and matrix 
models of Lie algebras"\ 
Adv.Math.
{\bf 180}
(2003)
596-647
\par
\noindent
[Be-S.03]:
Benkart,G.,Smirnov,O.;
"Lie algebras graded by
the root system $BC_{1}$"\ 
Jour Lie Theory {\bf 13}
(2003)
91-1132
\par
\noindent
[C-D-L,77]:
Cohen-Tannoudji,C., Bernard,D., Laloe,F.;
"Quantum mechanics I"\ 
John-Wiley \& Sons, N.Y.
(1977)
\par
\noindent
[E.97]:
Elduque,A.;
"symmetric Composition Algebra"\ 
J.Algebra
{\bf 196}(1997)
282-300
\par
\noindent
[E.04]:
Elduque,A.;
"The magic square 
and symmetric 
compositions"  
Rev.
Math.
\par
\noindent
Iberoamericana
{\bf 20} (2004)
479-493
\par
\noindent
[E.06]:
Elduque,A.;
"A new look of Freudenthal- magic square"\ 
In Non-associative Algebras and 
 its Applications\ 
ed.by L.Sabinin,
L.Sbitneva,
and I.Shestakov,
Chapman and Hall,
N.Y.
(2006),
149-165
\par
\noindent
[E.07]:
Elduque,A.;
"The $A_{4}$-action on the Tetrahedron
Algebra"\ 
Proc.Roy.Soc.Edinburgh
{\bf A 137}
(2007)
1227-1242
\par
\noindent
[E-K-O.03]:
Elduque,A.,Kamiya,N.,Okubo,S.;
"Simple $(-1,-1)$ Balanced Freudenthal Kantor Triple systems"\ 
Glasgow Math.J.
{\bf 45}
(2003)
353-392
\par
\noindent
[E-K-O.05]:
Elduque,A.,Kamiya,N.,Okubo,S.;
"$(-1,-1)$ Balanced Freudenthal Kantor triple systems 
and non-commutative Jordan algebras;\ 
J.Algebra
{\bf 294}(2005)19-40
\par
\noindent
[E-K-O.13]:
Elduque,A.,Kamiya,N.,Okubo,S.;
"Left unital Kantor triple system and 
structurable algebra",  Linear and Multi linear Algebra 
(2013)
online,
arXiv 1205-2489
(2013),
vol.62. Issue 10, October (2014) 1293-1313.
\par
\noindent 
[E-O.00]:
Elduque,A.,Okubo,S.;
"On algebras satisfying $x^{2}x^{2}=N(x)x,$"\ 
Math.Z.
{\bf 235}
(2000)
275-314
\par
\noindent
[E-O.07]:
Elduque,A.,Okubo,S.;
"Lie algebras with $S_{4}$-action and
structurable algebras"\ 
J.Algebras
{\bf 307}
(2007)
864-890
\par
\noindent
[E-O.08]:
Elduque,A.,Okubo,S.;
"$S_{4}$-symmetry and the Tits construction of Exceptional Lie algebra and
Superalgebras"\ 
Pub.Math.
{\bf 52}
(2008)
315-346
\par
\noindent
[E-O.09]:
Elduque,A.,Okubo,S.;
"Lie algebras with $S_{3}$-or $S_{4}$-action and
generalized Malcev algebra"\ 
Proc.Roy.Soc.
Edingburgh
{\bf 139A}
(2009)
321-352
\par
\noindent
[E-O.11]:
Elduque,A.,Okubo,S.;
"Special Freudenthal-Kantor Triple System and Lie algebras 
with dicyclic symmetry"\ 
Proc.Roy.Soc.
Edinburgh
{\bf 141A}
(2011)
1225-1262
\par
\noindent
[F.94]:
Faulkner,J.R.;
"Struturable triple,Lie triple,
and Symmetric Spaces"\ 
Forum Math.
{\bf 6}
(1994)
637-650
\par
\noindent
[H-T.07]:
Hartwig,B.,Terwilliger,P.;
"The tetrahedron algebra,the Onsager algebra
and the $sl_{2}$ loop algebra"\ 
J.Algebra
{\bf 308}
(2007)
840-863
\par
\noindent
[J.68]:
Jacobson,N.;
"Structure and representations of Jordan algebra"\ 
Ammer.Math.Soc.
Colloy.Pub.{\bf 39}
(1968)
(provode,R.I)
\par
\noindent
[Kam.87];
Kamiya.N.;
"A structure theory of Freudenthal-Kantor
triple systems,"\ 
J.Alg.
{\bf 110}
(1987)
no.1.108-123
\par
\noindent
[Kam.89]:
Kamiya,N.;
"A structure theory of Freudenthal-Kantor
triple 
system 
III"\ 
Mem.
Fac.
Sci.
Shimane 
Univ.
{\bf 23}
(1989)
33-51
\par
\noindent
[Kam.95]:
Kamiya,N.;
"On generalized structurable algebras and
Lie-related 
triple"\ 
Adv.
Clifford Algebras
{\bf 5}
(1995)
127-140
\par
\noindent
[Kan.73]:
Kantor,I.L.;
"Models of the exceptional
 Lie algebras"\ 
 Sov.
Math.
Dokl.
{\bf 14}(1973)
254-258
\par
\noindent
[K-K.03]:
Kantor,I.L.,Kamiya,N.;
"A Peirce decomposition for generalized Jordan triple
systems of second order"\ 
{\bf 31}(2003)
no.12.5875-5913
\par
\noindent
[K-M-O.10]:
Kamiya,N.,Mondoc,D.,Okubo,S.;
"Structure Theory of
$(-1,-1)$
\par
\noindent
Freudenthal-
Kantor triple system"\ 
Bull.Austr.
Math.Soc.{\bf 81}
(2010)
132-155
\par
\noindent
[K-O.00]:
Kamiya,N.,Okubo,S.;
"On $\delta$-Lie Super Triple System associated
with
$(\varepsilon,\delta)$ 
\par
\noindent
Freudenthal-Kantor
super triple systems"\ 
Proc. Edinburgh Math.Soc.
{\bf 43}
(2000)
243-260
\par
\noindent
[K-O.03]:
Kamiya,N.,Okubo,S;
"Construction of Lie Superalgebras
$D(2,1;\alpha),$
\par
\noindent
$G(3)$ and $F(4)$
from some triple system"\ 
Proc. Edinburgh Math.Soc.{\bf 46}(2003)
87-98
\par
\noindent
[K-O.10]:
Kamiya,N.,Okubo,S.;
"Representation of
$(\alpha,\beta,\gamma)$
triple system"\ 
Linear and Multilinear Algebra {\bf 58}
(2010)
617-643
\par
\noindent
[K-O.13]:
Kamiya,N.,Okubo,S;
"Symmetry of Lie algebras 
associated with 
$(\varepsilon,\delta)$
-Freuden-
thal
-Kantor 
triple systems"
\  arXiv 1303,0072
(2013) to appear in
 Proc.
Edinburgh.
Math.
Soc.
\par
\noindent
[K-O.14]:
Kamiya,N.,Okubo,S.;
"Triality of Structurable and
pre-structurable Algebras"\ 
J.Algebras
{\bf 416}
(2014)
58-83
\par
\noindent
[K-M-P-T.98]:
Knus,M.A.,Merkurjev,A.S.,Post,M.,Tignal,J.P;
"The Book of Involution"\ 
American Math.Soc.Coll.Pub.
{\bf 44}
Providence
(1998)
\par
\noindent
[M-O.93]:
Meyberg,R.,Osborn,J.M.;
"Pseudo-composition Algebra"\ 
Math.Z.
\par
\noindent
{\bf 214}
(1993)
67-77
\par
\noindent
[O.95]:
Okubo,S.;
"Introduction to Octonion and
other Non-associative 
Algebras 
in Physics"\ 
Cembridge Univ.
press.
Cambridge
(1995)
\par
\noindent
[O.05]:
Okubo,S.;
"Symmetric triality relations and 
structurable algebra"\ 
Linear Algebras and
its Applications,
{\bf 396}
(2005)
189-222
\par
\noindent
[O.06]:
Okubo,S.;
"Algebras satisfying symmetric 
triality relations in Non-associative
Algebras and its Applications"\ 
ed.by L.Sabinin,
L.Sbitneva,
and 
I.P.Shestakov,
Chapman 
and Hall.
N.Y.
(2006)
313-321
\par
\noindent
[O-O.81]:
Okubo,S.,Osborn,J.M.;
"Algebras with non-degenerate associative symmetric 
bi-linear form
permitting compositions"\ 
Comm.Algebra
{\bf 9}
(1981)
(I) 1233-1261,
(II) 2015-2073
\par
\noindent
[Y-O.84]:
Yamaguti,K.,Ono,A;
"On representation of Freudenthal-Kantor 
triple system
$U(\varepsilon,\delta)$"\ 
Bull.Fac.School.
Ed.Hiroshima Univ.
Part II,
{\bf 7}
(1984)
43-51
\end{document}